\documentclass[seceq]{ptptex}

\usepackage{graphicx}
\makeatletter

\newcommand{\beq}{\begin{equation}}
\newcommand{\eeq}{\end{equation}}
\newcommand{\beqn}{\begin{eqnarray}}
\newcommand{\eeqn}{\end{eqnarray}}

\newcommand{\cB}{{\cal{B}}}

\newcommand{\pa}{\partial}

\newcommand{\Ptot}{P_{\rm tot}}
\newcommand{\varep}{\varepsilon}

\def\agt{\mathrel{\raise.3ex\hbox{$>$}\mkern-14mu\lower0.6ex\hbox{$\sim$}}}
\def\alt{\mathrel{\raise.3ex\hbox{$<$}\mkern-14mu\lower0.6ex\hbox{$\sim$}}}
\def\hbar{\mathrel{\raise.7ex\hbox{--}\mkern-14mu\hbox{$~h$}}}

\title{
Magnetohydrodynamics of Neutrino-cooled Accretion Tori
around a Rotating Black Hole in General Relativity %
}

\author{
Masaru \textsc{Shibata}, Yu-ichiro \textsc{Sekiguchi} %
 and Rohta \textsc{Takahashi}
}

\inst{
Graduate School of Arts and Sciences, University of Tokyo,
Komaba, Meguro, Tokyo 153-8902, Japan
}

\abst{ We present our first numerical results of axisymmetric
magnetohydrodynamic simulations for neutrino-cooled accretion tori
around rotating black holes in general relativity. We consider tori of
mass $\sim 0.1$--0.4$M_{\odot}$ around a black hole of mass
$M=4M_{\odot}$ and spin $a=0$--$0.9M$; such systems are candidates for
the central engines of gamma-ray bursts (GRBs) formed after the
collapse of massive rotating stellar cores and the merger of a black
hole and a neutron star.  In this paper, we consider the short-term
evolution of a torus for a duration of $\approx 60$ ms, focusing on
short-hard GRBs. Simulations were performed with a plausible
microphysical equation of state that takes into account
neutronization, the nuclear statistical equilibrium of a gas of free
nucleons and $\alpha$-particles, black body radiation, and a
relativistic Fermi gas (neutrinos, electrons, and
positrons). Neutrino-emission processes, such as $e^{\pm}$ capture
onto free nucleons, $e^{\pm}$ pair annihilation, plasmon decay, and
nucleon-nucleon bremsstrahlung are taken into account as cooling
processes.  Magnetic braking and the magnetorotational instability in
the accretion tori play a role in angular momentum redistribution,
which causes turbulent motion, resultant shock heating, and mass
accretion onto the black hole. The mass accretion rate is found to be
$\dot M_* \sim 1$--$10 M_{\odot}$/s, and the shock heating increases
the temperature to $\sim 10^{11}$ K. This results in a maximum
neutrino emission rate of $L_{\nu}=$ several $\times 10^{53}$ ergs/s
and a conversion efficiency $L_{\nu}/\dot M_* c^2$ on the order of a
few percent for tori with mass $M_{\rm t} \approx 0.1$--0.4$M_{\odot}$
and for moderately high black hole spins. These results are similar to
previous results in which the phenomenological $\alpha$-viscosity
prescription with the $\alpha$-parameter of $\alpha_v=0.01$--0.1 is
used. It is also found that the neutrino luminosity can be enhanced by
the black hole spin, in particular for large spins, i.e., $a \agt
0.75M$; if the accretion flow is optically thin with respect to
neutrinos, the conversion efficiency may be $\agt 10\%$ for $a \agt
0.9M$. Angular momentum transport, and the resulting shock heating
caused by magnetic stress induce time-varying neutrino luminosity,
which is a favorable property for explaining the variability of the
luminosity curve of GRBs. }

\begin{document}
\maketitle
\section{Introduction}

There is growing evidence that gamma-ray bursts (GRBs) occur at
cosmological distances. Assuming the isotropic emission of gamma-rays,
the estimated absolute luminosity of many events is $\agt 10^{51}$
ergs/s.\cite{GRB} If the effect of anisotropy, such as the collimation of
emission, is taken into account, the luminosity is estimated to be
$\sim 10^{50}$ ergs/s. Furthermore, the duration of the emission is
very short, $\sim 0.01$--1000 s, indicating that the source is composed
of a compact object of stellar-mass size.\cite{GRB} Thus, the high
luminosity can be explained by considering the conversion of energy from 
gravitational energy to thermal energy during accretion processes onto
the compact object. Because a rotating black hole is the most efficient
generator, it is now widely believed that the central engines of GRBs
are composed of stellar-mass rotating black holes and massive, compact,
and hot accretion disks (or tori).

Over the past decade, many groups have studied the properties of
dense, hot accretion disks (or tori) around a black hole [e.g.,
Refs. \citen{GRB1,GRB2,GRB3,KM,GRB3.5,GRB4,SRJ,LRP}]. An often
employed method of study is to derive a stationary solution of
accretion flow around a black hole, taking into account the detailed
microphysics, but neglecting the effects of the thickness of the
disks.  \cite{GRB1,GRB2,GRB3,KM,GRB3.5,GRB4} Such studies have
clarified qualitative and semi-quantitative features of the accretion
disks, which are found to be dense, hot, and geometrically thick with
maximum density $\sim 10^{12}~{\rm g/cm^3}$, maximum temperature $\sim
10^{11}$ K, and geometrical thickness $\sim 100$ km for a black hole
of mass $\sim 3$--10$M_{\odot}$. Because of the high density and the
high temperature, neutrino emission plays a crucial role in the
dissipation of thermal energy in such accretion flows, resulting in a
high neutrino luminosity of $\agt 10^{53}$ erg/s. One important
property of this dense, hot accretion flow is that the density is so
high that a large fraction of the neutrinos may be trapped by the flow
and fail to escape before being swallowed by the black hole; i.e., the
accretion flow is neutrino-dominated accretion flow (NDAF).
\cite{GRB1,GRB2,GRB3}

Although the stationary solutions reveal qualitative properties, these
studies are not suitable for understanding the time variability of
accretion flows.  In the last three years, numerical simulations of
viscous and neutrino-cooled accretion tori have also been carried out
with detailed microphysics by two groups. \cite{SRJ,LRP} Setiawan,
Ruffert, and Janka have performed three-dimensional (3D)
pseudo-Newtonian simulations with a tabulated equation of state (EOS)
derived by Lattimer and Swesty \cite{LS} and with a neutrino cooling
employing a leakage scheme. As a transport mechanism of angular
momentum in the flow, they incorporated a phenomenological viscosity
using the so-called $\alpha$-prescription. \cite{ShSu} A general
relativistic gravitational field is mimicked using a pseudo-Newtonian
potential. They find that the efficiency of the conversion to neutrinos
(defined by the ratio of the neutrino luminosity to the
rest-mass-energy accretion rate) reaches typically to a few percent
with maximum value of $\sim 10\%$. Lee, Ramirez-Ruiz, and Page
performed simulations similar to that of Setiawan et al., but with
the assumption of axial symmetry, with slightly simplified microphysics,
and for a longer time scale of $\approx 0.5$ s. They reported 
results similar to those of Setiawan et al.

These simulations have provided a variety of nonstationary properties
which cannot be found from the study of stationary accretion
disks. They have found that the mass accretion rate and neutrino
luminosity exhibit a mild time variability. Lee et al. found that the
neutrino luminosity decreases on a time scale of 10--1000 ms, depending
on the viscous parameter. They also reported that the inhomogeneity of
the lepton number induces turbulent motion through a convective
instability. These properties can be found only by numerical
simulation. This implies that numerical simulation is a better
approach for the study of GRB accretion flows. 

Despite their success, there are still many issues that have not yet
been clarified by simulations.  First, the simulations carried out to
this time take into account general relativistic effects very crudely,
using a pseudo-potential prescription. Although this method may partly
capture the nature of general relativistic gravity around black holes,
there is no guarantee that this potential can provide quantitatively
accurate values concerning properties of black holes. Moreover,
special relativistic effects are not incorporated in this method. The
speed of matter around black holes is often a significantly large
fraction of the speed of light, reaching a high Lorentz factor of
$\sim 10$, and hence it is reasonable to expect that the matter motion
in the vicinity of a black hole cannot be followed accurately with the
pseudo-potential approach. Second, they add a so-called
$\alpha$-viscosity in their equations of motion to take into account
effects of angular momentum transport and resultant viscous heating.
In the $\alpha$-viscosity prescription, one has to {\it a priori}
determine a value for the $\alpha$-parameter (which is hereafter
denoted by $\alpha_v$). This parameter essentially fixes the mass
accretion rate, the viscous heating rate, and the resulting neutrino
luminosity.  This implies that one never find these important
quantities correctly, due to the unknown value of the
$\alpha$-parameter.  Most people in the astrophysics currently believe
that the magnitude of the viscosity in accretion flows is effectively
determined by the turbulent motion of fluid induced by magnetic
stress. If this is indeed the case, it implies that the actual
effective viscosity varies in time. To capture the nature of a
realistic accretion flow, a magnetohydrodynamics (MHD) simulation is
needed, instead of adding an $\alpha$-viscosity.

With the motivation described above, we have performed a general
relativistic magnetohydrodynamics (GRMHD) simulation using a code
recently developed \cite{SS05}, which has already been applied to
study the evolution of magnetized differentially rotating neutron
stars \cite{DLSSS,DLSSS2} and magnetized stellar core
collapse.\cite{DLSSS3} In the present work, this code is used for a
fixed black hole geometry but with detailed microphysics; a realistic
EOS and neutrino cooling. The assumption of the fixed geometry is
valid because we consider systems for which the mass of the torus is
much smaller than the mass of the black hole. In such systems, the
self-gravity of the torus does not play an important role.  In the
present context, magnetic stress naturally induces time-varying
angular momentum transport and shock heating. The rest-mass accretion
rate and shock heating rate are computed in a first-principles 
manner. Furthermore, general relativistic effects are taken into
account in a strict manner. Thus, it is possible to clarify the
dependence of the neutrino luminosity in the GRB accretion torus on the
spin parameter of black holes for the first time in dynamical
simulations. In addition, we show that the magnetic effects induce a
large time variability of the mass accretion rate and neutrino luminosity,
which is not found in simulations employing the $\alpha$-viscosity.

This paper is organized as follows. In \S 2, we describe the basic
equations, EOS, and neutrino processes that we take into account. In
\S 3, the initial conditions and set-up for the simulation are
described. In \S 4, numerical results are presented, focusing
particularly on accretion rates of the rest-mass, energy, and angular
momentum onto the black hole horizon and on the neutrino
luminosity. Varying the masses of the torus and the black hole spin
systematically, we clarify the dependence of the accretion rates and
neutrino luminosity on these parameters. Section 5 is devoted to a
summary and discussion. Throughout this paper, we adopt the
geometrical units in which $G=c=1$ where $G$ and $c$ are the
gravitational constant and the speed of light.  $k$ and $\hbar$ are
the Boltzmann and Planck constants.  Latin and Greek indices denote
spatial components and spacetime components,
respectively. $\eta_{\mu\nu}$ and $\delta_{ij}(=\delta^{ij})$ are the
flat spacetime metric (in the cylindrical coordinates) and the
Kronecker delta, respectively.

\section{Procedures for the numerical simulation}

We numerically solved GRMHD equations in the fixed geometry of a
rotating black hole, assuming that the ideal MHD relation holds; the
conductivity is assumed to be infinite.  The basic equations and
numerical methods are the same as those used in our first
paper,\cite{SS05} except for the EOS and microphysics that we employ. In
this section, after we concisely review our basic equations for GRMHD,
a detailed description of the EOS and microphysics is given.

\subsection{Definition of variables}

The fundamental fluid variables are $\rho$, the rest-mass density, 
$\varep$, the specific internal energy, $P$, the pressure, and 
$u^{\mu}$, the four velocity. For convenience, we also define the weighted 
rest-mass density $\rho_*$, three-vector $v^i$, specific enthalpy $h$, 
and a Lorentz factor $w$ as 
\beqn
&&\rho_* \equiv \rho u^t \sqrt{-g/\eta},\\
&&v^i \equiv {dx^i \over dt}={u^i \over u^t},\\
&&h \equiv 1+\varep + {P \over \rho},\\
&&w \equiv \alpha u^t,
\eeqn
where $g$ is the determinant of the spacetime metric $g_{\mu\nu}$,
$\eta$ is the determinant of $\eta_{\mu\nu}$, and
$\alpha$ is the lapse function equal to $(-g^{tt})^{-1/2}$
(see \S \ref{sec:gr}). 

In addition, we need to define the electron fraction $Y_e$ and the
temperature $T$ because these are used as arguments for tabulated EOSs 
(see \S \ref{sec:eos} for details). Here, $Y_e$ is defined by
\beqn
Y_e \equiv {m_u n_e \over \rho},
\eeqn
where $n_e$ is the number density of electrons ($n_-$) minus
that of positrons ($n_+$), and $m_u$ is the atomic mass unit.

The only fundamental variable in the ideal MHD is $b^{\mu}$, a
four-vector of magnetic field, that is perpendicular to $u^{\mu}$,
i.e., $b^{\mu}u_{\mu}=0$. $b^{\mu}$ is regarded as the magnetic field
observed in a frame comoving with the fluid. In the ideal MHD, the
electric field $E^{\mu}$ in the comoving frame is zero and thus, the
electric current $j^{\mu}$ is not necessary to determine the evolution
of the field variables. Using $b_{\mu}$, the electromagnetic tensor
$F^{\mu\nu}$ is defined by \cite{MTW}
\beqn
F^{\mu\nu}=\epsilon^{\mu\nu\alpha\beta}u_{\alpha}b_{\beta}, \label{eqFF}
\eeqn
where $\epsilon_{\mu\nu\alpha\beta}$ is the Levi-Civita tensor. 

From $b^{\mu}$, we define the three-magnetic field as
\beqn
\cB^i \equiv \alpha \sqrt{\hat \gamma} (u^t b^i - b^t u^i), \label{defB}
\eeqn
where $\hat \gamma=\gamma/\eta$ and $\gamma$ is the determinant of the
three-metric $\gamma_{ij}$.  $\cB^i$ is regarded as the magnetic field
observed in an inertial frame. We note that $\cB^t=0$,
$\cB_i=\gamma_{ij} \cB^j$, and
\beqn
b^t = {\cB^{\mu} u_{\mu} \over \sqrt{-g/\eta}}~~{\rm and}~~
b_i={1 \over w \sqrt{\hat \gamma}}\Big(\cB_i + \cB^j u_j u_i \Big). 
\eeqn

Using the hydrodynamic and electromagnetic variables,
the energy-momentum tensor is written as
\beqn
T_{\mu\nu}=(\rho h + b^2) u_{\mu} u_{\nu}+\Big(P + {1 \over 2}b^2\Big)
g_{\mu\nu}-b_{\mu}b_{\nu}, \label{eq2.9}
\eeqn
where
\beqn
b^2=b_{\mu} b^{\mu}={\cB^2 + (\cB^i u_i)^2 \over w^2 (-g/\eta)}.
\eeqn
Then, the projection of the energy-momentum tensor is defined by 
\beqn
&& \rho_{\rm H} \equiv T_{\mu\nu} n^{\mu}n^{\nu}=(\rho h +b^2) w^2
- \Ptot- (\alpha b^t)^2, \\
&& J_i \equiv  - T_{\mu\nu} n^{\mu}\gamma^{\nu}_{~i}
=(\rho h +b^2) w u_i -\alpha b^t b_i, \\
&& S_{ij} \equiv T_{\mu\nu} \gamma^{\mu}_{~i}\gamma^{\nu}_{~j}
=(\rho h + b^2) u_i u_j + \Ptot \gamma_{ij}- b_i b_j, 
\eeqn
where $n^{\mu}$ is the unit timelike vector normal to a spatial
hypersurface, i.e., $\gamma_{\mu\nu}n^{\nu}=0$. Also we define
the total pressure $\Ptot \equiv P+P_{\rm mag}$, where $P_{\rm
mag}=b^2/2$ is the magnetic pressure. 

Using $\rho_{\rm H}$, $J_i$, and $S_{ij}$, 
the energy-momentum tensor is rewritten in the form
\beqn
T_{\mu\nu}=\rho_{\rm H} n_{\mu} n_{\nu}
+ J_i \gamma^i_{~\mu} n_{\nu}+ J_i \gamma^i_{~\nu} n_{\mu}
+S_{ij} \gamma^i_{~\mu}\gamma^j_{~\nu}. \label{31t}
\eeqn
This form of the energy-momentum tensor is useful for deriving the
basic equations for GRMHD presented in \S \ref{sec:grmhd}.
For the following treatment, we define
\beqn
S_0 \equiv \sqrt{\hat \gamma} \rho_{\rm H} ~~{\rm and}~~
S_i \equiv \sqrt{\hat \gamma} J_i.
\eeqn
We determined the evolution of these variables together with $\rho_*$,
$Y_e$, and $\cB^i$ in the numerical simulation (see \S
\ref{sec:grmhd}).

\subsection{Gravitational field}\label{sec:gr}

The GRMHD equations are solved in a fixed stationary gravitational
field of a Kerr black hole. Following Refs. \citen{GMT} and
\citen{MG}, we choose the Kerr-Schild coordinates of the Kerr black
hole, because they have no coordinate singularity on the event
horizon, and hence, regular solutions can be obtained even on the event
horizon and slightly inside it.

In cylindrical coordinates $(\varpi,z,\varphi)$, the line element of
the Kerr-Schild solution is \cite{HE}
\beqn
&&ds^2=-dt^2+d\varpi^2+\varpi^2 d\varphi^2 + dz^2 \nonumber \\
&&~~~~~~+{2Mr^3 \over r^4+a^2 z^2}
\Biggl({r\varpi d\varpi-a\varpi^2d\varphi \over r^2 + a^2}
+{z dz \over r}+dt\Biggr)^2,
\eeqn
where $r$ is the radial coordinate in the Boyer-Lindquist coordinates, 
which is the positive solution of the equation 
\beqn
r^4 -(\varpi^2+z^2-a^2)r^2-a^2 z^2=0. 
\eeqn
Here, $M$ and $a$ are the mass and the spin parameter of the Kerr black hole.
A physical singularity of ring shape is located at $\varpi=a$ and $z=0$, 
and the event horizon is at $r=M+\sqrt{M^2-a^2} \equiv r_{\rm H}$. 
We note that the determinant of the spacetime metric $g_{\mu\nu}$ has the 
simple form 
\beq
g=-\varpi^2. 
\eeq

For the Kerr-Schild solution, the lapse function $\alpha$,
the shift vector $\beta^i$, and the three-metric $\gamma_{ij}$ are
\beqn
&&\alpha=\sqrt{{f \over f+2Mr^3}},\\
&&\beta^{\varpi}={2 M r^4 \varpi \over \sigma(f+2Mr^3)},~
\beta^{z}={2Mr^2 z \varpi \over f+2Mr^3},~
\beta^{\varphi}=-{2 M a r^3  \over \sigma(f+2Mr^3)},\\
&&\gamma_{\varpi\varpi}=1+{2 M r^5 \varpi^2  \over f \sigma^2},~
\gamma_{\varpi\varphi}=-{2 M a r^4 \varpi^3  \over f \sigma^2},~
\gamma_{\varpi z}={2 M r^3 \varpi z  \over f \sigma},\nonumber \\
&&\gamma_{\varphi\varphi}=\varpi^2
\Big(1+{2 M a^2 r^3 \varpi^2  \over f \sigma^2}\Big),~
\gamma_{\varphi z}=-{2 M a r^2 z \varpi^2  \over f \sigma}, ~
\gamma_{z z}=1+{2 M r z^2  \over f}, 
\eeqn
where $f \equiv r^4+a^2z^2$ and $\sigma \equiv r^2 + a^2$.
The determinant of $\gamma_{ij}$ is 
\beq
\gamma=-{g \over \alpha^2}=\varpi^2 \Big(1+{2Mr^3 \over f}\Big). 
\eeq

\subsection{GRMHD equations} \label{sec:grmhd}

Hydrodynamic equations to be solved are\footnote{The equation
$\nabla_{\mu} T^{\mu}_{~\nu}=-Q_{\nu}$ is derived from the
conservation equation of the total energy-momentum tensor
$T^{\rm T}_{\mu\nu}$, which is defined by the sum of energy momenta
of all the matter fields. The derivation and assumption for the
derivation is described in Appendix A.}
\beqn
&&\nabla_{\mu} (\rho u^{\mu})=0, \label{hyeq1}\\
&&\gamma_{i}^{~\nu} \nabla_{\mu} T^{\mu}_{~\nu}=-Q_{\mu} \gamma_i^{~\mu},
\label{hyeq2}\\ 
&&n^{\nu} \nabla_{\mu} T^{\mu}_{~\nu}=-Q_{\mu}n^{\mu}. \label{hyeq3}
\eeqn
The first, second, and third equations are the continuity, Euler, and
energy equations, respectively. Here, $Q_{\mu}$ denotes a four-dimensional
vector associated with neutrino cooling which is defined below, and 
$\nabla_{\mu}$ is the covariant derivative with respect to
$g_{\mu\nu}$. In the following, the equations are described in the
cylindrical coordinates $(\varpi, z, \varphi)$ with the 
assumption of axial symmetry, and we define $S_y \equiv
S_{\varphi}/\varpi$, $u_y=u_{\varphi} /\varpi$, and $\cB^y \equiv
\cB^{\varphi} \varpi$.

With the quantities defined above,
Eqs. (\ref{hyeq1})--(\ref{hyeq3}) are written 
\beqn
&&\pa_t \rho_* + {1 \over \varpi}\pa_{\varpi} (\rho_* v^{\varpi} \varpi)
+\pa_z (\rho_* v^z)=0, \label{conti2d}\\
&&\pa_t S_A
+ {1 \over \varpi}\pa_{\varpi} \Big[ \varpi \Big\{S_A v^{\varpi}
+ \Ptot \delta_A^{~\varpi}
-{1 \over (u^t)^2} \cB^{\varpi}
\Big(\cB_A+ u_A \cB^i u_i\Big)\Big\}\Big]
\nonumber \\
&&~~~~~~~~~+ \pa_z \Big[ S_A v^z + \Ptot \delta_A^{~z}
-{1 \over (u^t)^2}\cB^z (\cB_A+ u_A \cB^i u_i)\Big]
\nonumber \\
&&~~~~~~~~~=-S_0 \pa_A \alpha + S_k \pa_A \beta^k 
-{1 \over 2} S_{ik} \pa_A \gamma^{ik} + {\Ptot \over \varpi}\delta_{A}^{~\varpi}
-Q_{A},
\\
&&\pa_t S_y + {1 \over \varpi^2}\pa_{\varpi} \Big[\varpi^2
\Big\{S_y v^{\varpi}-{1 \over (u^t)^2}
\cB^{\varpi} \Big(\cB_y + u_y \cB^i u_i \Big)\Big\}\Big] \nonumber \\
&&~~~~~~~~~
+\pa_z \Big[S_y v^z-{1 \over (u^t)^2}
\cB^z \Big(\cB_y + u_y \cB^i u_i \Big) \Big]=-Q_{y},\\
&&\pa_t S_0
+ {1 \over \varpi}\pa_{\varpi} \Big[\varpi\Big\{ S_0 v^{\varpi}
+ \sqrt{\hat \gamma}\Ptot (v^{\varpi} + \beta^{\varpi})
-{1 \over u^t \sqrt{\hat \gamma}}\cB^i u_i \cB^{\varpi} \Big\} \Big] \nonumber \\
&&~~~~~~~~~
+ \pa_z \Big[ S_0 v^z + \sqrt{\hat \gamma}\Ptot (v^z + \beta^z)
-{1 \over u^t \sqrt{\hat \gamma}}\cB^i u_i \cB^z \Big] \nonumber \\
&&~~~~~~~~~~=
S_{ij} K^{ij} -S_k D^k \alpha+Q_{\mu}n^{\mu}, \label{ene2d}
\eeqn
where we use $g=-\varpi^2$. The subscript $A$ denotes $\varpi$ or $z$, 
and $i, j, k, \cdots$ are $\varpi$, $z$ or $\varphi$. Also, 
$K_{ij}$ is the extrinsic curvature, which is calculated 
in the stationary space from
\beqn
K_{ij}={1 \over 2\alpha}\biggl(D_i \beta_j + D_j \beta_i\biggr),
\eeqn
where $D_i$ is the covariant derivative with respect to $\gamma_{ij}$. 
Equations (\ref{conti2d})--(\ref{ene2d}) are solved in the method 
described in Ref. \citen{SS05}; we employ a high-resolution scheme
consisting of the central scheme proposed by 
Kurganov and Tadmor \cite{KT} with third-order cell-reconstruction. 
\cite{lucas}

In addition to the hydrodynamic equations, we solve the evolution
equation for $Y_e$, 
\beqn
{d Y_e \over dt}=-\gamma_e, \label{eqye0}
\eeqn
where $\gamma_e$ is the capture rate of the electron
whose definition is given in \S \ref{sec:eos}.
Using the continuity equation, Eq. (\ref{eqye0}) can be rewritten as 
\beqn
\nabla_{\mu} (\rho Y_e u^{\mu})=-\rho u^t \gamma_e,
\eeqn
or, more explicitly, 
\beqn
\pa_t (\rho_* Y_e) +
{1 \over \varpi}\pa_{\varpi} (\rho_* Y_e v^{\varpi} \varpi)
+\pa_z (\rho_* Y_e v^z)=-\rho_* \gamma_e. \label{eqye}
\eeqn
We numerically solved Eq. (\ref{eqye}) in the same manner as 
Eq. (\ref{conti2d}). 

When the ideal MHD relation holds,  
the Maxwell equations for the axisymmetric system are 
\beqn
&& {1 \over \varpi} \pa_{\varpi}(\varpi \cB^{\varpi})
+ \pa_z \cB^z= 0, \label{mag2d1}\\
&& \pa_t  \cB^{\varpi} =  - \pa_z (\cB^{\varpi} v^z - \cB^z v^{\varpi}),
\label{mag2d2}\\ 
&& \pa_t  \cB^z = {1 \over \varpi}\pa_{\varpi} \Big[\varpi
(\cB^{\varpi} v^z - \cB^z v^{\varpi})\Big],\\ 
&& \pa_t  \cB^y =  \pa_{\varpi} (\cB^{\varpi} v^y - \cB^y v^{\varpi})
+\pa_z (\cB^z v^y - \cB^y v^z). \label{mag2d4}
\eeqn
Equation (\ref{mag2d1}) is the no-monopoles constraint, and
Eqs. (\ref{mag2d2})--(\ref{mag2d4}) are induction equations, 
which are solved to obtain the evolution of
the magnetic field in the same manner as in Ref. \citen{SS05}, 
using the constraint transport scheme \cite{EH} with a
second-order accurate interpolation. 

The validity of the numerical code for the GRMHD equations was
verified with several test simulations. The numerical results for standard
test problems in relativistic MHD, including special relativistic
magnetized shocks, general relativistic magnetized Bondi flow in
stationary spacetime, and a long-term evolution for a magnetized disk
in full general relativity are presented in Ref. \citen{SS05}.

After solving for the evolution of
$\rho_*$, $S_i$, $S_0$, and $Y_e$ together with $\cB^i$, we
have to determine the primitive variables, such as $\rho$, $\varep$,
$u_i$, and $u^t$ (or $w=\alpha u^t$). The first step in this procedure
is to derive the following equation from the definition of $S_i$: 
\beqn
s^2 \equiv \rho_*^{-2} \gamma^{ij} S_i S_j =(B^2 + h w )^2(1-w^{-2})
-D^2 (h w)^{-2}(B^2 + 2 h w). \label{newton1}
\eeqn
Here, $B^2$ and $D^2$ are determined from the variables
$(\rho_*, S_i, \cB^i)$ as 
\beqn
B^2 = {\cB^2 \over \rho_* \sqrt{\hat \gamma}}~~~{\rm and}~~~
D^2 = {(\cB^i S_i )^2 \over \rho_*^3 \sqrt{\hat\gamma}},
\eeqn
and to obtain Eq. (\ref{newton1}), 
we use the relation $S_i \cB^i =\rho_* h \cB^i u_i$. 
Equation (\ref{newton1}) is regarded as a function of $hw$ and $w^{-2}$
for given data set of $s^2$, $B^2$, and $D^2$.
The definition of $S_0$ can also be regarded as a function of
$hw$ and $w^{-2}$: 
\beqn
{S_0 \over \rho_*}=h w - {P \sqrt{\hat \gamma} \over \rho_*}+B^2
-{1 \over 2}\Big[B^2 w^{-2} + D^2 (hw)^{-2}\Big]. \label{newton2}
\eeqn
Thus, Eqs. (\ref{newton1}) and (\ref{newton2}) constitute
simultaneous equations for $hw$ and $w$ for given values of 
$\rho_*$, $S_i$, $S_0$, $\cB^i$, and $Y_e$ at each grid point.

In our previous works \cite{SS05,DLSSS,DLSSS2,DLSSS3} in which the
$\Gamma$-law or hybrid EOSs are used, a single algebraic equation for
$hw$ can be constituted from Eqs. (\ref{newton1}) and (\ref{newton2}). 
Then, a Newton-Raphson-type method can be used to obtain 
$hw$.  In this work, we use tabulated EOSs for obtaining 
$P$, $\varep$, and $h$ (these are functions of $\rho$, $Y_e$,
and $T$; see \S \ref{sec:eos} for details). In such a case, it is not
easy to apply the same method, because of the complexity of the 
EOS. For this reason, we adopt a different iteration method. 

At each time step, $\rho_*$ and $Y_e$ are determined from their
evolution equations. Then, $\rho$ should be computed from
$\rho_*/(\sqrt{\hat \gamma}w)$, but $w$ is the variable determined by
solving Eqs. (\ref{newton1}) and (\ref{newton2}). Thus, we guess a
solution for $w$ as a first step and calculate a trial solution for
$\rho$. For the trial value of $w$, we chose the value of the previous
time step. For the resulting values of $\rho$ with $Y_e$,
Eq. (\ref{newton2}) can be regarded as an algebraic equation for $T$,
regarding $h$ and $P$ as functions of $T$. Then, searching for the
solution from the table, we determine the value of $T$ and,
subsequently, $P$, $\varep$, and $h$.

In the iteration, we have to determine a new trial value of $w$.
For this purpose, 
we derive an algebraic equation for $w^{-2}$ from
Eqs. (\ref{newton1}) and (\ref{newton2}):
\beqn
&&s^2=\Big(1-{1 \over w^{2}}\Big)\Big({S_0 \over \rho_*}
+{P \sqrt{\hat \gamma} \over
\rho_*}+{B^2 + D^2 h^{-2} \over w^2}\Big)^2 \nonumber \\
&&~~~~~~-{D^2 \over h^2 w^2}\Big(-B^2+
{S_0 \over \rho_*}+{P \sqrt{\hat \gamma} \over
\rho_*}+{B^2 + D^2 h^{-2} \over w^2}\Big).
\eeqn
If $P$ and $h$ are regarded as given values, this equation
is a 3rd-order algebraic equation for $w^{-2}$. Thus, to obtain 
new trial value of $w$, we solve this equation using the Cardano 
formula.

We repeat these two procedures until a convergent solution is
obtained.  In most cases, the solution is obtained within $\sim 10$
iterations.  However, in some cases, the solution is not obtained. The
problem that often arises is that the solution for $T$ is not found in
the EOS table. This happens when $\varep$ accidentally decreases to a
small value for which the corresponding value of $T$ is absent in the
EOS table.\footnote{In the realistic EOSs for high-density matter with
$\rho \geq 10^7~{\rm g/cm^3}$, $\varepsilon$ is determined primarily
by the radiation pressure for the high-temperature case, whereas it is
determined by the electron-degenerate pressure for the low-temperature
case. This implies that $\varepsilon$ is proportional to $T^{1/3}$ at 
high temperature, whereas it approaches a constant for $T \ll
\mu_e/k$, where $\mu_e$ denotes the chemical potential of
electrons. Therefore, if $\varepsilon$ drops below this limiting
constant, no solution for $T$ is found.}  In such a case, we set
$T$ to a minimum value that we choose arbitrarily. In the present
work, we choose the minimum value to be $10^{9+2/3}$ K (see \S
\ref{sec:eos}).

\subsection{Equation of state}\label{sec:eos}

Dense, hot accretion tori of mass $\sim 0.1$--1$M_{\odot}$ around a
black hole of mass 3--4$M_{\odot}$ are likely outcomes formed after
the gravitational collapse of a massive stellar core \cite{GRB0,Proga,SS07}
and after the merger of a low-mass black hole and neutron star, 
\cite{BHNS0,BHNS1,SU06} as indicated by numerical simulations.
According to the numerical results, the maximum density and
temperature of the tori are likely to be $\sim 10^{12}~{\rm g/cm^3}$
and $T\sim 10^{11}$ K. This temperature is high enough to
photo-dissociate heavy nuclei, and hence, the main components of the
baryon should be free protons, free neutrons, and
$\alpha$-particles. Because of this high temperature, electrons are
relativistic and electron-positron pair production is possible in a
low-density region.  Furthermore, the degeneracy of electrons is high, 
due to the high density; the chemical potential of the electrons $\mu_e$ is
comparable to or larger than $kT$.  Photons are strongly coupled to
the charged particles, and hence, are trapped and advect with the matter
flow. Neutrinos are also trapped in the high-density region with $\rho
\agt 10^{11}~{\rm g/cm^3}$, whereas, for the low-density region, they 
escape freely from the tori and this contributes to the cooling process.

We determine the EOS, considering the physical conditions mentioned
above. Our approach is similar to that of Ref. \citen{LRP}. We 
assume that the torus is composed of free protons, free neutrons,
$\alpha$-particles, electrons, positrons, neutrinos, and
radiation. Then, the pressure is written
\beqn
P=P_e + P_g + P_r + P_{\nu},
\eeqn
where $P_e$, $P_g$, $P_r$, and $P_{\nu}$ are the pressures of 
electrons and positrons, of baryons, of the radiation, and of neutrinos. 

In the following, we assume that the temperature is high  
enough ($T \agt 10^{10}$ K) that the electrons and positrons are
relativistic (i.e., $kT \gg m_e c^2$, where $m_e$ is the electron mass).
Then, the total pressure of the electrons and positrons can be
determined analytically: \cite{Bli}
\beqn
P_e = {(kT)^4 \over 12\pi^2 (\hbar c)^3} \biggl[
\eta_e^4 + 2\pi^2 \eta_e^2 +{7\pi^4 \over 15}\biggr], \label{prese}
\eeqn
where $\eta_e \equiv \mu_e/kT$.\footnote{We reintroduce $G$ and $c$
in \S.\ref{sec:eos} and \S \ref{sec:neu} to clarify the dimensional units.}
The number densities of electrons and positrons, $n_-$ and $n_+$,
are related so as to give an electron fraction $Y_e$ of 
\beqn
{\rho Y_e \over m_u}=n_- - n_+={(kT)^3 \over 3\pi^2(\hbar c)^3}
\Big[\eta_e^3 + \eta_e \pi^2 \Big]. \label{nume}
\eeqn
We note that the assumption $kT \gg m_e c^2$ is not always valid, 
but for the dense region of the torus in which we are interested,
it holds approximately.\cite{LRP}  Equations (\ref{prese}) and
(\ref{nume}) hold for arbitrary degeneracy as long as the temperature
is high enough. 

The gas and radiation pressure are written
\beqn
&& P_g = {\rho kT \over m_u}{1 + 3X_{\rm nuc} \over 4},\\
&& P_r = {a_r T^4 \over 3},
\eeqn
where $X_{\rm nuc}$ is the mass fraction of free nucleons and $a_r$
the radiation density constant ($\pi^2 k^4/15(\hbar c)^3$).
If the neutrinos of all species are in thermal equilibrium with the
matter, the neutrino pressure is given by
\beqn
P_{\nu}={7 \over 8} a_r T^4,
\eeqn
whereas the pressure is negligible if they are transparent. In this
paper, we follow the treatment of Ref. \citen{LRP} and set
\beqn
P_{\nu}={7 \over 8} a_r T^4 (1-e^{-\tau_{\nu}}),
\eeqn
where $\tau_{\nu}$ is the averaged optical depth of the neutrinos.

In the present context, the ratio of the radiation pressure to the
gas pressure is
\beqn
{P_r \over P_g}=3.0 \times 10^{-2} T_{11}^3 \rho_{12}^{-1}, 
\eeqn
where $T_{11}=T/10^{11}$ K, $\rho_{12}=\rho/10^{12}~{\rm g/cm^3}$, and
we assume that $X_{\rm nuc}=1$. For the case that the electron is degenerate
with $\eta_e \sim 10$, the ratio of the electron pressure to the gas
pressure is
\beqn
{P_e \over P_g} \approx 1.2 \times 10^{1} T_{11}^3 \rho_{12}^{-1}
\Big({\eta_e \over 10}\Big)^4 .
\eeqn
Thus, in the main body of the torus with $T_{11}=O(1)$ and $\rho_{12}=O(1)$,
the gas pressure is the dominant pressure source if the electrons are mildly
degenerate with $\eta_e \alt 5$, whereas for the case that the electrons 
are strongly degenerate, the degenerate pressure is dominant. 

The specific internal energy is written
\beqn
\varep=\varep_e + \varep_g + \varep_r + \varep_{\nu},
\eeqn
where each term is related to the pressure by 
\beqn
\varep_e = {3 P_e \over \rho},~~
\varep_g = {3 P_g \over 2\rho},~~
\varep_r = {3 P_r \over \rho},~~
\varep_{\nu} = {3 P_{\nu} \over \rho}.
\eeqn

The pressure and specific internal energy should be provided for given
values of $\rho$, $T$, and $Y_e$.\footnote{In Ref. \citen{LRP}, the
authors made the additional assumption that $Y_e$ is a function of
$\rho$ and $T$.  However, we do not make this assumption, as it does
not always hold in the optically thin region for dynamical systems.}
For this purpose, the chemical potential of electrons, $\mu_e$, and
the free nucleon fraction, $X_{\rm nuc}$, have to be written as
functions of $(\rho, T, Y_e)$.  Equation (\ref{nume}) is used to
obtain the functional form of the chemical potential.  Then, 
for simplicity, we use a fitting formula \cite{QW} for the mass
fraction of the free nucleon, 
\beqn
X_{\rm nuc}={\rm Min}
\Big(22.4 T_{10}^{9/8} \rho_{10}^{-3/4} \exp(-8.2 / T_{10}),1 \Big),
\label{Xnuc}
\eeqn
where $T_{10}=T/10^{10}$ K and $\rho_{10}=\rho/10^{10}~{\rm g/cm^3}$.

We tabulated an EOS table for the ranges $\rho_{\rm min} \leq \rho
\leq 10^{13}~{\rm g/cm^3}$ and $T_{\rm min} \leq T \leq 5 \times
10^{11}~{\rm K}$. The values $\rho_{\rm min}$ and $T_{\rm min}$ are
the minimum values that we arbitrarily chose to maintain the stability
of the numerical computation.  Because the temperature should not be much
smaller than $m_e c^2/k$, because of our assumption that the electrons
are relativistic, we set the value of $T_{\rm min}$ to $10^{9+2/3}$ 
K. The value of $\rho_{\rm min}$ is specified in \S \ref{sec:atmo}.

\subsection{Neutrino emission}\label{sec:neu}

Several processes contribute to the emission of neutrinos.
The most important ones in the present context are the
electron and positron captures onto free nucleons: 
\beqn
p + e^- \rightarrow n + \nu_e, \label{pennu1}\\
n + e^+ \rightarrow p + \bar \nu_e.\label{pennu2} 
\eeqn
These processes change the electron fraction. Assuming that
the electrons and positrons are relativistic, the electron capture rate 
in the optically thin region for neutrinos is approximately given by
(see Appendix B)
\beqn
\gamma_{e0} = K_c \biggl({kT \over m_e c^2}\biggr)^5
[F_{4-}(\eta_e)X_p - F_{4+}(\eta_e)X_n], \label{ecap1}
\eeqn
where $X_p$ and $X_n$ are the mass fractions of free protons and
free neutrons, $K_c=\ln 2/10^{3.035}~{\rm s}^{-1}$, and 
\beqn
&&F_{4-}(x)=45.59x+{2\pi^2 \over 3}x^3+{1 \over 5}x^5
+24\biggl(e^{-x}-{1 \over 32}e^{-2x}+{1 \over 243}e^{-3x}\biggr),\\
&&F_{4+}(x)=24\biggl(e^{-x}-{1 \over 32}e^{-2x}+{1 \over 243}e^{-3x}\biggr).
\eeqn
In the optically thick region for neutrinos, $\beta$-equilibrium
is assumed to be satisfied and the electron fraction does not change. 
In this paper, we simply multiply by $e^{-\tau_{\nu}}$ in order 
to take into account the opacity effect (diffusion effect) 
and define the electron capture
rate by 
\beq
\gamma_e=\gamma_{e0}e^{-\tau_{\nu}}. 
\eeq
We note that the actual optical depth depends on the neutrino
species and neutrino energy. In this paper, we ignore such a
dependence. 

As a result of the electron and positron captures, neutrinos are
emitted and they carry away thermal energy of the rate
\beqn
\dot Q_{\rm cap}=K_c {m_e \over m_u} \rho c^2 
\biggl({kT \over m_e c^2}\biggr)^6
[F_{5-}(\eta_e)X_p + F_{5+}(\eta_e)X_n], \label{ecap2}
\eeqn
where
\beqn
&&F_{5-}(x)=236.65 + {7\pi \over 6}x^2 + {5 \pi^2 \over 6}x^4
+{1 \over 6} x^6 \nonumber \\
&& ~~~~-120\biggl(e^{-x}-{1 \over 64}e^{-2x}+{1 \over 729}e^{-3x}\biggr),\\
&&F_{5+}(x)=120\biggl(e^{-x}-{1 \over 64}e^{-2x}+{1 \over 729}e^{-3x}\biggr).
\eeqn

In addition to electron and positron captures,  
we consider the annihilation of electron-positron pairs into
thermal neutrinos, nucleon-nucleon bremsstrahlung, and plasmon decay. 
For the neutrino emissivity due to electron-positron pair annihilation,
we use the fitting formula derived by Itoh et al. \cite{Itoh} 
For the other two, we write the emissivity as 
[e.g., Refs. \citen{KM,Itoh}, and \citen{RJ}]
\beqn
&& \dot Q_{\rm ff}=1.5 \times 10^{31}  T_{11}^{5.5} \rho_{12}^2
~~{\rm ergs/cm^3/s},\label{ffnu}\\
&& \dot Q_{\rm pla}=1.5 \times 10^{32}  T_{11}^{9} \gamma_p
e^{-\gamma_p}(2+2\gamma_p+\gamma_p^2)~~{\rm ergs/cm^3/s},\label
{planu}
\eeqn
where $\gamma_p=5.565\times 10^{-2} \sqrt{(\pi^2+3\eta_e^2)/3}$.\cite{RJ}
Then, we define the total emissivity as 
\beqn
\dot Q = (\dot Q_{\rm cap}+\dot Q_{\rm pair} + \dot Q_{\rm ff}
+\dot Q_{\rm pla}) e^{-\tau_{\nu}}. 
\eeqn
Because $\dot Q$ is the emissivity (cooling rate) measured in the
{\em fluid rest-frame}, we define $Q_{\mu}$ as \cite{Miha,AS}
\beqn
Q_{\mu}=\dot Q u_{\mu}. 
\eeqn
With this treatment, the effect of anisotropic emission associated with
the matter motion is taken into account.\footnote{As a result of
the anisotropic emission, angular momentum is dissipated, but the
magnitude of this effect is much smaller than the loss associated
with the infall into the black hole.} 
On the other hand, the effects of radiation transfer and
the heating due to the emitted neutrinos are ignored. 

The main source of opacity for neutrinos is scattering by free
nucleons and $\alpha$-particles. For these processes,
the order of the cross section is $\sim 10^{-42} T_{11}^2~{\rm cm^2}$.
\cite{ST}  Because the number
density of free nucleons is $\sim 10^{35}\rho_{11}~{\rm cm^{-3}}$, the
mean free path of the neutrinos is roughly
\beqn
\sim 10^7 ~T_{11}^{-2}\rho_{11}^{-1} ~{\rm cm},
\eeqn
which is $\sim 20GM/c^2$ for black holes of mass $4M_{\odot}$.  In
this work, we consider accretion tori of width and thickness $\sim
20$--$30GM/c^2$ and maximum temperature $\sim 10^{11}$ K. Hence,
we simply set the optical depth to 
\beq
\tau_{\nu}=\zeta \rho_{11},
\eeq
where $\zeta$ is a constant that we set to 1 or 1/3 [i.e.,
$\tau_{\nu}=\rho/(10^{11}~{\rm g/cm^3})$ or $\rho/(3\times
10^{11}~{\rm g/cm^3})$].  Previous works investigating dense accretion
tori (e.g., Ref. \citen{LRP}) shows that the neutrinos are optically thick
when the density satisfies the condition $\rho_{11} \agt 1$. Thus, the
present treatment can capture the neutrino-trapping effect, at least
qualitatively. 


\subsection{Diagnostics}

We monitor the total baryon rest-mass $M_*$, angular momentum $J$,
internal energy $E_{\rm int}$, rotational kinetic energy $T_{\rm
rot}$, and electromagnetic energy $E_{\rm EM}$, which are defined by
\begin{eqnarray}
&&M_*= \int \rho_* \sqrt{\eta} d^3x, \\
&&J = \int \rho_* h u_{\varphi} \sqrt{\eta} d^3 x, \\
&&E_{\rm int} = \int \rho_* \epsilon \sqrt{\eta} d^3x , \\
&&T_{\rm rot} = \int {1\over 2} \rho_* h \Omega u^t u_{\varphi}
\sqrt{\eta} d^3x, \label{eq:Trot} \\
&&E_{\rm EM}  = \int T_{\rm EM}^{t t} \alpha \sqrt{\eta} d^3x,
\end{eqnarray}
where $T_{\rm EM}^{\mu\nu}$ is the electromagnetic part of the
energy-momentum tensor, $\Omega$ is the angular velocity
defined by $u^{\varphi}/u^t$, 
and all the integrations are performed 
outside the event horizon. For the definitions of 
$E_{\rm int}$, $T_{\rm rot}$, and $E_{\rm EM}$, we follow Ref.
\citen{DLSSS2}. 

In stationary axisymmetric spacetime, the following
relations are derived from the conservation law of the
energy-momentum tensor in the absence of neutrino cooling: 
\beqn
\pa_{\mu} \sqrt{-g} T^{\mu}_{~t}=0,\label{tmunut}\\
\pa_{\mu} \sqrt{-g} T^{\mu}_{~\varphi}=0. \label{tmunup} 
\eeqn
From these relations, it is natural to define the energy and angular
momentum accretion rates by the surface integral at the event horizon as
\beqn
&&\dot E =\oint_{r=r_{\rm H}} T^r_{~t} \sqrt{-g} dS,\\
&&\dot J =-\oint_{r=r_{\rm H}} T^r_{~\varphi} \sqrt{-g} dS,
\eeqn
where $dS=d\theta d\varphi$. 
From the continuity equation, the rest-mass accretion rate is defined in the 
same manner as
\beqn
&&\dot M_*=-\oint_{r=r_{\rm H}} \rho_* v^r r_{\rm H}^2 dS. 
\eeqn
Because we adopt cylindrical coordinates in the Kerr-Schild
coordinates, grid points are not located at $r=r_{\rm H}$. To obtain the
values there, we use linear interpolations for all the necessary
quantities.

In the presence of neutrino cooling, Eq. (\ref{tmunut}) is written
\beqn
\pa_{\mu} \sqrt{-g} T^{\mu}_{~t}=-\sqrt{-g} \dot Q u_t. 
\eeqn
Thus, the rate of energy loss by the neutrinos is given by
\beqn
L_{\nu}=-\int_{r>r_{\rm H}} \sqrt{-g} u_t \dot Q d^3x. \label{lumi}
\eeqn
We refer to $L_{\nu}$ simply as the neutrino luminosity. Strictly speaking, 
$L_{\nu}$ is the sum of the energy emission rates of neutrinos toward
infinity and toward the black hole horizon. Thus, the luminosity observed
at infinity is smaller than this value, because a finite fraction of
neutrinos emitted at each radius is always swallowed by the black
hole. To roughly infer how much neutrino energy is likely to be
swallowed by black holes, we also compute the quantity 
\beqn
L_{\nu}(r_{\rm ph})
=-\int_{r>r_{\rm ph}} \sqrt{-g} u_t \dot Q d^3x. \label{lumi2}
\eeqn
where $r_{\rm ph}$ denotes the radius of the limiting circular
photon-orbit:\cite{ST}
\beqn
r_{\rm ph}=2M \Big[1+\cos\Big\{{2\over 3}\cos^{-1}(-a/M)\Big\}\Big]. 
\eeqn
This radius is characterized by the fact that 50\% of massless
particles from a stationary isotropic emitter at $r=r_{\rm ph}$ is
captured by the black hole \cite{ST}. To compute the luminosity more
accurately, it is necessary to multiply by an escape probability that
depends on the position and velocity of the emitter. In the present paper,
we do not take into account such probability but simply calculate the
luminosity using Eq. (\ref{lumi}) or (\ref{lumi2}). 


From $\dot M_*$ and $L_{\nu}$, the total rest-mass swallowed
by a black hole and the emitted neutrino energy are defined by
\beqn
\Delta M_*(t) = \int_0^{t} \dot M_* dt',\\
\Delta E_{\nu}(t) = \int_0^{t} L_{\nu} dt'.
\eeqn
In the following, we refer to $\Delta E_{\nu}/\Delta M_*$ as
the average energy conversion efficiency to neutrinos. 

\section{Initial conditions and setting for computation}

\subsection{Initial condition for torus}

As the initial conditions, we first prepare equilibrium states of a
torus rotating around a black hole with no magnetic field and with no
neutrino cooling.  Such equilibria are determined from the first
integral of the Euler equation, which has already been derived for
axisymmetric stationary spacetime.\cite{FM,AJS} We adopt the
prescription of Ref. \citen{AJS} because of its simplicity. In
Kerr-Schild coordinates, the derivation of the basic equations is
slightly more complicated than in Boyer-Lindquist coordinates.  For
example, in Boyer-Lindquist coordinates, we have $u_{\varpi}=u_{z}=0$,
whereas they are nonzero in Kerr-Schild coordinates. We have to be
careful regarding this point in setting the initial conditions. For
completeness, here, we describe the basic equations for obtaining
equilibria.

First, we assume that the four-velocity has the following components:
\beqn
u^{\mu}=(u^t, 0, 0, u^{\varphi}).
\eeqn
Note that this does not imply that $u_{\varpi}$ and $u_{z}$ are zero, 
because of the presence of nonzero off-diagonal components of $g_{\mu\nu}$.
Defining the angular velocity $\Omega \equiv u^{\varphi}/u^t$, the
Euler equation in the stationary axisymmetric system is written in the
well-known form
\beqn
{\pa_k u^t \over u^t} - u^t u_{\varphi}\pa_k \Omega - {\pa_k P \over \rho h}
=0. \label{firsteuler}
\eeqn
In the method of Ref. \citen{AJS}, one assumes
\beqn
\ell=-{u_{\varphi} \over u_t}={\rm const}.
\eeqn
Then, Eq. (\ref{firsteuler}) can be rewritten as 
\beqn
(\pa_k u_t) u^t(1 - \ell \Omega) - {\pa_k P \over \rho h}
=0. \label{firsteuler2}
\eeqn
Now, $u^t u_t + u^{\varphi}u_{\varphi}=-1$
gives $u^t (1-\ell\Omega)=-1/u_t$.
Thus, we obtain
\beqn
\pa_k \ln(-u_t) + {\pa_k P \over \rho h}=0. \label{firsteuler3}
\eeqn
If we assume that the fluid is isentropic, the first law of
thermodynamics is used to rewrite the second term
as $\pa_k \ln(h)$. Thus, we obtain the first integral
of the Euler equation in the same form as that in Ref. \citen{AJS}, 
\beqn
h u_t = C, \label{huc}
\eeqn
where $C$ is an integration constant. Then, using the relation
\beqn
\Omega=-{g_{tt}\ell + g_{t\varphi} \over g_{t\varphi}\ell
+ g_{\varphi\varphi}},
\eeqn
we obtain
\beqn
u_t =-\biggl({g_{t\varphi}^2-g_{tt}g_{\varphi\varphi} \over
g_{\varphi\varphi}+2g_{t\varphi}\ell+g_{tt}\ell^2}\biggr)^{1/2}. \label{ut}
\eeqn
Hence, $h$ at each point is determined from Eq. (\ref{huc}) for given values 
of $\ell$ and $C$. 

Specifically, the configuration of tori is determined by fixing the
inner and outer edges on the equatorial plane. We denote the
cylindrical coordinate radii of these edges by $\varpi_1$ and
$\varpi_2~(>\varpi_1)$. The condition $h=1$ at $\varpi=\varpi_1$ and
$\varpi_2$ determines the values of $C$ and $\ell$ using
Eqs. (\ref{huc}) and (\ref{ut}). Subsequently, the profile for $h$ is
determined from Eq. (\ref{huc}).

To derive $\rho$ and $P$ from $h$, the EOS table is used. In doing so,
we have to further assume relations among $\rho$, $Y_e$, and $T$, because
$h$ is a function of these three arguments in the table. In the
present paper, we assume that $T=10^{10}$~K uniformly in all 
regions.  Because the tori and its atmosphere formed after stellar core
collapse and the merger of compact objects are likely to be moderately hot
at their formation, this assumption is reasonable. To determine 
$Y_e$, we use the prescription of Ref. \citen{LRP}: For optically
thick regions, we assume that $\beta$-equilibrium holds, i.e.,
\beqn
p + e^- \leftrightarrow n + \nu_e,
\eeqn
and hence, the ratio of the proton fraction to the neutron fraction is assumed
to be
\beqn
{X_p \over X_n}=e^{(Q-\mu_e)/kT},
\eeqn
where $Q=1.293$ MeV is the difference between the neutron and proton
masses. Here, we assume that the chemical potential of neutrinos is
zero for simplicity.

For optically thin regions, we assume that the reaction rates of the
processes given in Eqs. (\ref{pennu1}) and (\ref{pennu2}) are
identical and that the number densities of protons and neutrons are
unchanged.\cite{Belo} Under these assumptions, the electron fraction
is approximately given by [see Eq. (12) of Ref. \citen{LRP}]
\beqn
&&Y_e = {1 - X_{\rm nuc} \over 2} +X_{\rm nuc}
\biggl[ \biggl( {1 \over 2}+0.487 {Q/2-\mu_e \over kT}
\biggr) e^{-\tau_{\nu}} \nonumber \\
&&~~~~~~~~~~~~~~~~~~~~~~~~~~~~~
+ {1 \over 1+e^{(\mu_e-Q)/kT}} (1-e^{-\tau_{\nu}})
\biggr], 
\eeqn
where for $X_{\rm nuc}$, Eq. (\ref{Xnuc}) is used. 
In this rule with $T_{10}=1$, $Y_e$ is smaller than 0.5 for
high-density regions with $\rho \agt 10^{11}~{\rm g/cm^3}$, whereas
for low-density regions, it is larger than 0.5. 

The initial conditions obtained with the present method are not isentropic
although we assume isentropic conditions in deriving Eq. (\ref{huc}).
Thus, the torus is not exactly in equilibrium. However, in the test
simulations with no magnetic field and no neutrino cooling, we found
that the torus approximately remains in the initial state even after
the density maximum orbits the black hole for 10 rotational periods. 
With this test, we confirmed that the initial conditions adopted in this
method are approximately in equilibrium.

There are several free parameters for determining the initial
conditions: The black hole mass, $M$, the black hole spin, $a$, the
mass of the torus, $M_*$, and the density profile of the torus. We fix
the black hole mass to $4M_{\odot}$.  Such a black hole is a plausible
outcome, forming soon after the stellar core collapse of massive star
and after the merger of black hole-neutron star binaries. Because we
do not know a plausible value of the black hole spin, numerical
computations were systematically performed for a wide range of $a$. A
plausible mass of a torus formed in the vicinity of a black hole is a
few tenths of $M_{\odot}$. In the present work, we choose $M_* \approx
0.1$--$0.4M_{\odot}$.  Because we consider compact tori rotating
around a black hole, the inner edge of the torus is chosen to be
slightly outside of the radius of the innermost stable circular orbit
(ISCO) as $\varpi_1 = 5$--$7M$. Then, for a fixed value of $M_*$, the
location of the outer edge is automatically determined to be $\approx
30$--40$M$ for an $\ell=$constant angular momentum profile.

In Table I, we list parameter values for selected models adopted in
the numerical simulation. For models A--E, the torus mass is
approximately given by $M_* \approx 0.25M_{\odot}$, whereas the black
hole spins are all different. For models F--J, the black hole spin is
fixed to $a/M=0.75$, whereas the mass and inner edges of the torus are
different.  In Table II, we list the parameters for a test particle
orbiting a black hole at the ISCO for several values of $a/M$. These
are key quantities for determining the neutrino luminosity and the
conversion efficiency of the rest-mass energy to neutrinos (see \S
\ref{sec:3} for discussion).

\begin{table}[t]
\caption{Parameter set for the adopted models. The spin parameter of
the black hole, the rest-mass of tori, the coordinate radius of the
inner and outer edges, $\varpi_1$ and $\varpi_2$, of tori, the maximum
density of tori, $E_{\rm int}/M_*$, $T_{\rm rot}/M_*$, $J/M_*$, and
the rotation period at the density maximum, $P_c$, of tori.  For every
model, the black hole mass is fixed to be $M=4M_{\odot}$, and the ratio
of the electromagnetic energy to $E_{\rm int}$ is 1/200. $\varpi_1$
and $\varpi_2$ are given in units of $M$, and $E_{\rm int}$ and $T_{\rm
rot}$ are in units of $M_*/100$.}
\begin{center}
\begin{tabular}{cccccccccc}
\hline \hline
Model & $a/M$ & $M_*~(M_{\odot})$
& $\varpi_1,\varpi_2$ & $\rho_{\rm max}$(g/cm$^{3}$)
& $E_{\rm int}$ & $T_{\rm rot}$ & $J/M_*$ & $P_c/M$ & $\zeta$
\\ \hline
A & 0    & 0.244  & 6.4,~41 & $3.5 \times 10^{11}$ &
$0.85$ & $3.3$ & $4.03$ & 248 & 1 \\
A2& 0    & 0.248  & 6.4,~41 & $3.6 \times 10^{11}$ &
$0.85$ & $3.3$ & $4.03$ & 248 & 1/3 \\
B & 0.25 & 0.240  & 6  ,~36 & $4.7 \times 10^{11}$ &
$0.94$ & $3.7$ & $3.82$ & 219 & 1 \\
B2& 0.25 & 0.243  & 6  ,~36 & $4.7 \times 10^{11}$ &
$0.93$ & $3.7$ & $3.82$ & 219 & 1/3 \\
C & 0.5  & 0.249  & 6  ,~34 & $5.4 \times 10^{11}$ &
$0.99$ & $3.7$ & $3.70$ & 207 & 1 \\
C2& 0.5  & 0.253  & 6  ,~34 & $5.5 \times 10^{11}$ &
$0.99$ & $3.7$ & $3.69$ & 207 & 1/3 \\
D & 0.75 & 0.247  & 6  ,~32 & $6.1 \times 10^{11}$ &
$1.0$ & $3.8$ & $3.58$ & 203 & 1 \\
D2 & 0.75 & 0.251  & 6 ,~32 & $6.1 \times 10^{11}$ &
$1.0$ & $3.8$ & $3.58$ & 203 & 1/3 \\
E & 0.9  & 0.247  & 6  ,~31 & $6.4 \times 10^{11}$ &
$1.1$ & $3.8$ & $3.51$ & 197 & 1 \\
E2& 0.9  & 0.251  & 6  ,~31 & $6.5 \times 10^{11}$ &
$1.1$ & $3.8$ & $3.51$ & 197 & 1/3  \\\hline
F & 0.75 & 0.237  & 4.8 ,~24.6 & $1.2\times 10^{12}$ &
$1.3$ & $4.9$ & $3.27$ & 146 & 1 \\
G & 0.75 & 0.132  & 6.9 ,~33 & $2.8\times 10^{11}$ &
$0.82$ & $3.5$ & $3.74$ & 241 & 1 \\
H & 0.75 & 0.146  & 6.6,~31.8 & $3.4\times 10^{11}$ &
$0.87$ & $3.6$ & $3.68$ & 227 & 1 \\
I & 0.75 & 0.366  & 5.4,~31.2 & $1.0\times 10^{12}$ &
$1.2$ & $4.1$ & $3.45$ & 173 & 1 \\
J & 0.75 & 0.397  & 6.0,~36 & $7.8\times 10^{12}$ &
$1.1$ & $3.5$ & $3.61$ & 209 & 1 \\
\hline
\end{tabular}
\end{center}
\end{table}

\begin{table}[t]
\caption{Quantities for a test particle orbiting a black hole at the
ISCO for various values of the black hole spin. The value of the $r$
coordinate of the ISCO, $r_{\rm ISCO}$, the specific energy at the
ISCO, $E_{\rm ISCO}$, and the specific angular momentum at ISCO,
$J_{\rm ISCO}$.  The fifth and sixth columns list $r_{\rm H}$ and $r_{\rm
ph}$.  Note that the $\varpi$ coordinate on the equatorial plane is
related to $r$ by $\varpi=\sqrt{r^2+a^2}$.  In this paper, we refer to
$1-E_{\rm ISCO}$ as the specific gravitational binding energy at the
ISCO. }
\begin{center}
\begin{tabular}{cccccc}
\hline \hline
$a/M$ & $r_{\rm ISCO}/M$ & $E_{\rm ISCO}$ & $J_{\rm ISCO}/M$
& $r_{\rm H}/M$ & $r_{\rm ph}/M$ 
\\ \hline
0    & 6.000 & 0.9428 & 3.464 & 2.000 & 3.000 \\
0.25 & 5.156 & 0.9331 & 3.221 & 1.968 & 2.695 \\
0.50 & 4.233 & 0.9179 & 2.917 & 1.866 & 2.347 \\
0.75 & 3.158 & 0.8882 & 2.489 & 1.661 & 1.916 \\
0.90 & 2.321 & 0.8442 & 2.100 & 1.436 & 1.558 \\
\hline
\end{tabular}
\end{center}
\end{table}

An important feature found from Table I is that for (approximately)
fixed values of $M$, $M_*$, and the radius of the torus, the maximum
density is higher for larger values of $a$ (compare models A--E). It
is also found that for fixed values of $M$, $M_*$, and $a$, the
maximum density is higher for smaller values of the rotation radius of
the torus. For larger values of $a$, the radius of the ISCO on the
equatorial plane is smaller (see Table II), and hence, the torus can
be more compact, thereby increasing the maximum density. All these
features imply that for larger values of $a$, the density can be
higher. Because the optical depth of neutrinos depends strongly on the
density for $\rho \agt 10^{11}~{\rm g/cm^3}$, this dependence of the
maximum density on the black hole spin plays an important role in
determining the neutrino luminosity.

\begin{figure}[t]
\vspace{-13mm}
\begin{center}
\begin{minipage}[t]{0.47\textwidth}
(a)\hspace{-8mm}\includegraphics[width=8.5cm]{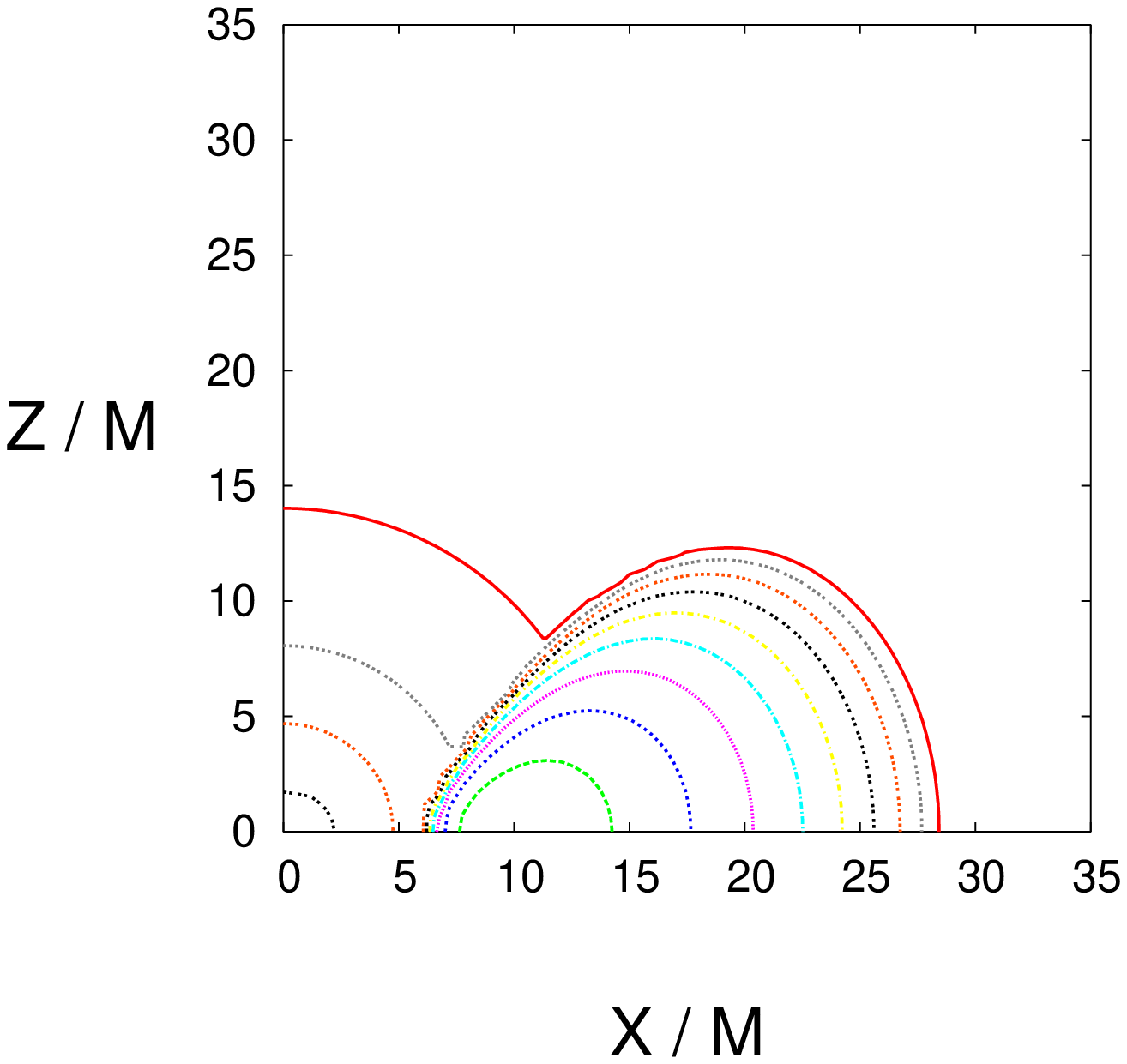}
\end{minipage}
\begin{minipage}[t]{0.47\textwidth}
(b)\hspace{-8mm}
\includegraphics[width=8.5cm]{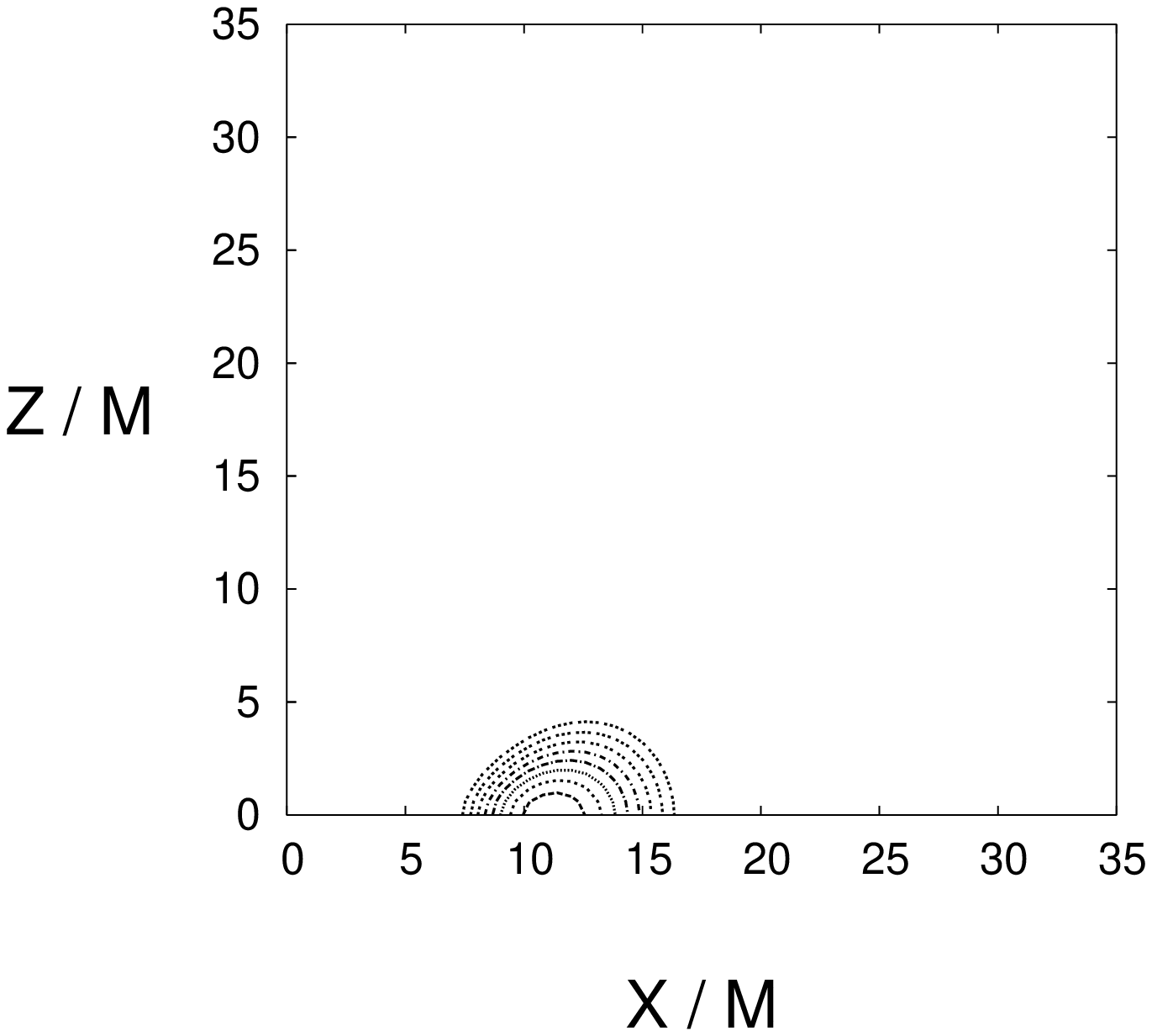}
\end{minipage}
\end{center}
\caption{(a) Density contour curves of initial conditions for model
D. The contour curves are drawn for $\rho=10^{12-0.5 i}~{\rm g/cm^3}$ with
$i=1$--10.  The atmosphere is included in these initial conditions. (b) The
same as (a) but for magnetic field lines.
\label{FIG1}}
\end{figure}

In Fig. \ref{FIG1}(a), we plot density contour curves for model D (cf. 
Table I). It is seen that the torus has a geometrically thick
structure in which the width $\varpi_2-\varpi_1$ is comparable to the
maximum thickness, $\sim 30M$. This is a universal property
that holds for all the models. 

\begin{figure}[t]
\begin{center}
\begin{minipage}[t]{0.47\textwidth}
\includegraphics[width=6cm]{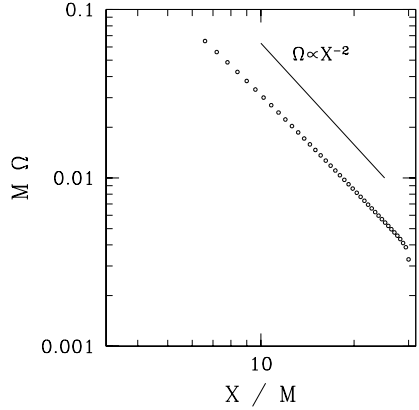}
\end{minipage}
\end{center}
\vspace{-2mm}
\caption{Angular velocity as a function of the cylindrical radius
on the equatorial plane for model D. 
\label{FIG2}}
\end{figure}

Figure \ref{FIG2} plots the angular velocity $\Omega$ as a function of
the cylindrical radius on the equatorial plane. It is found that
the torus has a differentially rotating velocity field with 
the steepness of the differential rotation $|d\ln \Omega /d\ln \varpi| 
\approx 2$, steeper than the Keplerian law. This is also a feature
of the $\ell=$constant velocity profile. 

\subsection{Magnetic field}

To induce angular momentum transport during the evolution, 
a weak poloidal magnetic field is initially added to the equilibrium torus.
In the present work, the profile of the magnetic field is chosen 
following Refs. \citen{GMT,MG,VH} as 
\beqn
A_{\varphi}=\Big\{
\begin{array}{ll}
A_0 (\rho-\rho_{0}) & {\rm for}~\rho \geq \rho_0,\\
0                 & {\rm for}~\rho < \rho_0,
\end{array}
\eeqn
where $A_0$ is a constant that specifies the magnetic field strength
and $\rho_0$ is a cut-off density; the magnetic field is initially
non-zero only in the region with $\rho > \rho_0$. Here, $A_0$ is
chosen such that the ratio of the magnetic energy to the thermal
energy is $1/200$, and $\rho_0$ is chosen as $0.2\rho_{\rm max}$, 
where $\rho_{\rm max}$ is the maximum density of the torus. Because the
magnetic field is weak, the equilibrium configuration is not
essentially modified by the magnetic force.

In Fig. \ref{FIG1}(b), we show the magnetic field lines for model D
(cf. Table I). Comparing Fig. \ref{FIG1}(a), it is seen that the
magnetic field is initially present in a localized region near the
density maximum because the density of the torus decreases
rapidly in the direction of the surface. 

Tori both with the poloidal magnetic field and with differential rotation
are subject to winding of the field lines and subsequent magnetic
braking \cite{MB}. In addition, they are subject to the 
magnetorotational instability (MRI).\cite{MRI} Due to the winding of the
magnetic field lines, the toroidal magnetic field $\cB^T$ grows
linearly with time in the early phase of the numerical simulation as
[e.g., Ref. \citen{DLSSS2}]
\beqn
\cB^T \approx t \cB^{\varpi} {\pa
\Omega \over \pa \ln \varpi} \sim -2 \Omega t \cB^{\varpi}.
\label{winding}
\eeqn
This amplification continues until the magnetic energy of the toroidal
field becomes a few tenths of the rotational energy (cf. 
\S\ref{sec:D1}). After the amplification stops, the magnetic braking
transports the angular momentum outwards. For the present initial
conditions with which $E_{\rm EM}/T_{\rm rot} \sim 0.001$, the field
growth saturates when $\cB^T$ becomes $\sim 30 \cB^{\varpi}$. This
implies that the field growth stops at $t \sim 15 \Omega^{-1} \sim 
2P_{\rm o}$, where $P_{\rm o}$ denotes the local rotational
period. Because the magnetic field is confined near the density
maximum, the expected saturation time is $\sim 2$--$3P_c$, where $P_c$
is the rotation period at the density maximum.

The growth time ($e$-folding time) and wavelength of the fastest-growing
mode of the MRI in the Newtonian theory are approximately given by
[e.g., Refs. \citen{MRI} and \citen{DLSSS3}]
\beqn
&&t_{\rm MRI} = 
2 \left|{\partial \Omega \over \partial \ln \varpi }\right|^{-1} 
\ ,  \label{tmri} \\
&&\lambda_{\rm max}  = {2\pi v_A^z \over \Omega}
\biggl[1-\biggr({\kappa \over 2\Omega}\biggr)^4\biggr]^{-1/2}, 
\eeqn
where $\kappa$ is the epicyclic frequency of Newtonian theory, 
\beq
\kappa^2 \equiv \frac{1}{\varpi^3}\frac{\partial(\varpi^4\Omega^2)}
{\partial \varpi},
\eeq
and $v_A^z \equiv \sqrt{\cB^z \cB_z /\rho \hat \gamma}$ is the
approximate Alfv\'en velocity of the $z$ component. In general
relativity, the growth rate is modified, but it agrees with
Eq. (\ref{tmri}) within $\sim 30\%$.\cite{Gammie} As in the case of
the field winding, the growth time scale of the MRI is of order $P_c$. 
With the present initial conditions, we have $|v_A^z| \approx 10^{-2}$
and $P_c \sim 200M$ at the density maximum (see Table I), resulting in
$\lambda_{\rm max} \sim 2M$. Because this is smaller than the width of
tori, the MRI should play an important role in transporting the
angular momentum and driving the turbulent motion in the accretion
torus. We also note that the grid spacing is chosen to be typically
$0.15M$, so that the MRI can be resolved in numerical simulations. We
note that with the present models as described above, we have $E_{\rm
EM}/E_{\rm int}=1/200$.  If the magnetic field strength is smaller
than its present initial strength by a factor of more than 5--10, then
the wavelength $\lambda_{\rm max}$ is too small to be resolved.  If
the magnetic field strength is chosen to be too large, $\lambda_{\rm
max}$ is so large that the MRI does not play an important role. The
present choice of the field strength is appropriate for taking into
account both the effects of the magnetic braking and the MRI.

It is known that non-magnetized tori with $\ell=$constant is often
unstable with respect to the runaway instability (see, e.g.,
Ref. \citen{Font} and references cited therein).  However, in the
present model, the angular momentum of tori is redistributed by the
magnetic braking and the MRI on a short time scale of $\sim
2$--$3P_c$, and then a quasistationary accretion state, which has a
less steep velocity profile, and hence, is stable with respect to the
runaway instability, is established.  The resulting quasistationary
state is expected to depend only weakly on the initial profile of
the angular momentum distribution, and thus, the numerical results 
should not depend strongly on the given velocity profile.

\subsection{Preparation of grid and boundary conditions}

A numerical simulation is performed in a nested grid zone; we divide
the computational domain into three zones and follow each zone with
a different grid spacing.  The computation in innermost zone is
carried out with the finest grid spacing $\Delta$. This zone covers
the square domain satisfying $0 \leq X \leq 48M$, except for the inner
square patch with $0 \leq X \leq 0.8r_{\rm H}$, which is excised. Here, $X$
denotes $\varpi$ and $z$.  The second and third zones cover $0 \leq X
\leq 96M$ and $0 \leq X \leq 192M$, with the grid spacings $2\Delta$
and $4\Delta$, respectively.  At the outer boundaries of the first and
second zones, values linear-interpolated from the 1-level larger zone
are assigned as the boundary values. Specifically, the interpolation
is performed for $\rho_*$, $\hat u_i \equiv h u_i$, $T$, $Y_e$, and
$\cB^i$. Then, $P$ and $h$ are determined using the EOS table, and
finally, $S_i$ and $S_0$ are determined. At the outer boundaries of
the third zone, we use an outflow boundary condition. The inner
boundary condition at $X=0.8r_{\rm H}$, which is located well inside the
event horizon, is arbitrary because it is inside the black hole, and
backflow are automatically prohibited. 

The numerical simulation for the innermost zone is performed with
$\Delta = 0.15M$ and $0.2M$. For models D and E (see Table I),
short-term simulations are performed with smaller grid spacings of
$\Delta=0.1M$ or $0.12M$ to check the convergence.  We find that even
for $\Delta = 0.2M$, the torus is well resolved with the present model
and parameter values. Thus, quantities such as the total rest-mass,
the rotational kinetic energy, and the electromagnetic energy depend
weakly on the grid resolution. The accretion rates at each time step,
$\dot M_*$, $\dot E$, and $\dot J$, which are evaluated at the event
horizon of radius $\leq 2M$, depend more strongly on the grid
resolution.  However, an approximately convergent solution is obtained
with $\Delta=0.15M$ (see \S \ref{sec:D}). The neutrino luminosity and
the total emitted energy also depend strongly on the grid resolution, 
because the luminosity depends on a high power of the temperature, and hence
a slight change in the temperature results in a significant change in the
luminosity. However, we found that the total emitted energy converges
within 20--30\% for $\Delta =0.15M$. The convergence tends to be slower
for larger values of $a$. The reason for this is that a larger value of $a$
results in smaller radii of the event horizon and the ISCO, and thus a
better grid resolution is required. 

\subsection{Atmosphere}\label{sec:atmo}

Because any conservation scheme of hydrodynamics is unable to evolve a
vacuum, we have to introduce an artificial atmosphere outside the
tori.  In particular, in the MHD simulation, an artificial atmosphere
of moderately large density is necessary because the computation often
crashes when the ratio of the magnetic energy density to the rest-mass
energy density exceeds a critical value. (In our present code, the
critical value is $\sim 10^4$--$10^5$.) As the rule for introducing
the atmosphere, we adopt the floor method proposed in Refs. \citen{MG}
and \citen{VH}.  Specifically, when the density drops below a critical
value, it is reset immediately. We choose a position-dependent
critical density $\rho_{\rm crit}$ as
\beqn
\rho_{\rm crit}={\rm Max}\Bigl[\rho_{\rm min},
{\rm Min}\Big(\rho_{\rm min}(r/50M)^{-2}, 300 \rho_{\rm min})\Bigr]. 
\eeqn
Our fiducial value for $\rho_{\rm min}$ is $10^{19/3}~{\rm g/cm^3}$.
The floor-density prescription sacrifices the
exact conservation of energy, rest-mass, and angular momentum.
To elucidate the dependence of the numerical results on
the value of the floor, we performed test simulations for model D using 
\beqn
&&\rho_{\rm crit}={\rm Max}\Bigl[\rho_{\rm min},
{\rm Min}\Big(\rho_{\rm min}(r/50M)^{-2}, 150 \rho_{\rm min})\Bigr],
\eeqn
with $\rho_{\rm min}=10^{20/3}~{\rm g/cm^3}$, and 
\beqn
&&\rho_{\rm crit}={\rm Max}\Bigl[\rho_{\rm min},
{\rm Min}\Big(\rho_{\rm min}(r/50M)^{-2}, 100 \rho_{\rm min})\Bigr],
\eeqn
with $\rho_{\rm min}=10^{19/3}~{\rm g/cm^3}$. We found that because
the floor density is much smaller than that of the main body of the
tori, the energy, rest-mass, and angular momentum are conserved to
within $\sim 10\%$.  We also found that the total energy emitted by
neutrinos and the total accreted rest-mass vary by no mor than $\sim
10\%$ as the floor values are varied.  However, the density of
atmosphere is still high enough to prevent formation of an outflow of
a high Lorentz factor: During its outward propagation, the outflow
carries a sufficiently large rest-mass and is thus decelerated.  In
other words, the velocity of the outflow depends strongly on the
density of the atmosphere. In the present paper, we do not focus on
the properties of the outflow. 

\section{Numerical results}\label{sec:3}

Numerical simulations were performed for a wide variety of models, 
listed in Table I, and for two or three grid resolutions in each case. The
computations were continued to $t \approx 60$ ms ($3000M$). The total
emitted neutrino energy and the total rest-mass swallowed by the black hole
for $0 \leq t \leq 50$ ms are presented in Table III. As mentioned in
\S \ref{sec:atmo}, these results depend on the density of the atmosphere
weakly, varying by $\sim 10\%$.

Magnetized accretion torus evolves qualitatively similarly for a
choice of the parameter values. In \S\ref{sec:D1}, we summarize
the general features and present the results for model D. In the
subsequent subsections, we clarify the quantitative dependence of
several quantities, such as the neutrino luminosity, maximum density,
and maximum temperature, on the mass and the initial rotation radius of
the torus, the spin of the black hole, and the optical depth.

\subsection{General features: Results for model D}\label{sec:D}

\subsubsection{Evolution of torus}\label{sec:D1}

\begin{figure}[t]
\begin{center}
\begin{minipage}[t]{0.47\textwidth}
\includegraphics[width=6.9cm]{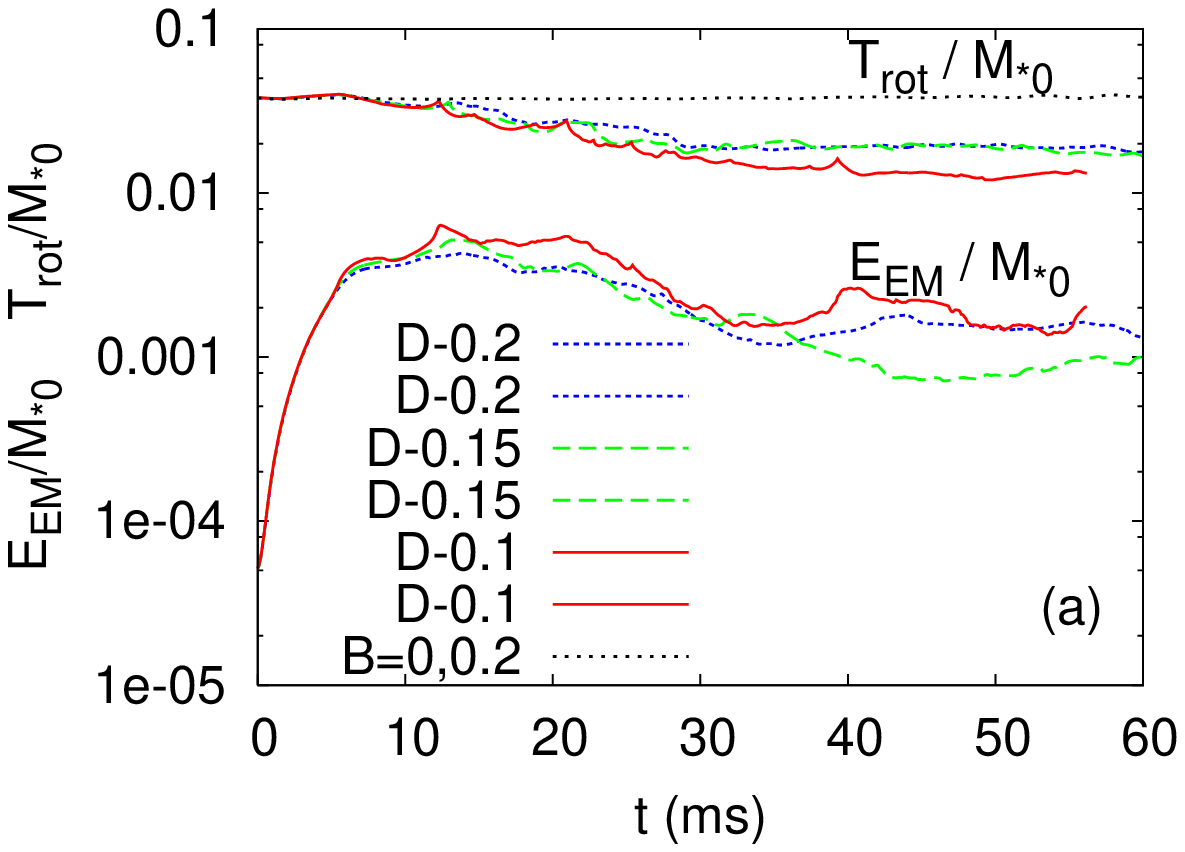}
\end{minipage}
\begin{minipage}[t]{0.47\textwidth}
\includegraphics[width=6.9cm]{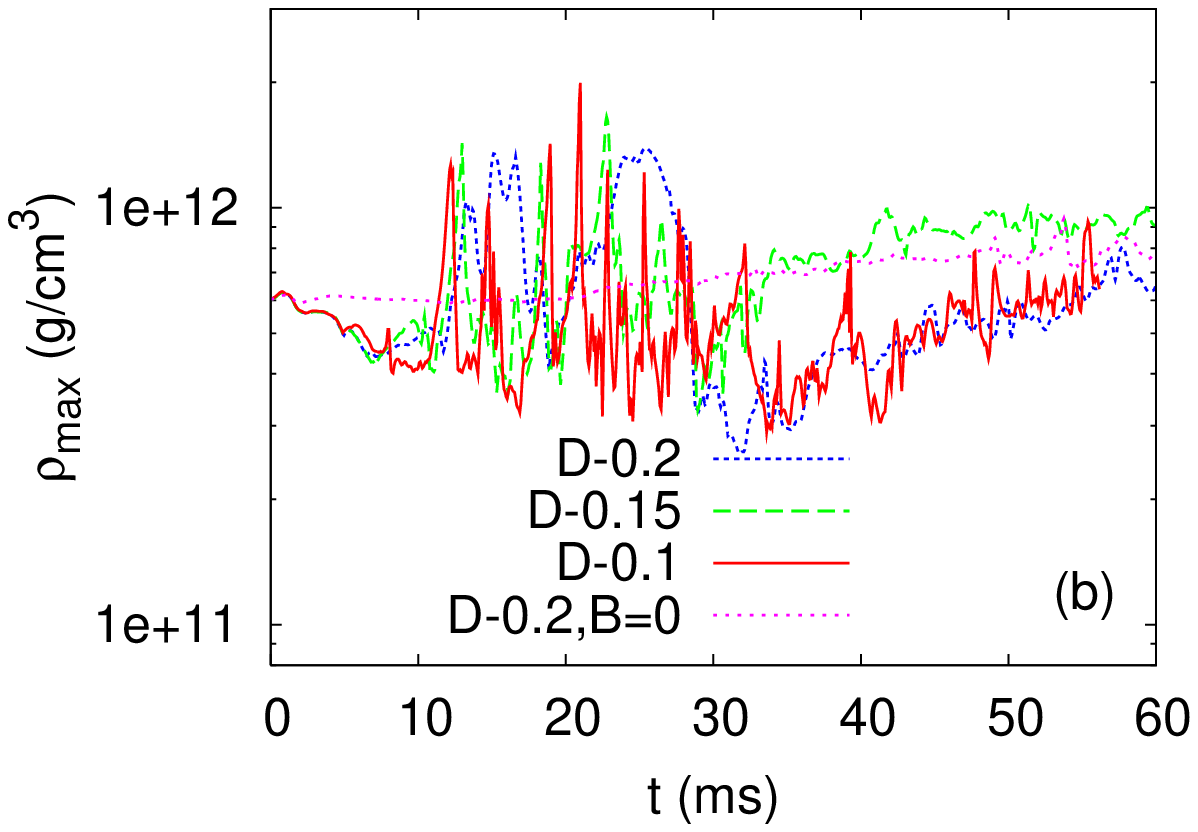}
\end{minipage}\\
\begin{minipage}[t]{0.47\textwidth}
\includegraphics[width=6.9cm]{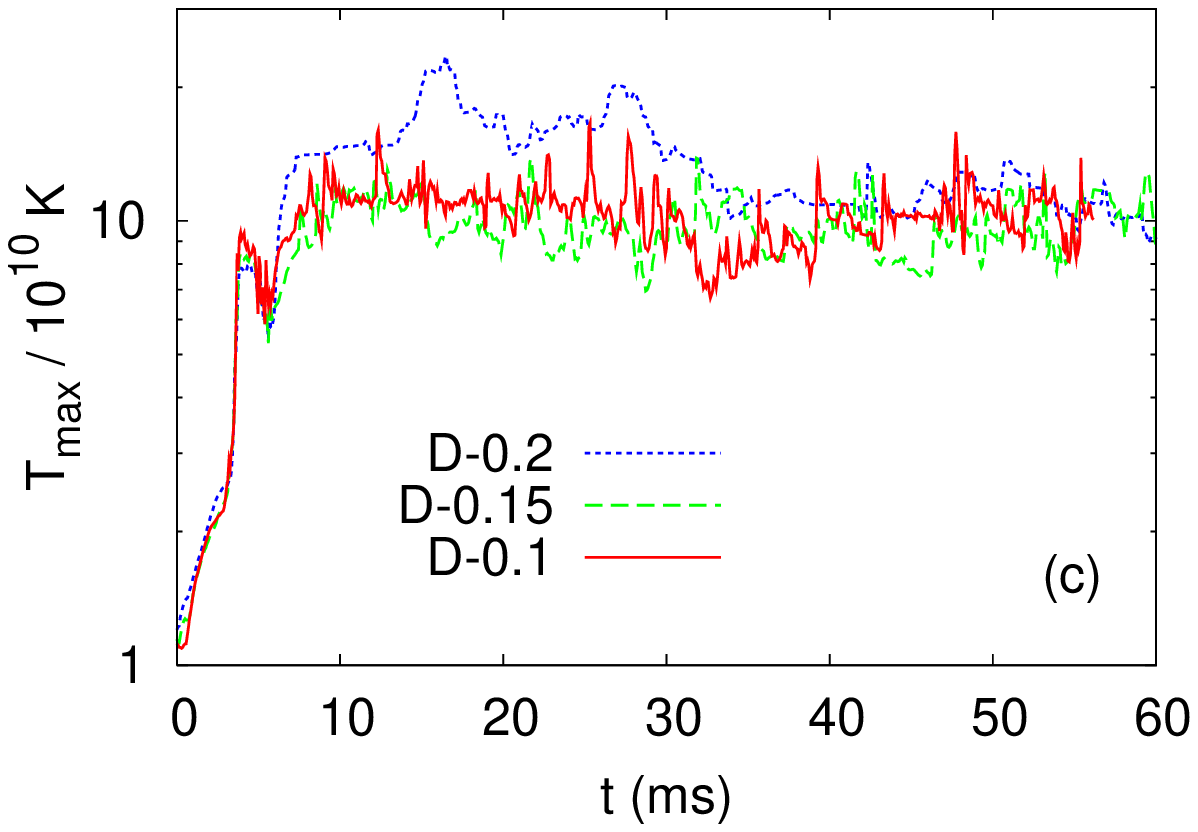}
\end{minipage}
\begin{minipage}[t]{0.47\textwidth}
\includegraphics[width=6.9cm]{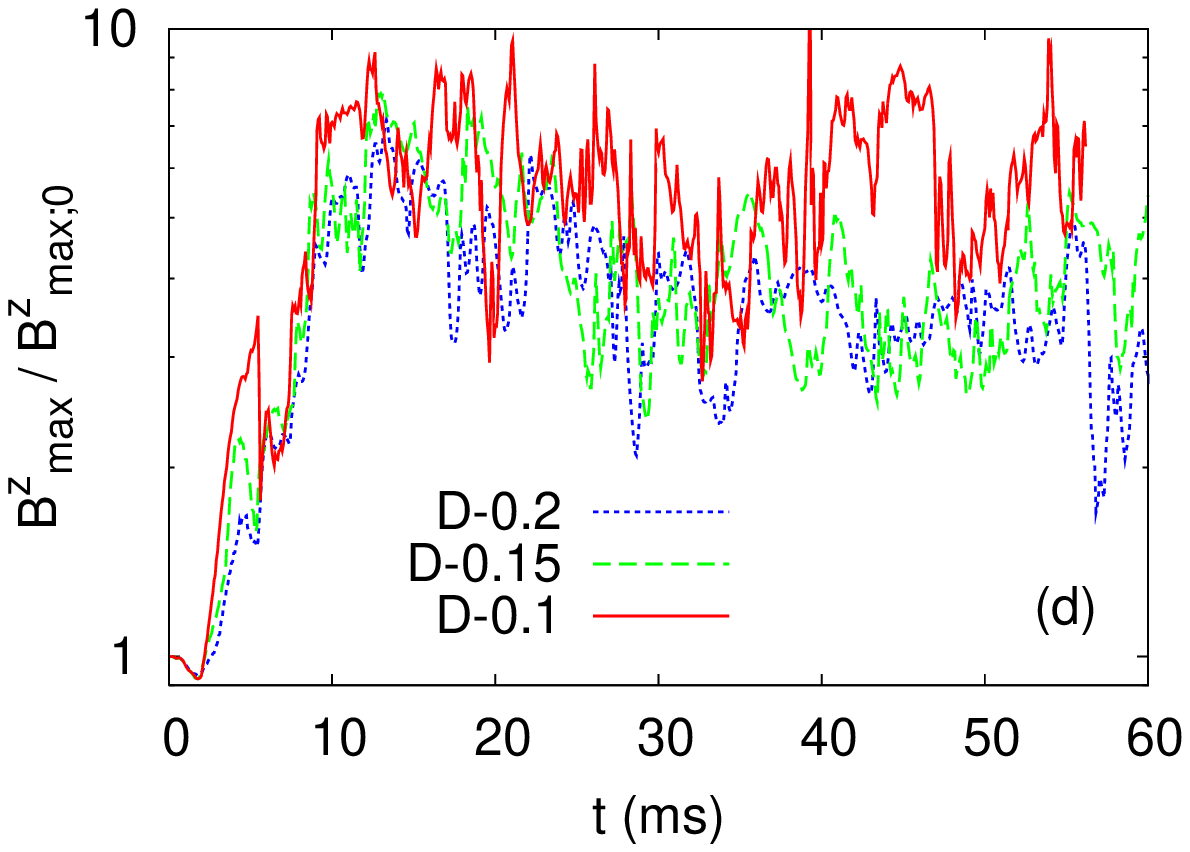}
\end{minipage}\\
\begin{minipage}[t]{0.47\textwidth}
\includegraphics[width=6.9cm]{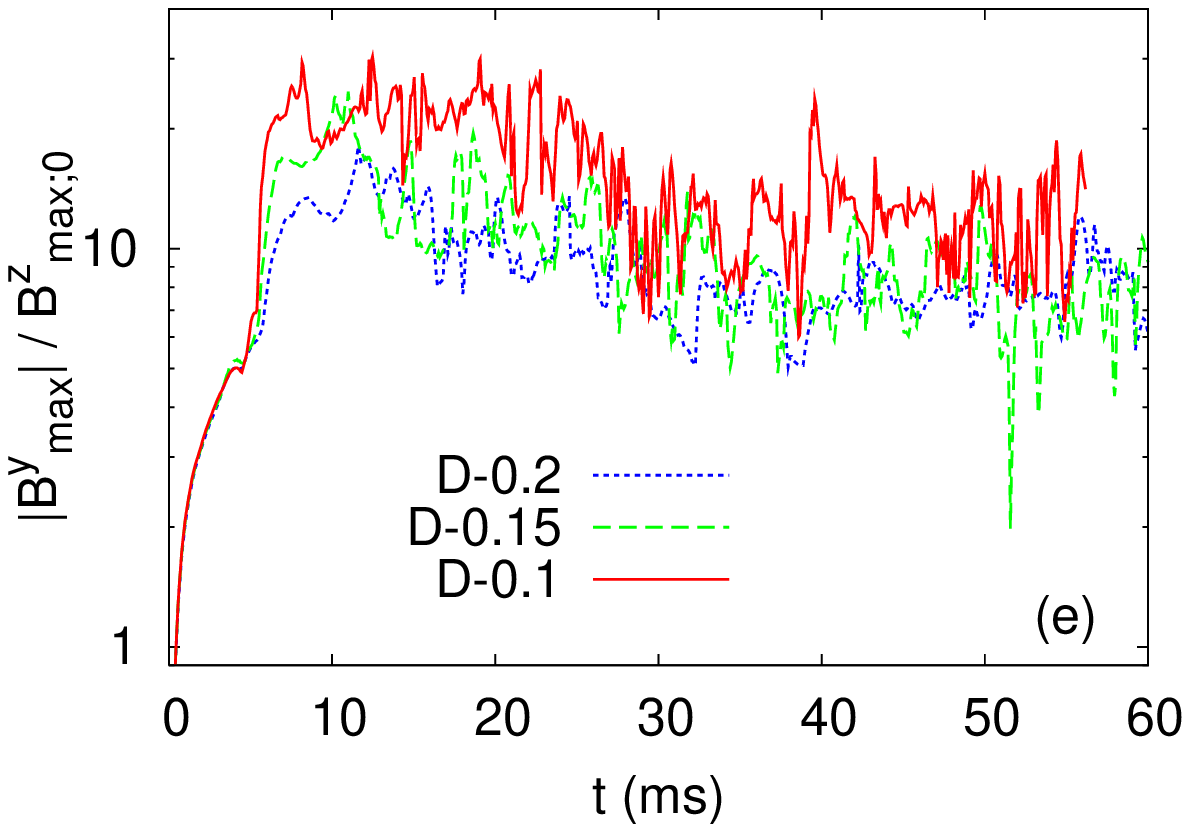}
\end{minipage}
\begin{minipage}[t]{0.47\textwidth}
\includegraphics[width=6.9cm]{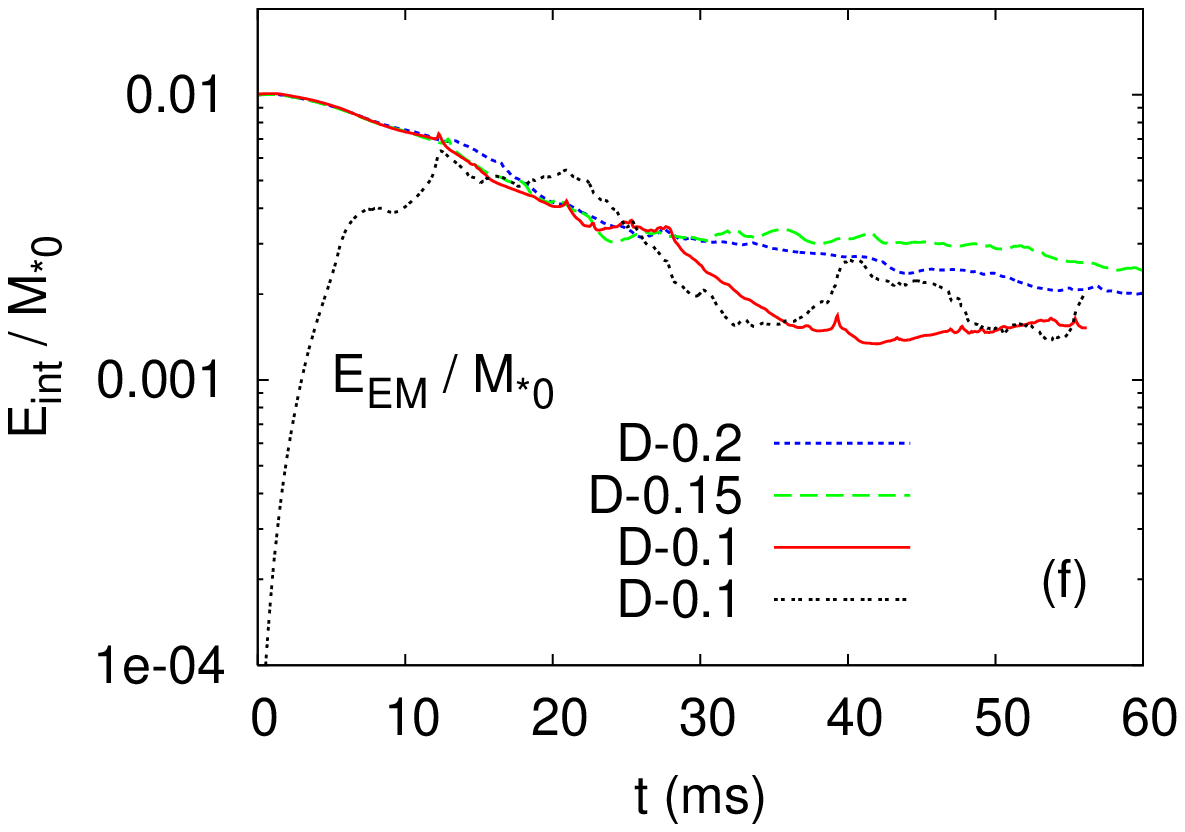}
\end{minipage}
\end{center}
\vspace{0mm}
\caption{(a) Evolution of the rotational kinetic energy and
electromagnetic energy in units of the initial value of the rest-mass, 
$M_{*0}$, for model D with $\Delta/M=0.1$, 0.15, and 0.2. For
comparison, the results with no magnetic field and with $\Delta/M=0.2$
are also shown. (b) The same as (a), but for the maximum
density. (c) The same as (a), but for the maximum temperature. (d) The
same as (a), but for the maximum value of $\cB^z$ in units of the
initial maximum value of $\cB^z$. (e) The same as (d), but for the
maximum value of $\cB^y$. (f) The same as (a), but for the internal
energy. For comparison, the electromagnetic energy for $\Delta=0.1M$ is
also shown.
\label{FIG3}}
\end{figure}

In Fig. \ref{FIG3}(a), we plot the evolution of the rotational kinetic energy
and electromagnetic energy for model D. For the first $\sim 10$ ms, 
magnetic field strength is amplified by the winding of the field lines
and the MRI due to the presence of the poloidal magnetic fields and
differential rotation. Here, the winding primarily contributes to
the increase of the electromagnetic energy.  This is found from the
fact that the electromagnetic energy increases approximately in
proportion to $t^2$ for $t \alt 5$ ms [see Eq. (\ref{winding})]. The
amplification continues until the electromagnetic energy reaches 
$\sim 10\%$ of the rotational kinetic energy. After the field growth
saturates, transport of angular momentum and the resulting mass accretion
onto the black hole are enhanced.  The strong magnetic field 
subsequently induces turbulent matter motion driven by the MRI. As a
result, shocks are generated and heat the matter, in particular for the
inner part of the torus, up to $\sim 10^{11}$ K. This contributes to quick
rise of the neutrino luminosity in the first $\sim 10$ ms (see
Fig. \ref{FIG5}). At the same time, rotational kinetic energy is
converted to thermal energy through the magnetic effects. Due to this
effect, together with accretion, the rotational kinetic energy (as well
as the internal energy) quickly decreases for $t \alt 30$ ms. For
comparison, we show the evolution of the rotational kinetic energy obtained
in the simulation of a non-magnetized torus. This reveals that the torus
does not change much only in the presence of neutrino cooling. 

For $t \agt 30$ ms, no sharp decrease of the rotational kinetic
energy nor sharp increase of the electromagnetic energy is observed. 
This indicates that the torus relaxes to a quasistationary
accretion state. In this phase, the rotational kinetic energy
gradually decreases as a result of the mass accretion. 

\begin{table}[t]
\caption{The total energy emitted by neutrinos in units of $10^{51}$
ergs, the total accreted rest-mass, and the conversion efficiency.
The time integration was performed for $0 \leq t \leq 50$ ms. The
results with $\Delta/M=0.2$ and 0.15 are shown.  For models with
``---'', simulations were not performed.}
\begin{center}
\begin{tabular}{c|ccc|ccc}
\hline \hline
    &  &$\Delta/M=0.2$ & 
& & $\Delta/M=0.15$  & \\ 
Model & $\Delta E_{\nu}$
& $\Delta M_*(M_{\odot})$ 
& $\Delta E_{\nu}/\Delta M_*(\%)$ 
& $\Delta E_{\nu}$
& $\Delta M_*(M_{\odot})$
& $\Delta E_{\nu}/\Delta M_*(\%)$ 
\\ \hline
A & 2.2 & 0.11 &1.1& 2.0 &0.12 & 0.9\\
A2& 2.8 & 0.11 &1.4& --- & --- & ---\\
B & 2.4 & 0.12 &1.1& 2.5 & 0.12 & 1.2\\
B2& 3.3 & 0.12 &1.5& ---   & --- & ---\\
C & 2.7 & 0.13 &1.2& 2.5 & 0.11 & 1.3\\
C2& 4.0 & 0.13 &1.7 & --- & --- & ---\\
D & 3.4 & 0.13 &1.5& 3.7 & 0.10 & 2.0\\
D2& 6.2 & 0.13 &2.7& 5.3 & 0.12 & 2.4\\
E & 5.7 & 0.11 &3.0& 4.2 & 0.10 & 2.3\\
E2& --- & ---  & --- & 12& 0.12 & 5.8\\ 
F & 4.1 & 0.16 &1.4& 3.7 & 0.12 & 1.7 \\
G & 1.6 & 0.054 &1.6& 1.5  & 0.050 & 1.7\\
H & 2.0 & 0.061 &1.9& 1.6  & 0.062 & 1.4\\
I & 6.2 & 0.22 &1.6& 5.0  & 0.17 & 1.6\\
J & 6.5 & 0.23 &1.6& 5.7  & 0.18 & 1.8\\
\hline
\end{tabular}
\end{center}
\end{table}

Figure \ref{FIG3}(b) plots the evolution of the maximum density. For
10 ms $\alt t \alt 30$ ms, we see that it varies violently in the
range between $\sim 3 \times 10^{11}~{\rm g/cm^3}$ and $\sim 2 \times
10^{12}~{\rm g/cm^3}$.  This indicates that in such an early phase,
the turbulent motion is strongly excited. For comparison, the result
for the non-magnetized case is also shown. It is seen that the maximum
density increases only gradually with time due to neutrino cooling,
showing that the torus is approximately in a quasistationary
state. This indicates that the violent time variation of the maximum
density in the early phase is induced by the magnetic effects. By 
contrast, the maximum density gradually increases for $t \agt 30$
ms. This is due to the fact that the rotation radius of the torus
gradually decreases in the quasistationary mass accretion phase 
making the torus compact. [The density maximum is initially located at
$\varpi \approx 10M$, whereas it is at $\varpi \approx 5M$ at $t=40$
ms; cf. Fig. \ref{FIG6}(a)].

Because the region with $\rho \agt 10^{11}~{\rm g/cm^3}$ is optically
thick with respect to neutrinos, a fraction of the neutrinos are trapped by
the matter flow, \cite{GRB3,LRP} and this reduces the fraction of neutrinos
that escape to infinity. This fact is clarified in \S\ref{sec:3.2}, 
which displays the dependence of the neutrino luminosity on $\zeta$. The
neutrino-trapping effect always plays a role for tori of mass $\agt
0.1M_{\odot}$ in the model studied presently because $\rho_{\rm max}$ of such
tori is $\geq 10^{11}~{\rm g/cm^3}$ (cf. \S \ref{sec:mass}).

Figure \ref{FIG3}(c) plots the evolution of the maximum temperature.
Because of shock heating, the temperature quickly rises in the first
$\sim 10$ ms. Then, the maximum temperature reaches $\sim 10^{11}$ K
and relaxes approximately to a constant value. This is a qualitatively 
universal feature for the tori considered in this paper. The value of
the maximum temperature depends on the spin of the black hole and
the mass of the torus. (cf. \S\ref{sec:mass} and \S\ref{sec:spin}).

Figures \ref{FIG3}(d) and (e) plot the evolution of the maximum values of
$|\cB^z|$ and $|\cB^y|$ in units of the initial maximum value of
$|\cB^z|$. In the early phase, the poloidal component of the magnetic
field grows in an exponential manner, indicating that the MRI occurs. 
By contrast, the toroidal component $|\cB^{y}|$ increases
linearly with time due to the winding of the magnetic field
lines. This property is also seen in Fig. \ref{FIG3}(a), where the
electromagnetic energy increases in proportion to $t^2$ for $t \alt
5$ ms. After the amplification of the magnetic field, the
electromagnetic energy density becomes comparable to the internal
energy density [see Fig. \ref{FIG3}(f)].  Then the growth of the field
saturates. This is also one of the universal features for the
evolution of magnetized tori, irrespective of the mass of the tori and 
the black hole spin. 

Figures \ref{FIG3}(b)--(e) show that numerical results for the maximum
density, maximum temperature, and maximum magnetic field strength do
not depend strongly on the grid resolutions. This indicates that the
simulation with $\Delta=0.15M$, which is a typical choice for the grid
resolution, provides a result of good convergence. We note that this
is also the case for any value of $a \leq 0.9M$, although higher grid
resolution is required for larger values of $a$, because the
coordinate radius of the event horizon is smaller in this case.

\subsubsection{Evolution of accretion rates}\label{sec:D2}

\begin{figure}[t]
\begin{center}
\begin{minipage}[t]{0.47\textwidth}
\includegraphics[width=6.9cm]{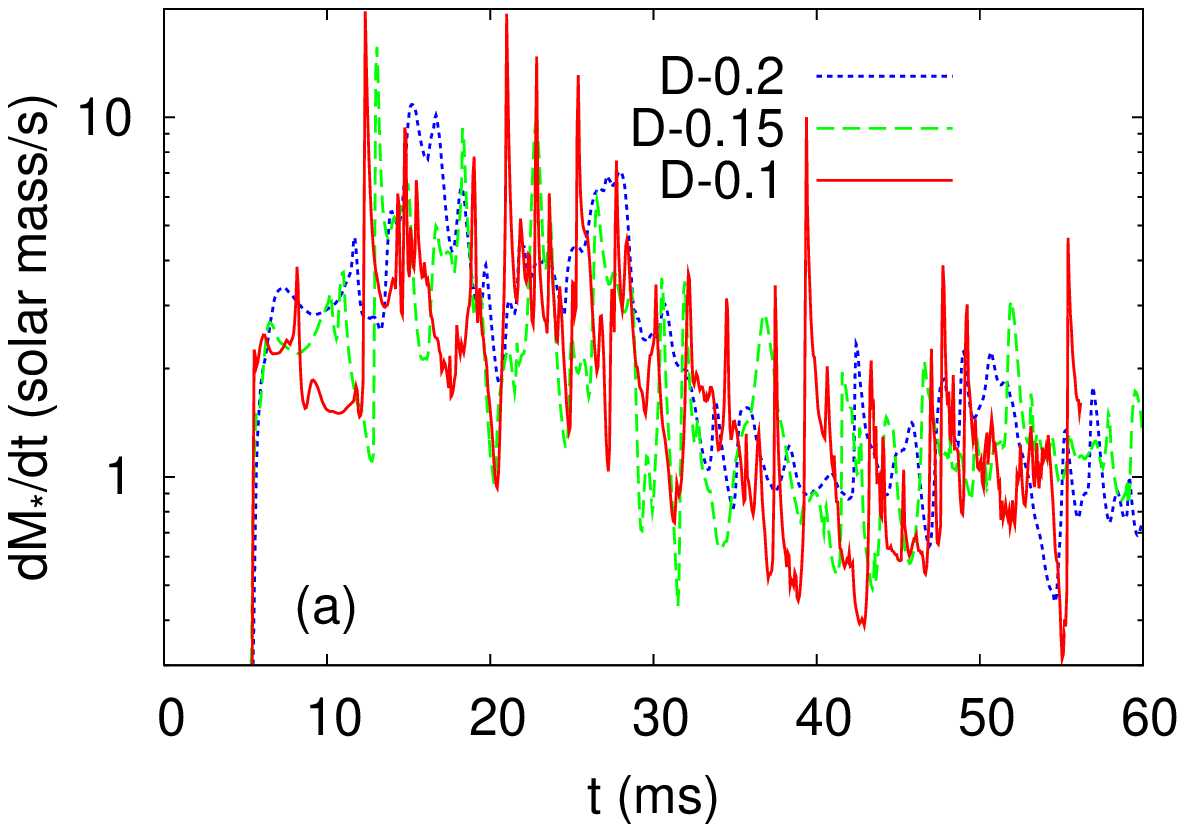}
\end{minipage}
\begin{minipage}[t]{0.47\textwidth}
\includegraphics[width=6.9cm]{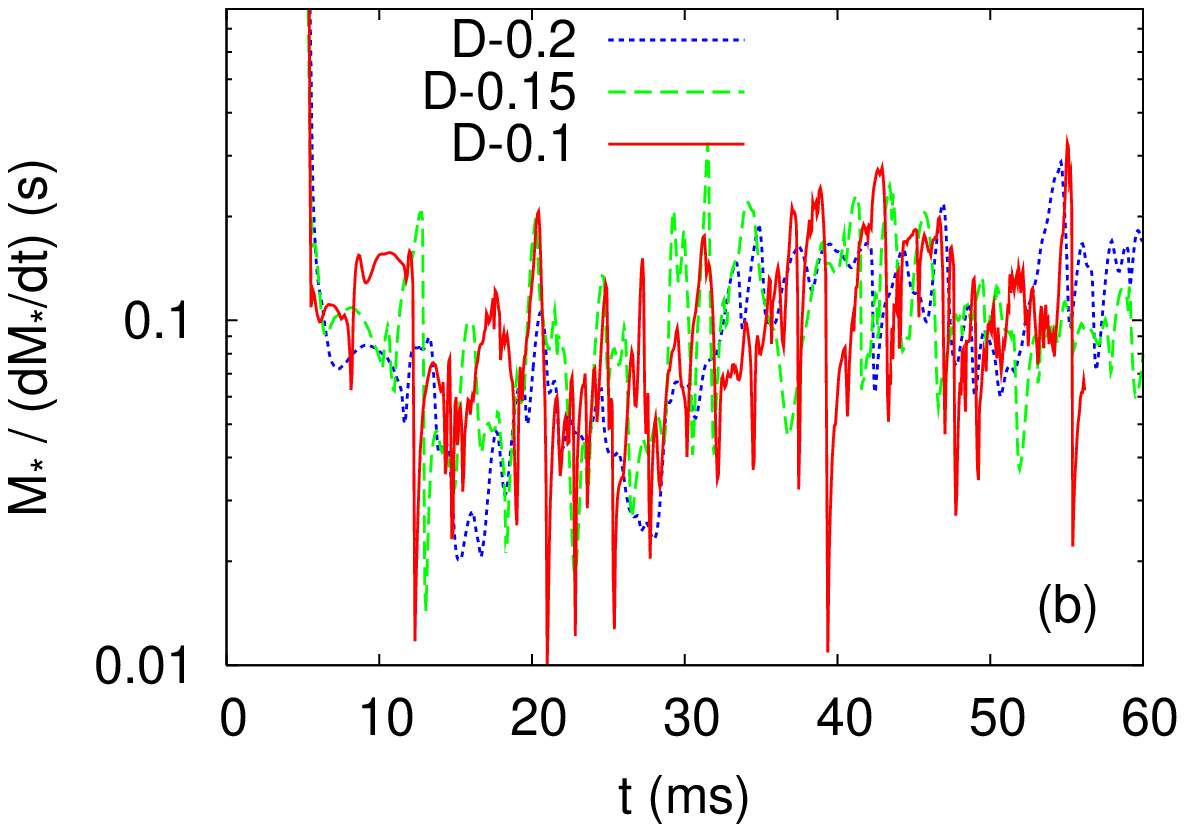}\\
\end{minipage}
\begin{minipage}[t]{0.47\textwidth}
\includegraphics[width=6.9cm]{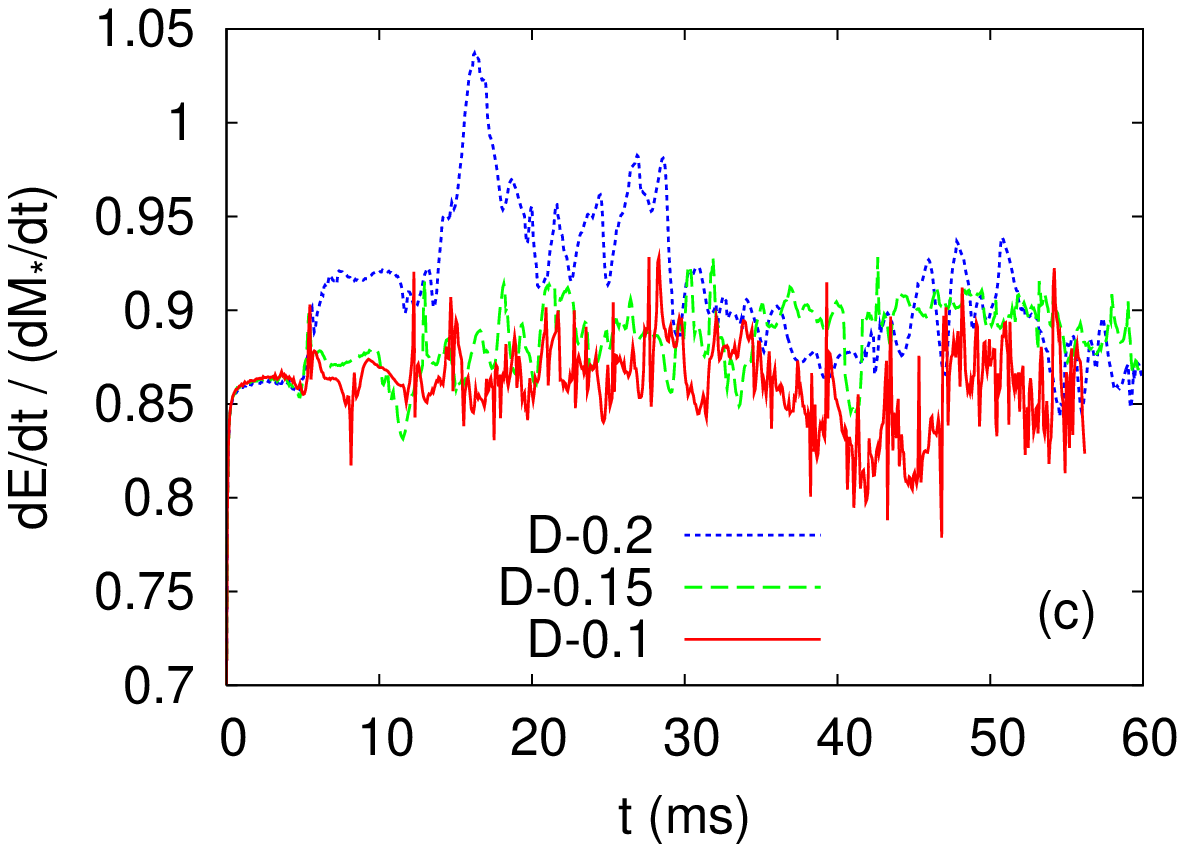}
\end{minipage}
\begin{minipage}[t]{0.47\textwidth}
\includegraphics[width=6.9cm]{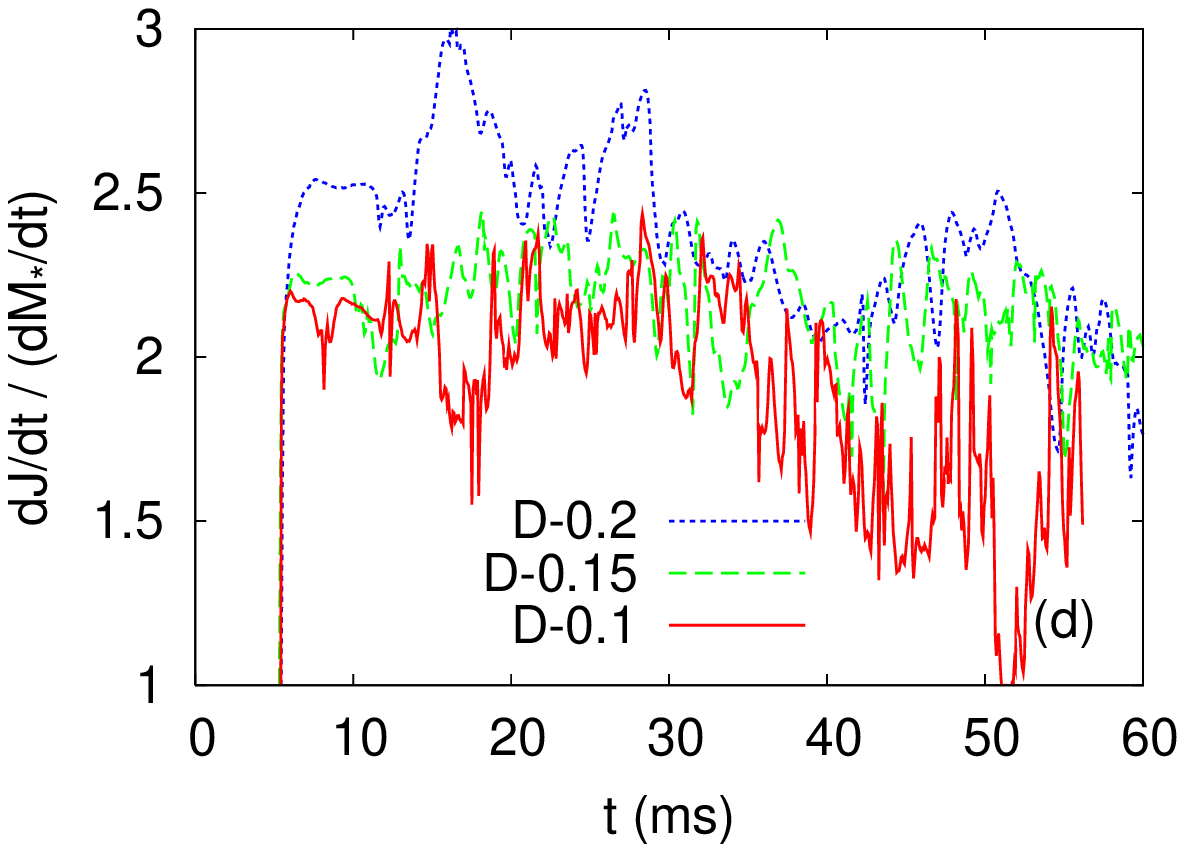}
\end{minipage}
\end{center}
\vspace{0mm}
\caption{Evolution of (a) the rest-mass accretion rate, $\dot M_*$,
(b) the accretion time scale, $M_*/\dot M_*$, (c) the energy accretion
rate in units of the rest-mass accretion rate, $\dot E/\dot M_*$, and
(d) the angular momentum accretion rate in units of the rest-mass
accretion rate, $\dot J/\dot M_*$, for model D with $\Delta/M=0.1$,
0.15, and 0.2.
\label{FIG4}}
\end{figure}

In Fig. \ref{FIG4}, we plot the evolution of the rest-mass accretion rate, 
$\dot M_*$, accretion time scale, $M_*(t)/\dot M_*$, and energy and
angular momentum accretion rates, $\dot E$ and $\dot J$, in units of
$\dot M_*$. It is seen that as the electromagnetic energy increases [see
Fig. \ref{FIG3}(a)], the mass accretion rate quickly increases, and
when the growth of the magnetic field saturates, it reaches $\sim
10M_{\odot}$/s. Then, it gradually decreases and eventually relaxes to
$\sim 1$--$2M_{\odot}$/s in a quasistationary state. We note that
the accretion rate depends on the mass of torus and on the black hole
spin (see \S\ref{sec:mass} and \S\ref{sec:spin}).

The associated accretion time scale for $t \alt 30$ ms is violently
time-varying with the average value of $\sim 50$ ms, 
but it relaxes to $\sim 0.1$--0.2 s in the quasistationary
state. In a previous work \cite{LRP} in which the $\alpha$-viscosity is
used for the evolution of a neutrino-cooled torus of mass $\approx
0.3M_{\odot}$, the accretion time scale is $\sim 50$ ms for the
$\alpha$-viscous parameter value $\alpha_v=0.1$ and $\sim 0.5$ s for
$\alpha_v=0.01$. The present numerical results approximately
correspond to the case that $\alpha_v \approx 0.1$ for the early
phase and $\alpha_v \approx 0.04$ for the quasistationary phase. This
is a universal feature for any values of the mass of the torus and the
black hole spin.

The ratio of the energy accretion rate to the mass accretion rate, $\dot
E/\dot M_*$, is $\sim 0.85$--0.9 for the high-resolution runs. One may
think that this is a reasonable result, because the ratio of the
specific energy to the rest-mass energy of a test particle rotating
around a Kerr black hole of $a/M=0.75$ at the ISCO is $\approx 0.89$
(see Table II). However, this is an accidental coincidence, as can be 
understood from the following two points. (i) The torus is not composed of test
particles but fluid with a significant internal energy obtained 
by shock heating. Due to the contribution of the specific
internal energy, the value of $\dot E$ should increase beyond that
of the test particles. (ii) The accreting matter loses kinetic energy
as it falls into the black hole from the ISCO due to the magnetic
stress. This effect should reduce the value of $\dot E$.  In the
present case, these two effects approximately cancel each other.

The ratio $\dot J/\dot M_*$ is found to be $\sim 1.5$--2.3, which is
smaller than the value of the specific angular momentum for a test
particle orbiting at the ISCO, $\approx 2.5$ (see Table II). This
indicates that during the infall of matter into the black hole from
the ISCO, the magnetic stress extracts angular momentum, which is
transported outwards along the field lines, as in the case of $\dot E$. 
This effect was pointed out in Ref. \citen{MG}. 

\subsubsection{Neutrino luminosity}\label{sec:D3}

\begin{figure}[t]
\begin{center}
\begin{minipage}[t]{0.47\textwidth}
\includegraphics[width=6.9cm]{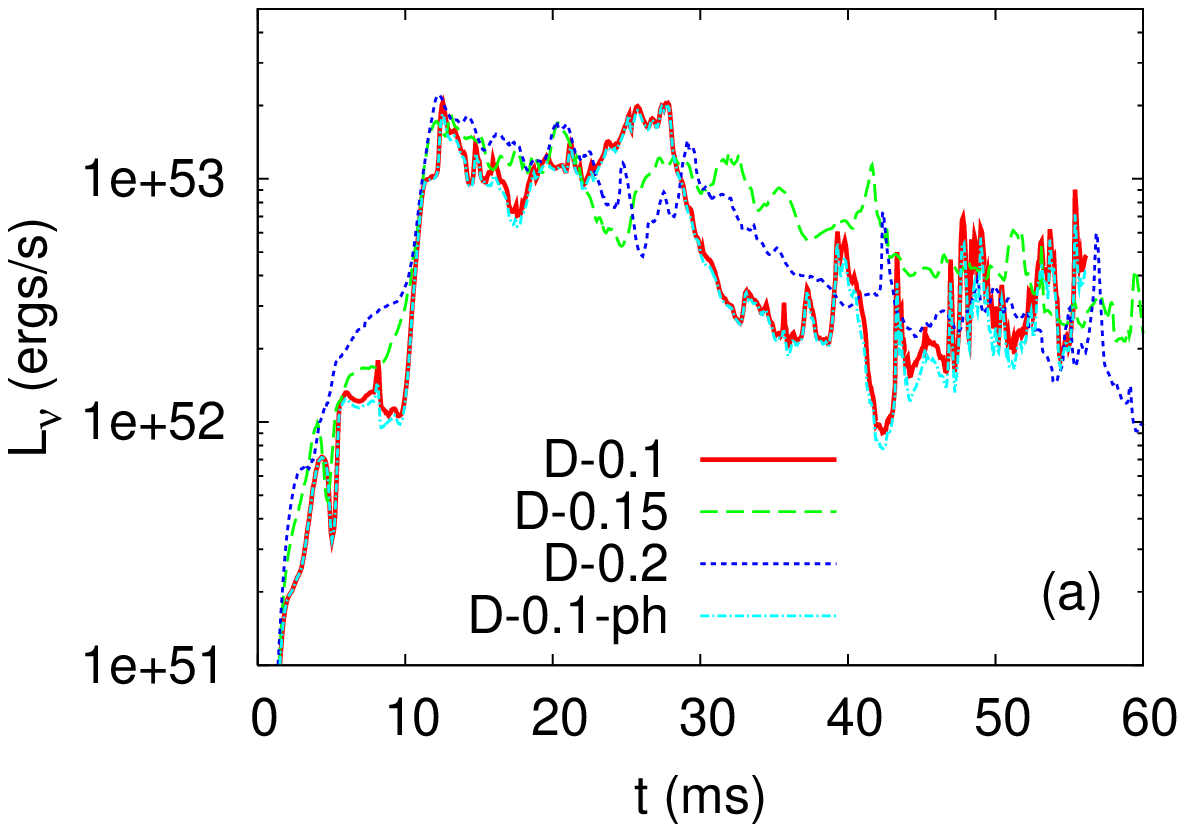}
\end{minipage}
\begin{minipage}[t]{0.47\textwidth}
\includegraphics[width=6.9cm]{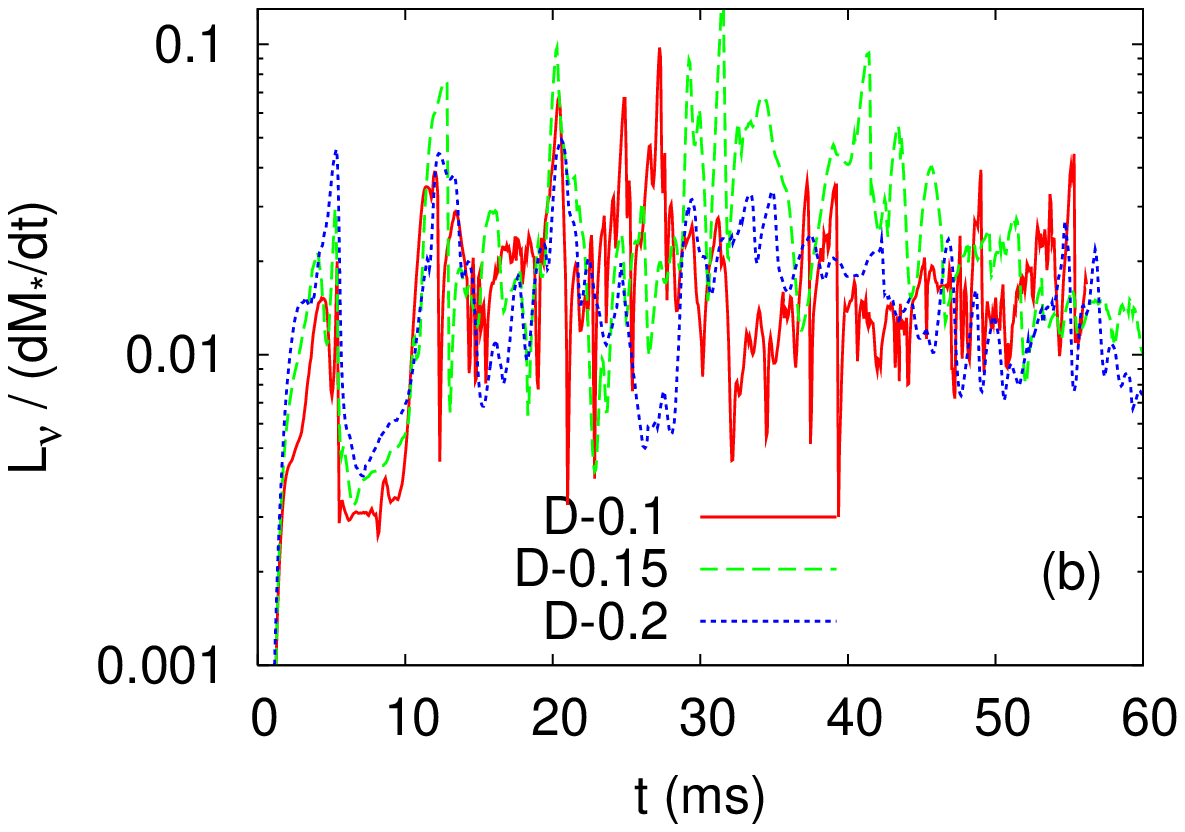}
\end{minipage}
\end{center}
\vspace{0mm}
\caption{Evolution of (a) the neutrino luminosity, $L_{\nu}$, and (b)
the efficiency of the conversion to neutrinos, $L_{\nu}/\dot M_*$,
for model D with $\Delta/M=0.1$, 0.15, and 0.2. Here, ``ph'' denotes
the result for the case that the luminosity is given by 
Eq. (\ref{lumi2}), which approximately overlaps with $L_{\nu}$, as the
difference is less than 10\%.
\label{FIG5}}
\end{figure}

Figure \ref{FIG5} plots neutrino luminosity $L_{\nu}$ and the
efficiency of the conversion to neutrinos (defined by the ratio of the
neutrino luminosity to the rest-mass energy accretion rate,
$L_{\nu}/\dot M_* c^2$).  Soon after the magnetic field growth
saturates at $t \sim 10$ ms, the neutrino luminosity reaches a maximum
of $\sim 2 \times 10^{53}$ ergs/s as a result of shock heating by
violent matter motion induced by magnetic stress. Then, for 10 ms
$\alt t \alt 30$ ms, during which there is significant turbulent
matter motion (cf. \S \ref{sec:D1}), it remains $\agt 10^{53}$ ergs/s.
For $t \agt 30$ ms, the accretion rate gradually decreases, and so
does the neutrino luminosity. In the quasistationary phase for $t \agt
30$ ms, we have $L_{\nu}\sim 1$--$5 \times 10^{52}$ ergs/s.  As shown
in subsequent sections, these values for the neutrino luminosity
depends on the mass of the torus and the black hole spin.

We derive the neutrino luminosity by simply integrating the emissivity
outside the event horizon. In an actual system of this kind, a
fraction of the neutrinos emitted in the vicinity of black hole should
be swallowed by it, because of the strong gravity. To estimate the
dependence of the luminosity on the chosen domain for the integration,
we also compute the luminosity, choosing the inner boundary of
integration for the luminosity at $r=r_{\rm ph}$ (see the dotted curve
in Fig. \ref{FIG5}). Note that only half of the neutrinos emitted by a
stationary emitter at $r=r_{\rm ph}$ can escape. In this case, the
luminosity is smaller by $\sim 10\%$ than $L_{\nu}$. Thus, the real
luminosity may be $\sim 90\%$ of the value of $L_{\nu}$ obtained
from the integration for $r > r_{\rm H}$.

From the time integration, the total emitted neutrino energy, $\Delta
E_{\nu}$, and the total accreted rest-mass for $t \leq 50$ ms are
calculated, giving 3--$4 \times 10^{51}$ ergs and $\sim
0.1$--$0.13M_{\odot}$ for model D. (Note that these values depend on
the grid resolution.) Thus, $\sim 1.4$--2.0\% of the total accreted
rest-mass energy, $\Delta M_*c^2$, is converted into neutrinos. (For
$\Delta/M=0.1$, 0.15, and 0.2, it is 1.6\%, 2.0\%, and 1.4\%,
respectively).  The order of magnitude of this value agrees with the
results presented in Refs. \citen{SRJ} and \citen{LRP}. Our efficiency
is smaller by a factor of $\sim 2$--3 than those obtained in the
previous works. This is probably because of the differences in the
treatments of the gravitational fields, neutrino opacity, and angular
momentum transport process. Figure \ref{FIG5}(b) also shows that the
conversion efficiency is in a narrow range (between $\sim 1\%$ and
$\sim 3\%$), close to the average value of $L_{\nu}/\dot M_*c^2 \sim
1$--2\%, for the entire time. 

For $a/M=0.75$, the maximum hypothetical conversion efficiency is
about 11\% (cf. Table II). The results here imply that the conversion
efficiency is not as large as the maximum. Note that the accretion
time scale, $M_*/\dot M_* \sim 100$ ms, is approximately as long as
the neutrino emission time scale, $E_{\rm int}/L_{\nu}$. This implies
that a part of the thermal energy is trapped by the matter flowing
into the black hole and fails to be converted into neutrinos. Indeed,
we find that the conversion efficiency depends strongly on the value
of $\zeta$ (see \S \ref{sec:3.2}). This is one reason for the
relatively small conversion efficiency.  Another reason is that the
initial radius of the torus is $\approx 10M$ at the density maximum
and at most $30M$ in the present model.  The difference between the
specific binding energy at $r=10M$ and at the ISCO is $\approx 6\%$,
and hence much smaller than 11\%. We note that the radii of the ISCO
are larger for smaller values of the black hole spin. Thus, the
conversion efficiency is even suppressed for smaller values of the
black hole spin, as shown in \S \ref{sec:spin}. On the other hand, the
suppression factor for larger spin with $a/M=0.9$ is smaller.

Numerical simulations indicate that the formation of a black
hole-torus system from stellar core collapse and the merger of a black
hole and neutron star is divided into two stages [e.g.,
Refs. \citen{GRB0,Proga,SS07,BHNS0,BHNS1,SU06}].  In the first stage,
the black hole and accretion torus are formed dynamically. According
to our present numerical results, in such a stage, the amplification
of magnetic fields and the subsequent redistribution of angular
momentum proceed violently in the presence of magnetic fields of
appreciable magnitude. As a result, shocks are generated, increasing
the temperature of the torus, and the neutrino luminosity thereby
reaches $\agt 10^{53}$ ergs/s. In the second stage, the system relaxes
to a quasistationary state. In such a stage, the mass and spin of the black
hole become approximately constant and the accretion proceeds with
a time scale longer than that of the first stage. The present
numerical results indicate that for this stage, the neutrino
luminosity is of order $10^{52}$ ergs/s.

A characteristic feature found in the MHD simulation is that the
luminosity curve is not smooth, in contrast to that reported in
Refs. \citen{SRJ} and \citen{LRP}. This is due to the fact that in the
present simulation, shock heating is associated with turbulent
(i.e. irregular) matter motion driven by highly variable magnetic
stress, whereas in the previous simulations, the heating is induced by
$\alpha$-viscosity, which leads to a nearly stationary accretion. 

The light curve of GRBs is often not smooth.\cite{GRB} If the GRBs are
driven by pair annihilation of neutrino-antineutrino pairs emitted
from the torus \cite{GRB}, the luminosity curve of neutrinos should
also be highly variable. The scenario in which the original thermal
source of neutrinos is attributed to shock heating driven by the
chaotic magnetic activity is thus favorable.

The luminosity and total emitted neutrino energy, $\Delta E_{\nu}$, do
not depend strongly on the grid resolution for $a/M \leq 0.75$. (The
values of $\Delta E_{\nu}$ for $\Delta/M=0.15$ and 0.2 agree within 
$\sim 10$--20\%; cf. Table III.) This indicates that for computing the
luminosity, a grid resolution with $\Delta=0.2M$ is acceptable for $a
\leq 0.75M$.\footnote{The mass accretion and neutrino emission rates
are induced by magnetic stress, which causes turbulent motion.  As a
result, the numerical results for different grid resolutions do not
agree at each moment of time. However, the average values over a
duration $\sim 10$ ms do not depend strongly on the grid resolution.} 
By contrast, for $a=0.9M$, the luminosity and total energy for
$\Delta/M=0.15$ and 0.2 do not agree well. (The relative error is by
20--30\%; cf. Table III.) The reason for this difference is that the
radius of the ISCO is much smaller than in the other models. (The ISCO
is located at $\varpi \approx 2.5M$; cf. Table II.)  Because a large
fraction of neutrinos are emitted near the ISCO, the luminosity
depends strongly on the resolution near there. For $a \geq 0.9M$, a
grid resolution with $\Delta=0.2M$ is not acceptable for obtaining a
result with good convergence.  For $a=0.9M$, we performed a simulation
with $\Delta=0.12M$ and found that the results with $\Delta=0.15M$ are
in good agreement with those with $\Delta =0.12M$. This indicates that
a grid resolution of $\Delta=0.15M$ is acceptable even for $a=0.9M$.

\subsubsection{Structure of the torus in the quasistationary phase}
\label{sec:D4}

\begin{figure}[p]
\vspace{-5mm}
\begin{center}
\begin{minipage}[t]{0.47\textwidth}
(a)\includegraphics[width=7.cm]{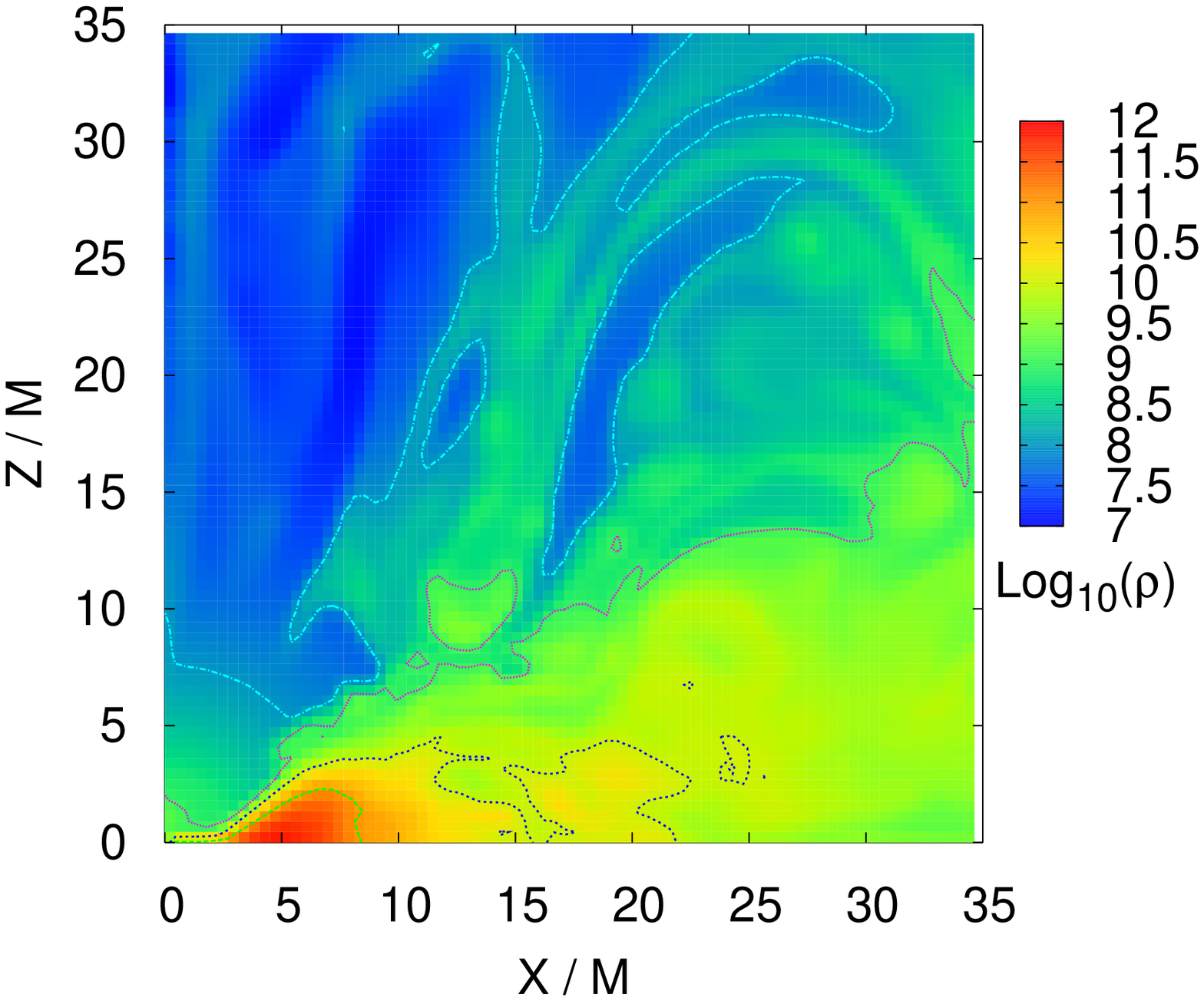}
\end{minipage}
\begin{minipage}[t]{0.47\textwidth}
(b)\includegraphics[width=7.cm]{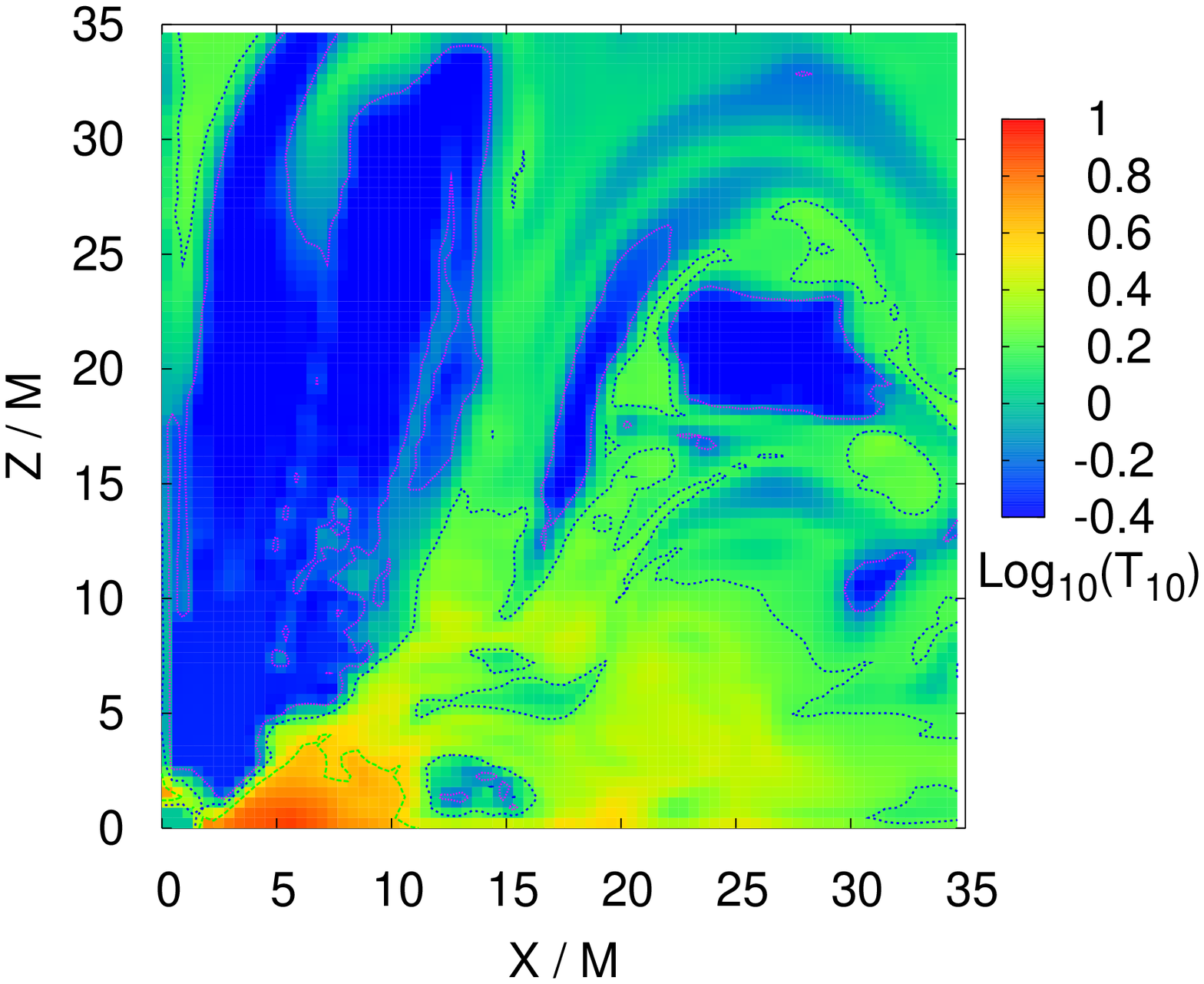}
\end{minipage}
\\
\vspace{-10mm}
\begin{minipage}[t]{0.47\textwidth}
(c)\includegraphics[width=7.cm]{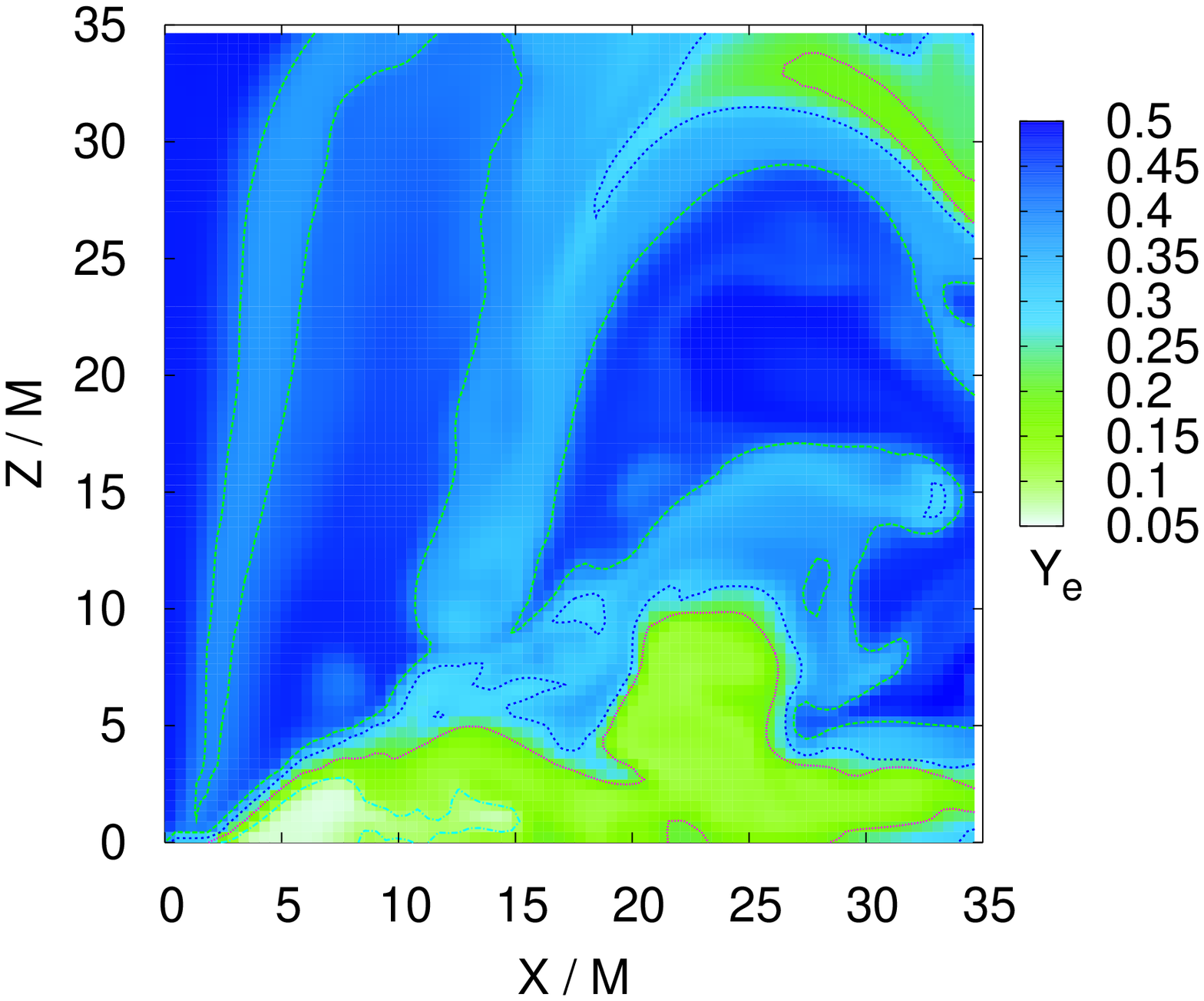}
\end{minipage}
\begin{minipage}[t]{0.47\textwidth}
(d)\includegraphics[width=7.cm]{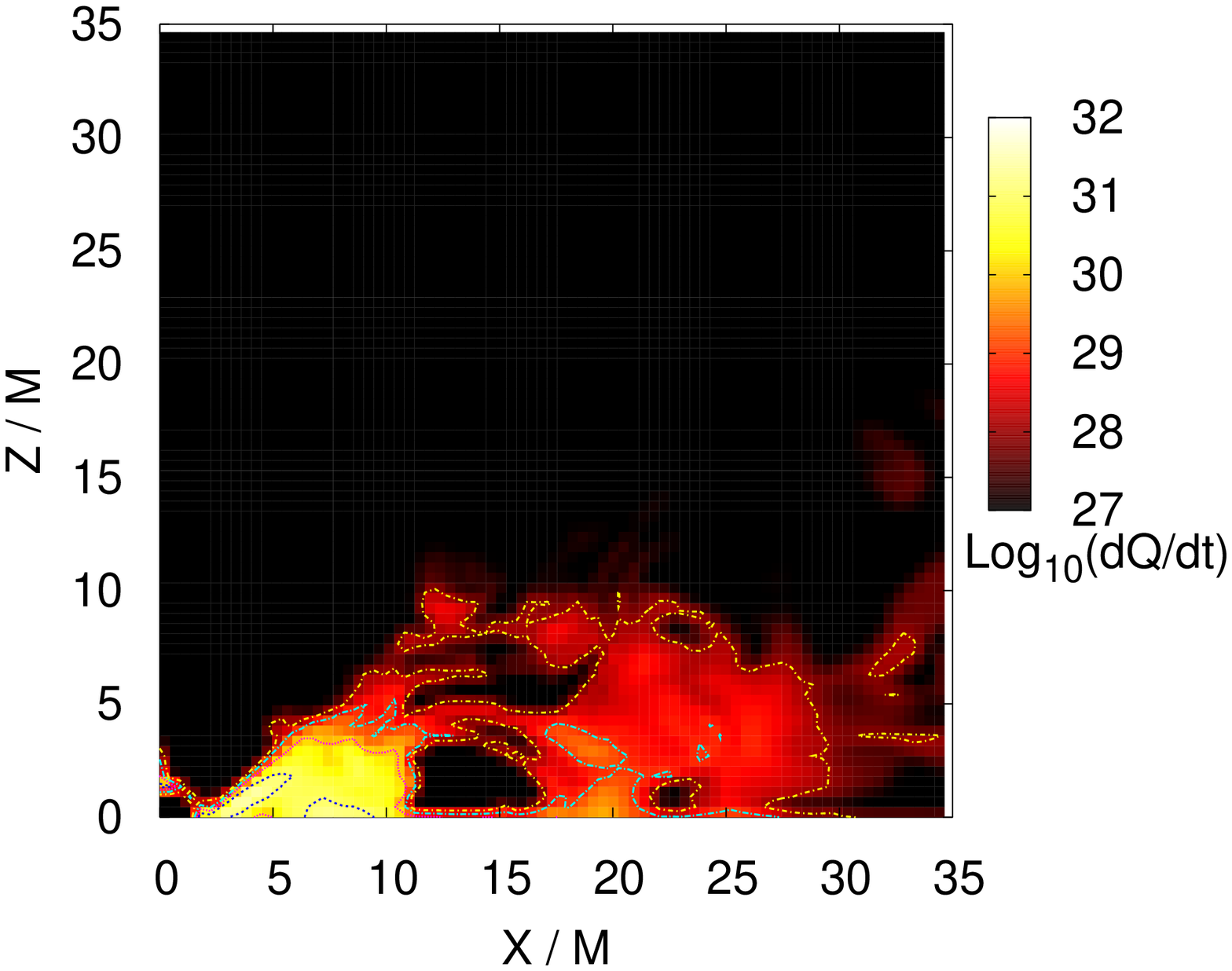}
\end{minipage}
\\
\vspace{-10mm}
\begin{minipage}[t]{0.47\textwidth}
(e)\includegraphics[width=7.cm]{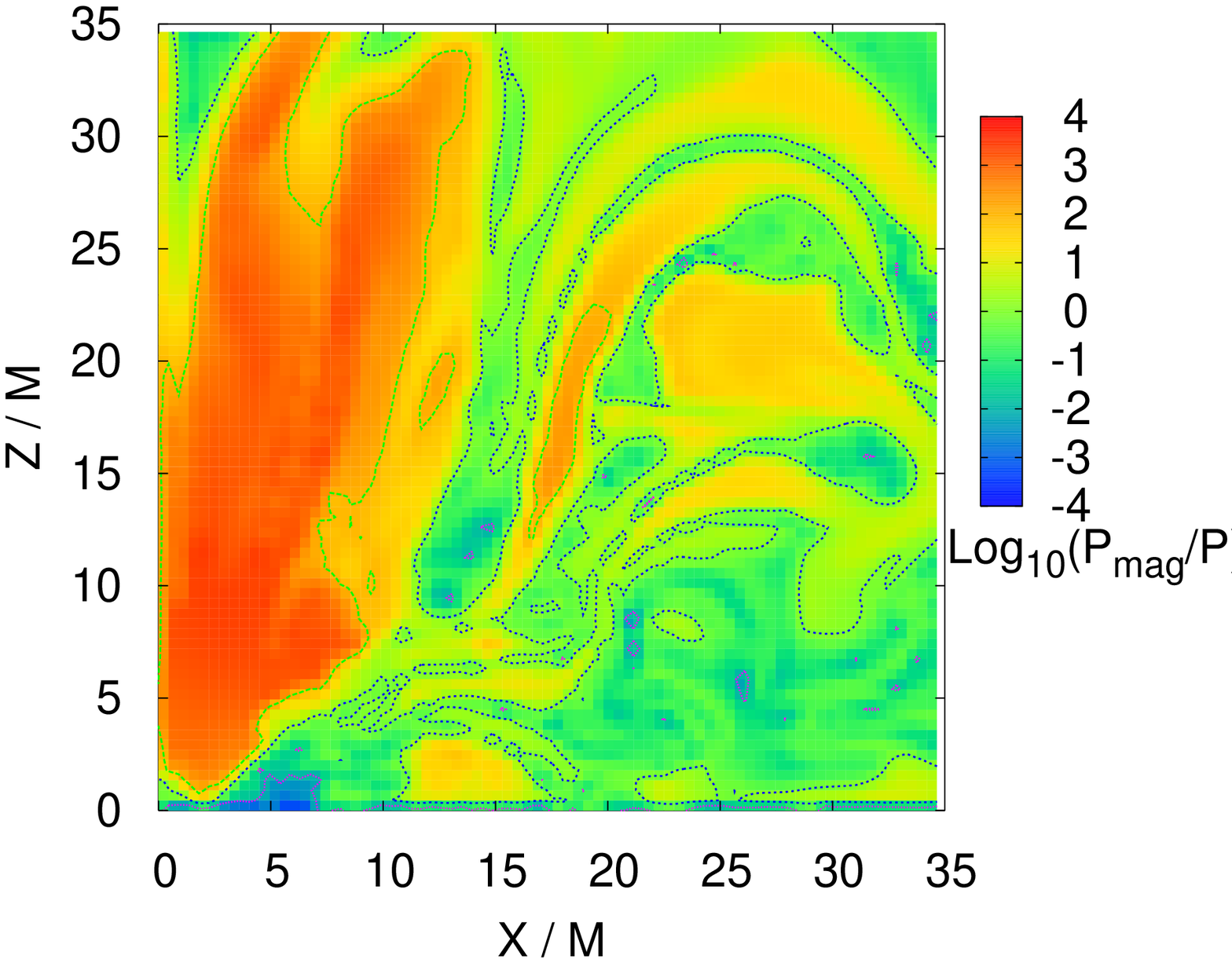}
\end{minipage}
\begin{minipage}[t]{0.47\textwidth}
(f)\includegraphics[width=7cm]{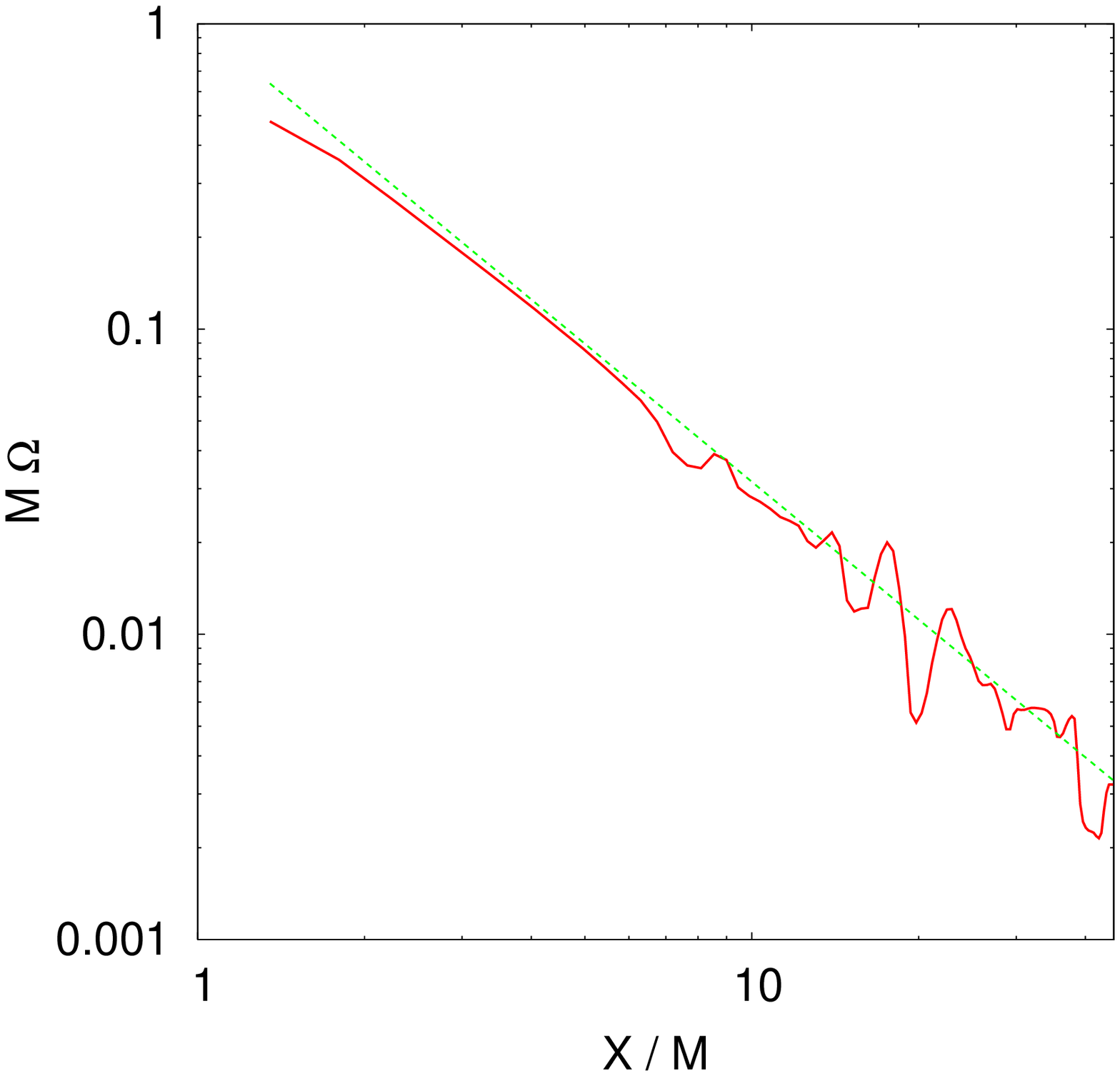}
\end{minipage}
\end{center}
\caption{Color plots for (a) the density, (b) the temperature, (c) the
electron fraction, (d) the neutrino emissivity, and (e) the ratio of
the magnetic pressure to the gas pressure at $t \approx 40$ ms for
model D with $\Delta/M=0.15$.  For (d), the region with $\dot Q <
10^{27}~{\rm ergs/s/cm^3}$ is shaded black. The contour curves are
plotted for (a) $\rho=10^{8+i}~{\rm g/cm^3}~(i=0$--3), (b)
$\log(T_{10})=0$, 0.4, and 0.8, (c) $Y_e=0.1 \times i~(i=1$--5, (d)
$\dot Q=10^{28+i}~{\rm ergs/s/cm^3}~(i=0,3)$, and (e) $P_{\rm
mag}/P=10^{-2+2i}~(i=0$--2).  (f) The angular velocity as a function
of the cylindrical radius on the equatorial plane (solid curve). The
dotted line denotes $M \Omega=(M/\varpi)^{3/2}$. Note that $M \approx
5.9$ km in the geometrical units.
\label{FIG6}}
\end{figure}

In Fig. \ref{FIG6}, we display color plots and contour curves for the
density, temperature, electron fraction, neutrino emissivity, and
ratio of the magnetic pressure to the gas pressure at $t \approx 40$
ms for model D.  At $t=40$ ms, the accretion torus relaxes to an
approximately quasistationary state.  The density maximum of the torus
with $\rho_{\rm max} \sim 10^{12}~{\rm g/cm^3}$ is located at $\varpi
\approx 5M$ on the equatorial plane. Only a small part of the inner
region with $\varpi \alt 10M$ and with $|z| \alt 2M$ has a density
larger than $10^{11}~{\rm g/cm^3}$ and is optically thick. The
temperature is also highest, $\sim 10^{11}$ K, for this high-density
region, because the cooling is not efficient for such a region, due to
its high opacity . The temperature of the optically thin region with
$\rho < 10^{11}~{\rm g/cm^3}$ is $\sim 3 \times 10^{10}$ K near the
equatorial plane. An interesting feature is that the temperature is
not uniformly low in the region above the torus. The reason for this
is that the magnetic stress in the torus induces an outflow which
ejects gas with high temperature. This dilute gas subsequently emits
neutrinos, and the temperature thereby quickly decreases. 

As reported in Ref. \citen{LRP}, the electron fraction inside the
torus is low, $\alt 0.1$. This reflects the fact that the density is
so high that the electrons are highly degenerate. A typical value of
the degeneracy parameter is $\eta_e\sim 2$--4 in the region with $\rho
\agt 10^{11}~{\rm g/cm^3}$.  Around the envelope of the torus in the low
density region ($\rho \alt 10^9~{\rm g/cm^3}$), by contrast, the
degeneracy is low, with $\eta_e <1$ for a large fraction of the fluid.

Neutrinos are efficiently emitted in the region satisfying $\varpi
\alt 10M$ and $|z| \alt 5M$. The neutrino emissivity is highest near
the surface of the torus (not at the density maximum), because the
region in the vicinity of the density maximum is so dense that the
neutrinos cannot escape. In particular, the region near the inner edge
of the torus has high emissivity.  A large fraction of the neutrinos
emitted from there will propagate toward the symmetry axis, where the
neutrinos can escape freely, and hence, the pair annihilation rate of
neutrinos and antineutorinos is expected to be highest in the vicinity
and around the symmetry axis of a rotating black hole.

The distribution of $P_{\rm mag}/P$ exhibits a clear contrast. In the
torus, the value of $P_{\rm mag}/P$ is much smaller than unity,
implying that the gas pressure is dominant. By contrast, the value
near the symmetry axis is much larger than unity, and hence, the gas
pressure is negligible there. This structure has already been found in
previous GRMHD simulations with a simple $\Gamma$-law EOS 
\cite{MG,VH}.  The present results show that the formation of such a 
distribution of $P_{\rm mag}/P$ does not depend on the EOS and the
neutrino cooling.

Figure \ref{FIG6}(f) plots the angular velocity along the cylindrical
radius in the equatorial plane.  This illustrates that the velocity of
the torus is approximately Keplerian in the quasistationary phase. At
the beginning of the simulation, the profile of the angular velocity is
steeper than this, as shown in Fig. \ref{FIG2}. Due to the magnetic
braking and the MRI, the angular momentum is redistributed, and the
profile of the angular velocity is modified and becomes Keplerian.  The
resulting torus is likely to be stable with respect to nonaxisymmetric
instabilities and the runaway instability because of the Keplerian
velocity profile.

\subsection{Dependence on the mass and the initial radius of the torus}
\label{sec:mass}

\begin{figure}[t]
\begin{center}
\begin{minipage}[t]{0.47\textwidth}
\includegraphics[width=6.9cm]{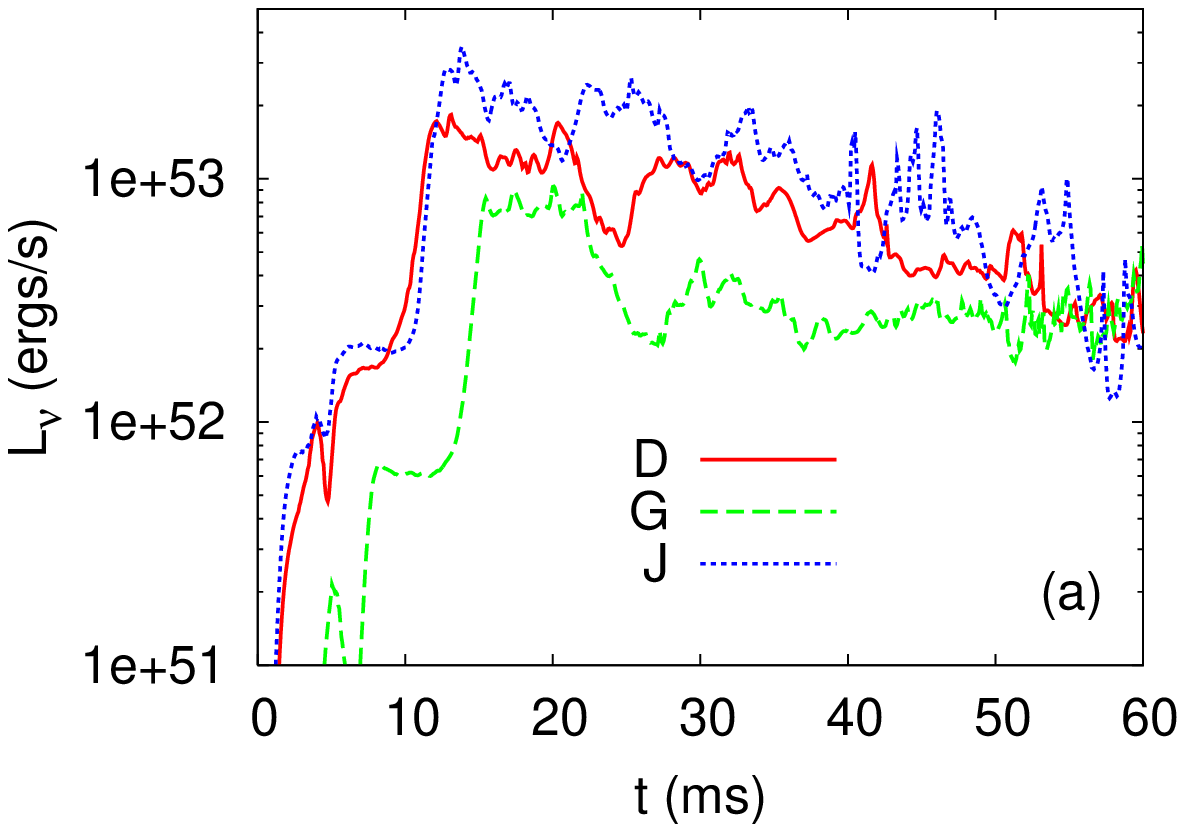}
\end{minipage}
\begin{minipage}[t]{0.47\textwidth}
\includegraphics[width=6.9cm]{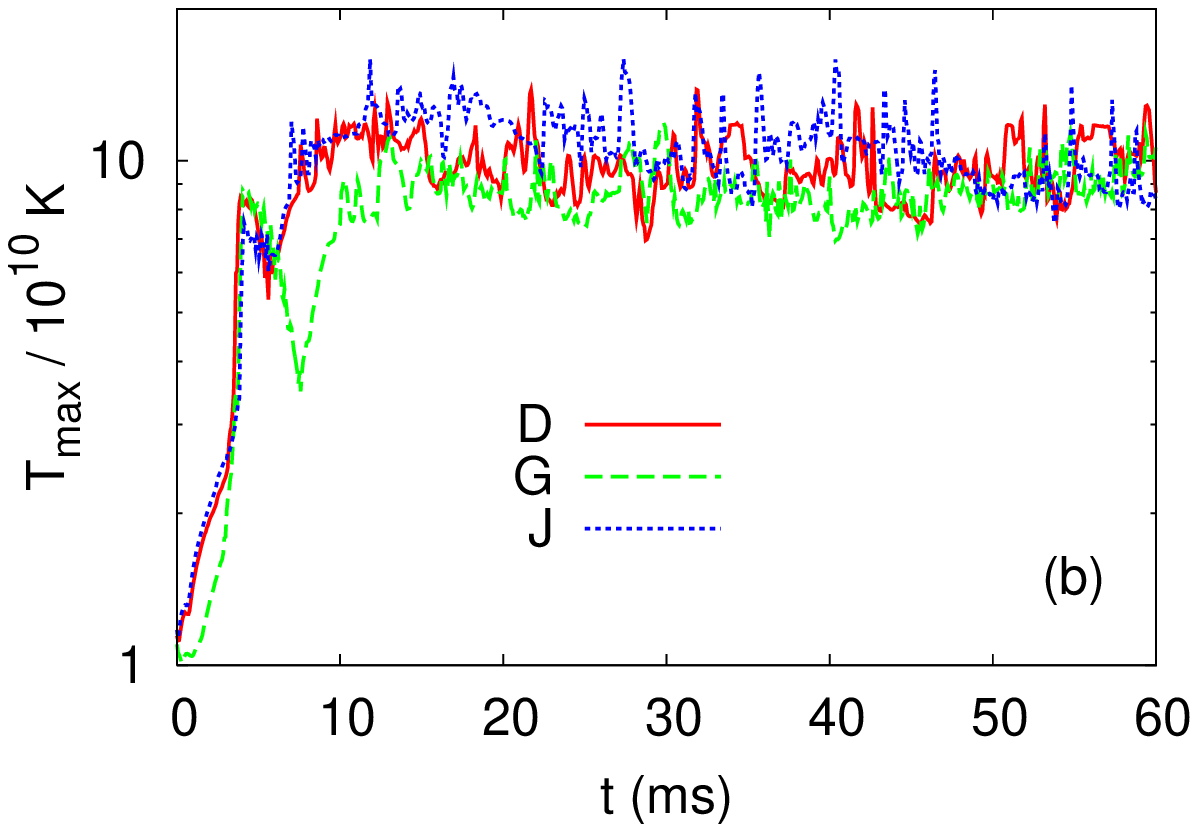}
\end{minipage}\\
\begin{minipage}[t]{0.47\textwidth}
\includegraphics[width=6.9cm]{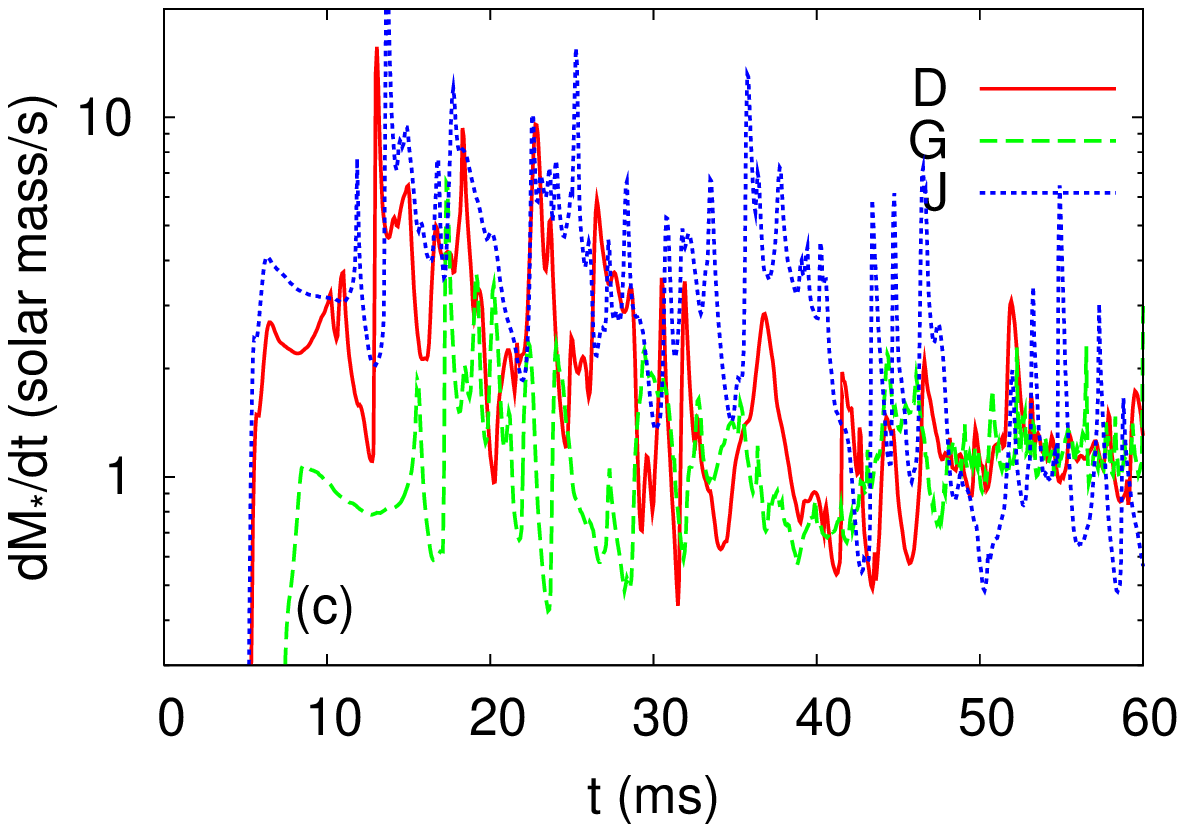}
\end{minipage}
\begin{minipage}[t]{0.47\textwidth}
\includegraphics[width=6.9cm]{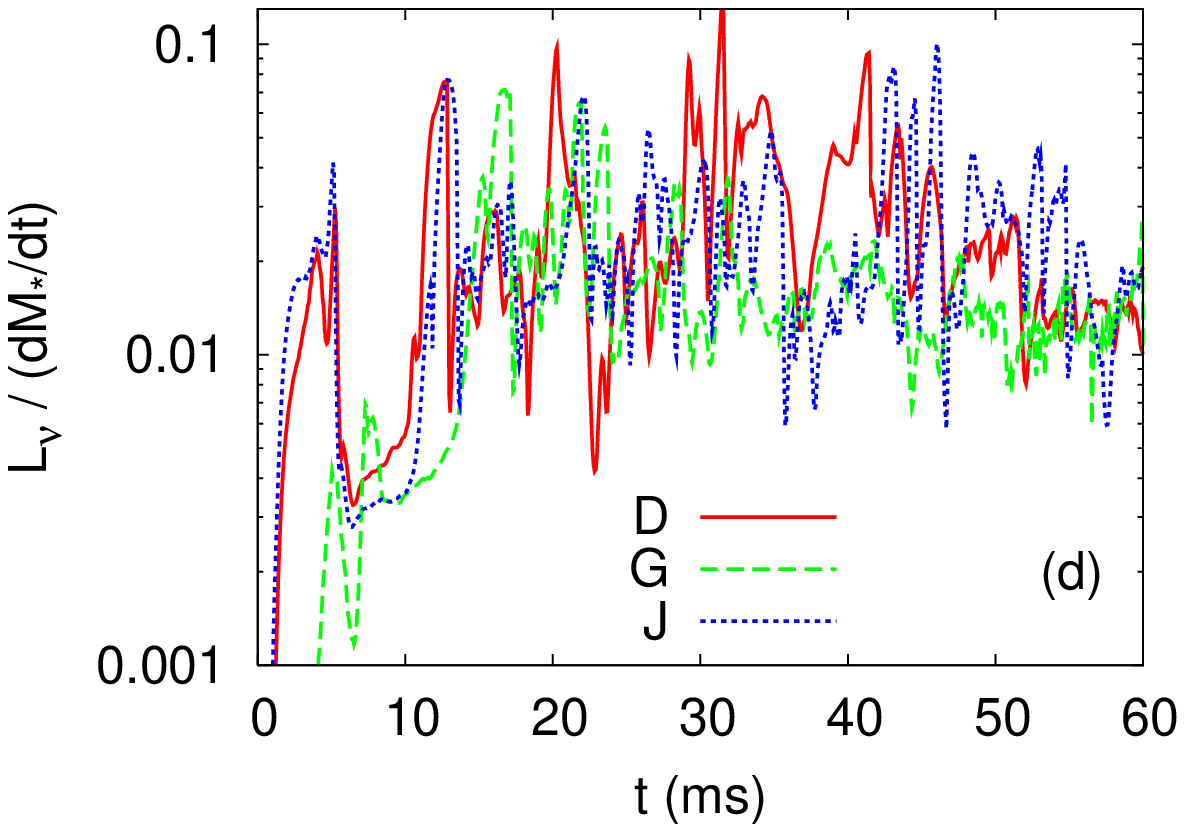}
\end{minipage}
\end{center}
\vspace{0mm}
\caption{Evolution of (a) the neutrino luminosity,
(b) the maximum temperature, 
(c) the mass accretion rate, and 
(d) the efficiency of the conversion to neutrinos $L_{\nu}/\dot M_*$,
for models D, G, and J with $\Delta/M=0.15$. Because the initial
rotation radii are slightly larger, the luminosity and temperature for 
model G rise at slightly later time than for other models. 
\label{FIG7}}
\end{figure}

In Fig. \ref{FIG7}(a), we plot the neutrino luminosity as a function
of time for models D, G, and J.  The mass of torus varies among these
models, while the spin parameter of the black hole is the same and the
initial rotation radii of the torus are approximately equal at
$t=0$. Figure \ref{FIG7}(a) shows that the luminosity systematically
increases as the mass of the torus increases. This is simply due to
the fact that there are more emitters for the larger-mass
models. Because the maximum density is slightly higher (see Table I),
there should be more trapped neutrinos for the larger-mass models,
suppressing the neutrino luminosity. However, this does not decrease
the neutrino luminosity significantly. The temperature is slightly
higher for the larger-mass models [see Fig. \ref{FIG7}(b)]. This is
one reason that the neutrino luminosity increases with the torus mass.

The mass accretion rate is larger for the larger-mass models in the
early phase, i.e., for $t \alt 40$ ms [see Fig. \ref{FIG7}(c)]. As a
result, at late times ($t \agt 40$ ms), the differences among the
mass accretion rates of the three models are relatively small, resulting in
relatively small differences among the neutrino luminosities. The
authors of Ref. \citen{SRJ} report that the neutrino luminosity is
approximately proportional to $M_*$ for the case of nonzero
$\alpha$-viscosity. In the present result, the maximum luminosity and
total energy carried by neutrinos are approximately proportional to
$M_*$.

Figure \ref{FIG7}(d) plots the conversion efficiency, $L_{\nu}/\dot
M_*$, as a function of time for models D, G, and J. Although the
neutrino luminosity differs significantly among three models, the
conversion efficiency, varying between $\sim 1$\% and $\sim 5\%$, does
not depend on the mass of the torus as strongly as the luminosity. The
average conversion efficiency, $\Delta E_{\nu}/\Delta M_*$, is also in
a narrow range between 1.4\% and 1.8\% for models G--J (see Table
III).

\begin{figure}[th]
\begin{center}
\begin{minipage}[t]{0.47\textwidth}
\includegraphics[width=6.9cm]{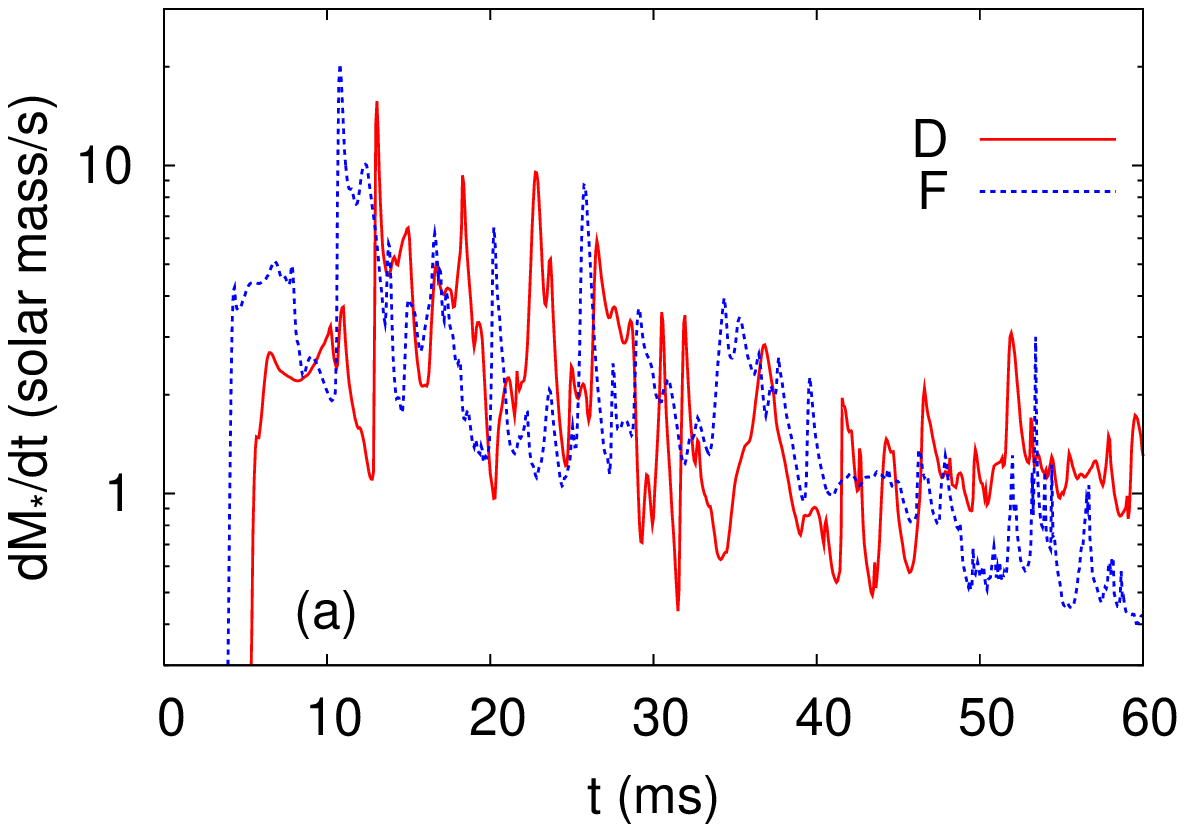}
\end{minipage}
\begin{minipage}[t]{0.47\textwidth}
\includegraphics[width=6.9cm]{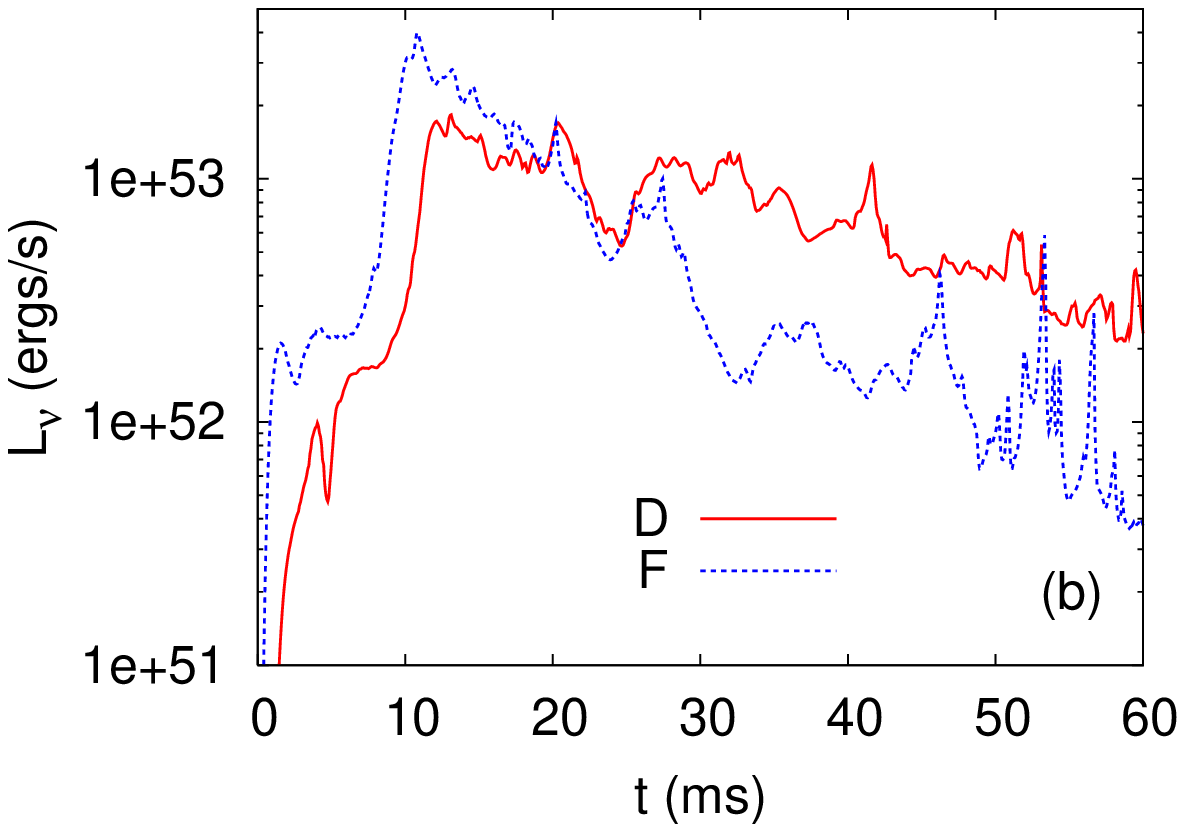}
\end{minipage}
\end{center}
\vspace{0mm}
\caption{Evolution of (a) the mass accretion rate, $\dot M_*$, and (b)
the neutrino luminosity, $L_{\nu}$, for models D and F with
$\Delta/M=0.15$.
\label{FIG8}}
\end{figure}

To see the dependence of the rest-mass accretion rate and neutrino
luminosity on the initial radius of the torus, we performed a
simulation for model F.  This model has the same black hole spin and
approximately the same torus rest-mass as model D, but it has a
smaller initial rotation radius. In Fig. \ref{FIG8}, we plot the mass
accretion rate and the neutrino luminosity for models D and F. Because
model F has a smaller radius initially, the mass accretion rate and
the neutrino luminosity quickly increase at earlier times. Also, the
peak value of the luminosity for model F is larger by a factor of
$\sim 2$ than that for model D, whereas the luminosity at late times
is smaller for model F.  However, the total emitted neutrino energy
and the efficiency of the conversion to neutrinos are similar 
(see Table III).

\subsection{Dependence on the black hole spin}\label{sec:spin}

\begin{figure}[t]
\begin{center}
\begin{minipage}[t]{0.47\textwidth}
\includegraphics[width=6.9cm]{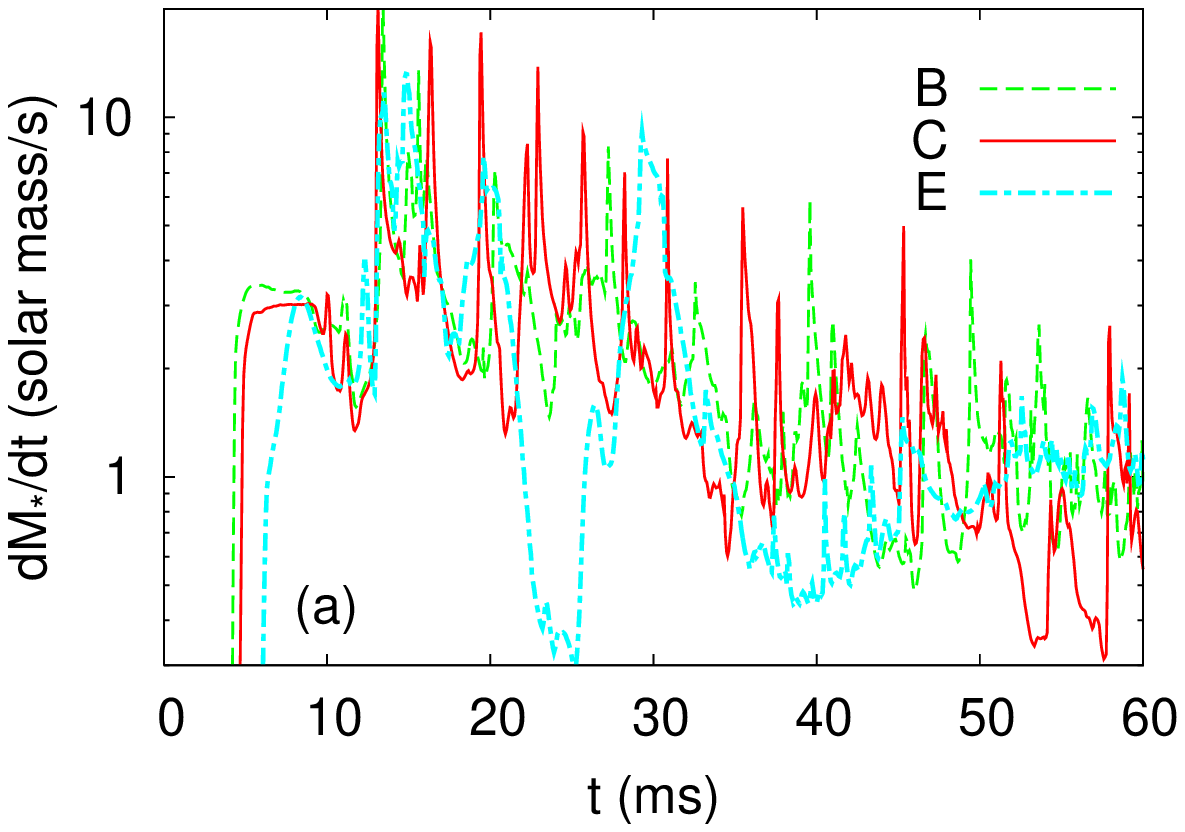}
\end{minipage}
\begin{minipage}[t]{0.47\textwidth}
\includegraphics[width=6.9cm]{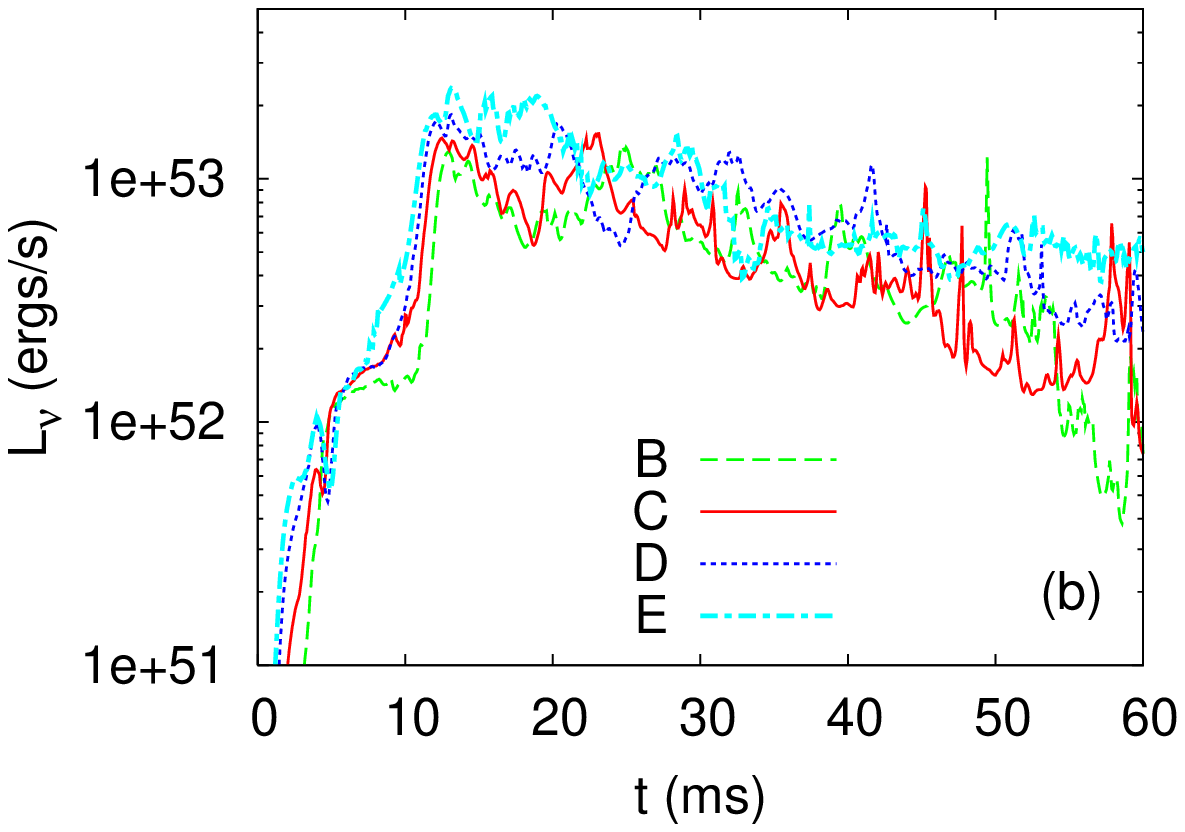}
\end{minipage}
\end{center}
\vspace{0mm}
\caption{Evolution of (a) the mass accretion rate
for models B, C, and E with $\Delta/M=0.15$.
(b) the neutrino luminosity, $L_{\nu}$,  
for models B, C, D, and E with $\Delta/M=0.15$.
\label{FIG9}}
\end{figure}

Figure \ref{FIG9} plots the mass accretion rate and the neutrino
luminosity as functions of time for models B--E. The black hole spin is
different for each of these models, but the mass and rotation radii of
the torus are approximately equal at $t=0$. We find the following. (i)
The luminosity for model E is largest among the four models, whereas
the mass accretion rate is smallest for model E, probably because its
event horizon has the smallest area. (ii) The luminosity increases
with $a$ by a significant amount only for $a \geq 0.5M$ (see also
Table III).  The magnitude of the luminosity varies only slightly for
$a/M=0$--0.5, while the luminosities for models B and C are
approximately the same. (iii) The decay time of the luminosity
increases with $a$ for $a/M \geq 0.5$.  The facts (i)--(iii) indicate
that for a sufficiently large value of the spin, $a \agt 0.75M$, the
effect of the black hole spin enhances the neutrino luminosity, but
for smaller values, the luminosity is not significantly enhanced by
the spin effect. In the following, we describe the reasons for these
types of behaviors. 

\begin{figure}[t]
\begin{center}
\begin{minipage}[t]{0.47\textwidth}
\includegraphics[width=6.9cm]{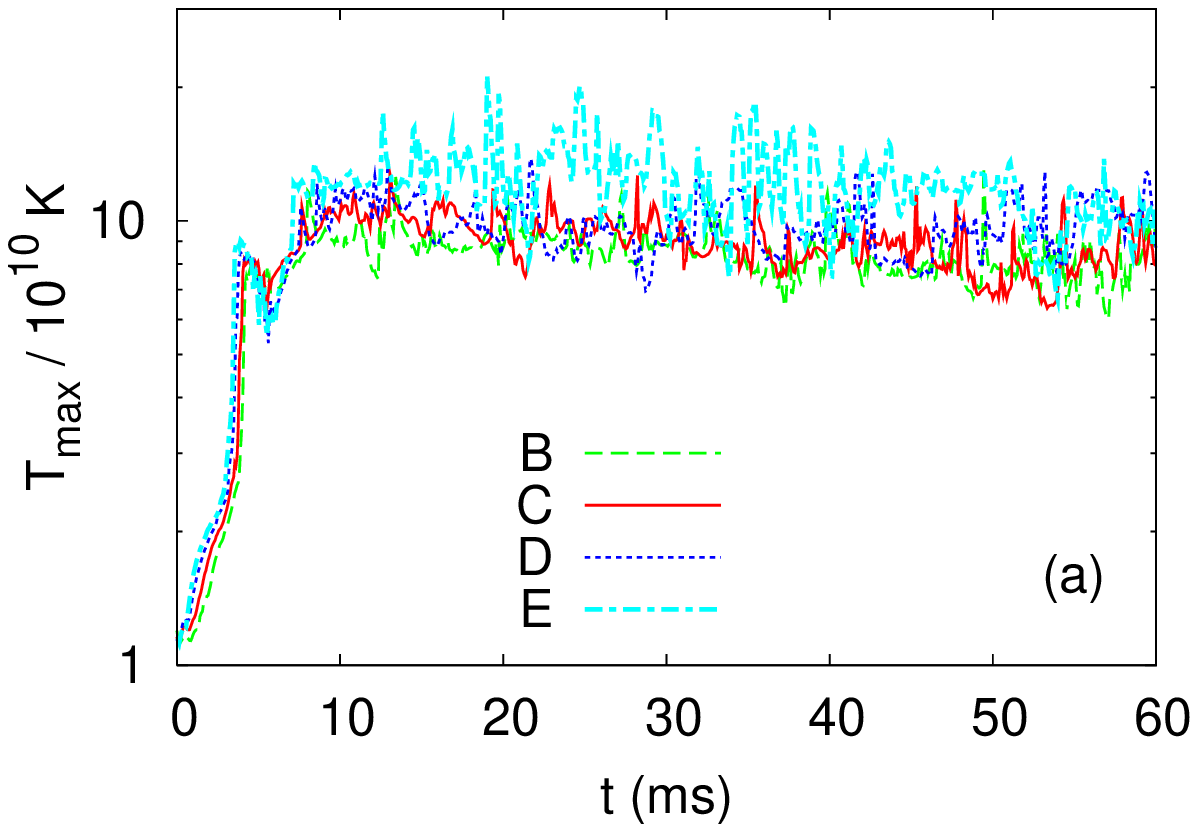}
\end{minipage}
\begin{minipage}[t]{0.47\textwidth}
\includegraphics[width=6.9cm]{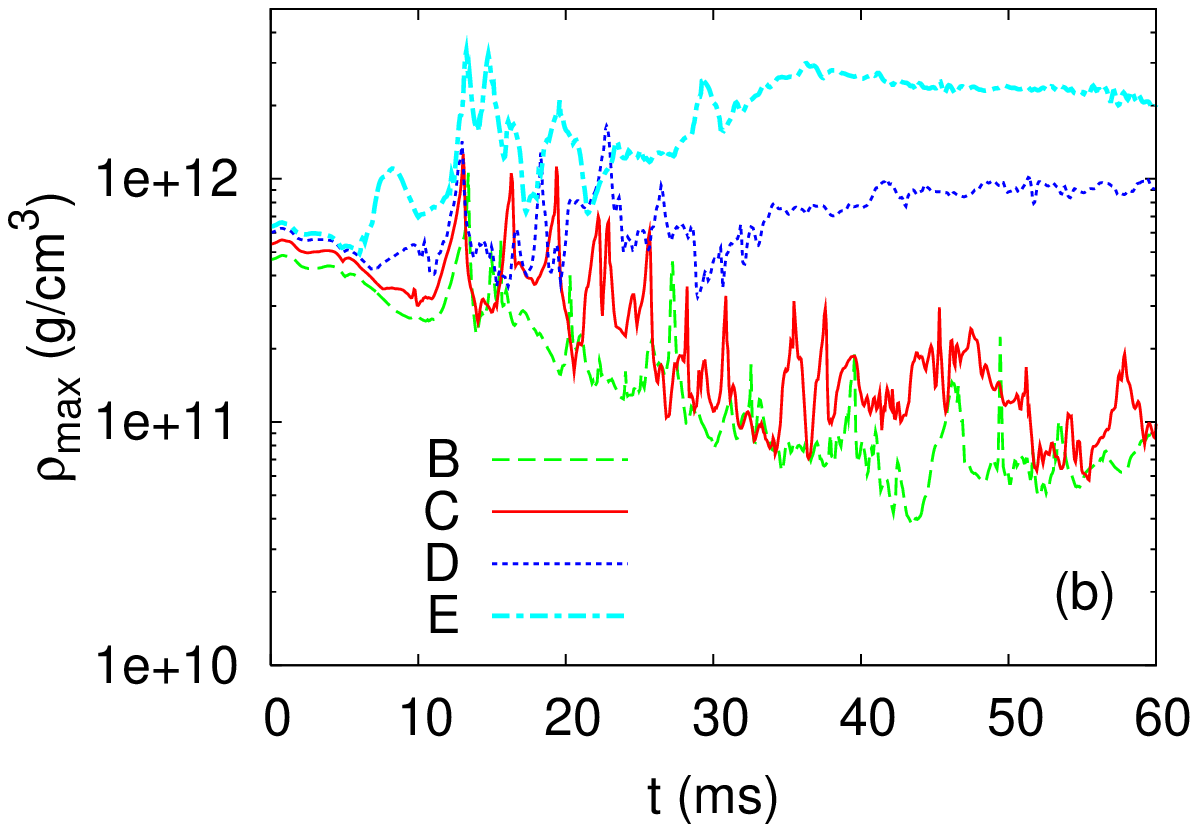}
\end{minipage}
\end{center}
\vspace{0mm}
\caption{Evolution of the maximum temperature, $T_{\rm max}$, and
density, $\rho_{\rm max}$, for models B--E with $\Delta/M=0.15$. 
\label{FIG10}}
\end{figure}

There are two factors which determine the dependence of the luminosity
on the black hole spin. One is the fact that with larger spin, the
radius of the ISCO decreases as the spin increases (cf. Table II).
Then, the gravitational binding energy at the ISCO ($1-E_{\rm ISCO}$),
which could be converted to thermal energy, increases as a function of
$a$ (cf. Table II).  Indeed, the maximum temperature also increases as
$a$ increases [see Fig. \ref{FIG10}(a)].  This results in an
enhancement of the efficiency for converting gravitational binding
energy into neutrinos.  Note that the gravitational binding energy at
the ISCO changes slowly as a function of $a$ for small values of
$a$. This implies that this effect is not very important for $a \ll
M$.

The small radii of the ISCO for larger values of the spin also result
in the decrease of the mass accretion rate and in the increase of the
accretion time.  To achieve a high neutrino luminosity, a longer
accretion time is favorable, because a fraction of the thermal energy
trapped with the matter flowing into black hole is reduced. Thus, the
case of a large spin has an advantage with regard to enhancing the
neutrino luminosity due to (i) the large binding energy at the ISCO
and (ii) the long accretion time.

The other factor is that for a larger value of $a$, the rotation
radius of the torus can be smaller. This results in a more compact
torus which has a higher temperature and larger density. As a result
of the larger density, (i) the trapped fraction of neutrinos is
increased, whereas (ii) the neutrino luminosity could be increased
because of the higher temperature and larger density. Figure
\ref{FIG10} plots the evolution of the maximum temperature and density
for models B--E. For model B, the maximum density gradually decreases
below $10^{11}~{\rm g/cm^3}$, which implies that no trapped neutrinos
exist in the late phase. For model C, $\rho_{\rm max}$ is slightly
larger than $10^{11}~{\rm g/cm^3}$, and hence, a small fraction of
neutrinos are trapped. For models D and E, in which $\rho_{\rm max}
\sim 10^{12}~{\rm /cm^3}$, a large fraction of neutrinos are trapped,
whereas the emissivity is enhanced by the large density.\footnote{The
increase of the maximum density as a function of the spin is partly
due to the fact that the mass accretion rate decreases as a function
of the spin.}

For models B and C, the above two factors seem to play an accidently
identical role, resulting in similar luminosities.\footnote{In the
discussion of this section, we use numerical results obtained with
$\zeta=1$. If we instead used $\zeta=1/3$, the situation would be
different. In this case, the fraction of trapped neutrinos is small
for model C2, and hence, the luminosity for this model is larger than
that for model B2.}  For models D and E, on the other hand, the
enhancement of the luminosity at high temperature and high density
plays a stronger role than the effect of the opacity. Because the
luminosity for models D and E is much larger than for models A--C, the
longer accretion time and high temperature resulting from the large
spin play the most significant role in enhancing the luminosity.

In the present numerical work, there is the additional reason that the
conversion efficiency increases rapidly with the spin for $a/M \agt
0.75$. As mentioned above, the radius of the ISCO for black holes of
larger spins is smaller, and a larger fraction of the rest-mass energy
can be converted into thermal energy (cf. Table II). In the present
model, the density maximum for the torus is initially located at $\sim
10M$. For smaller values of the spin, then, the possible conversion
rate for matter near the density maximum is much smaller than the
maximum allowed value given in Table II, because the difference between
the specific binding energy at the ISCO and at the density maximum is
much smaller than the specific binding energy at the ISCO. For large
values of the spin with $a \rightarrow M$, by contrast, the difference
is close to the specific binding energy at the ISCO.  This is one
reason that the conversion efficiency for models D and E is much
larger than those for other models.

\subsection{Dependence on $\zeta$} \label{sec:3.2}

\begin{figure}[t]
\begin{center}
\begin{minipage}[t]{0.47\textwidth}
\includegraphics[width=6.9cm]{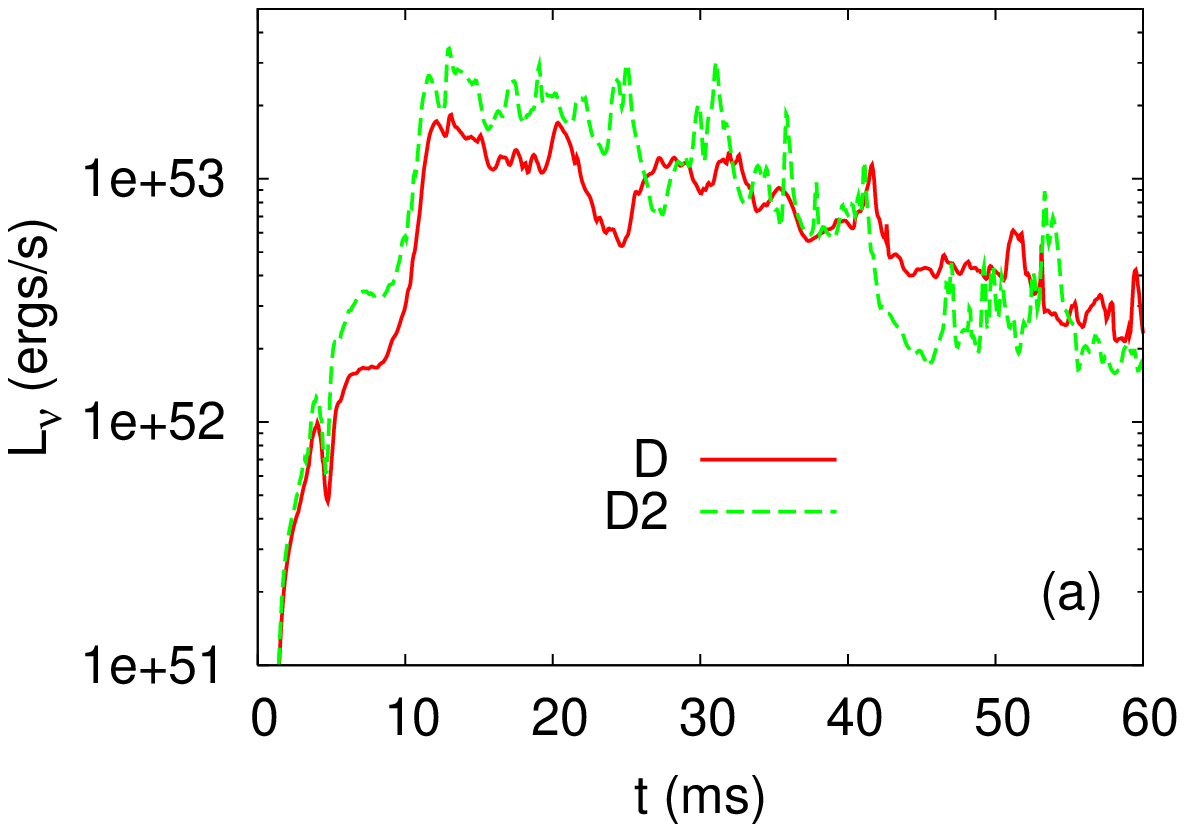}
\end{minipage}
\begin{minipage}[t]{0.47\textwidth}
\includegraphics[width=6.9cm]{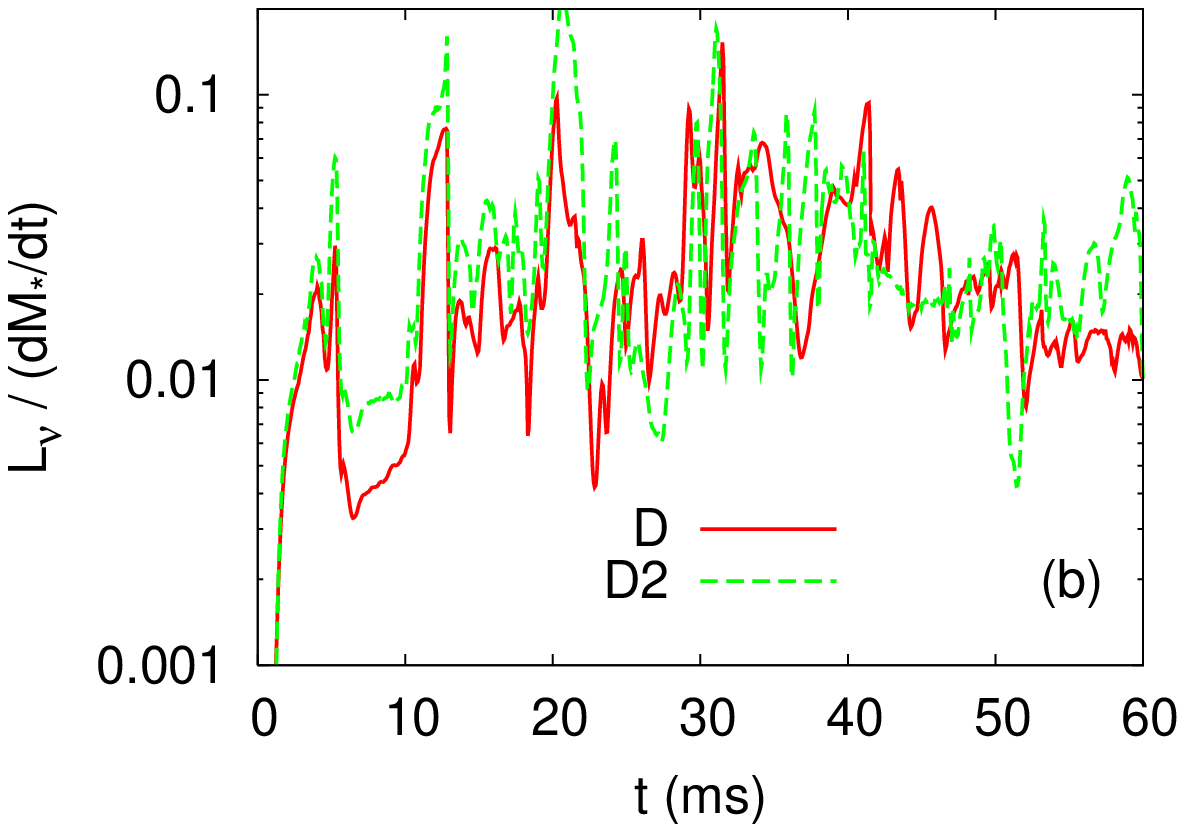}
\end{minipage}\\
\begin{minipage}[t]{0.47\textwidth}
\includegraphics[width=6.9cm]{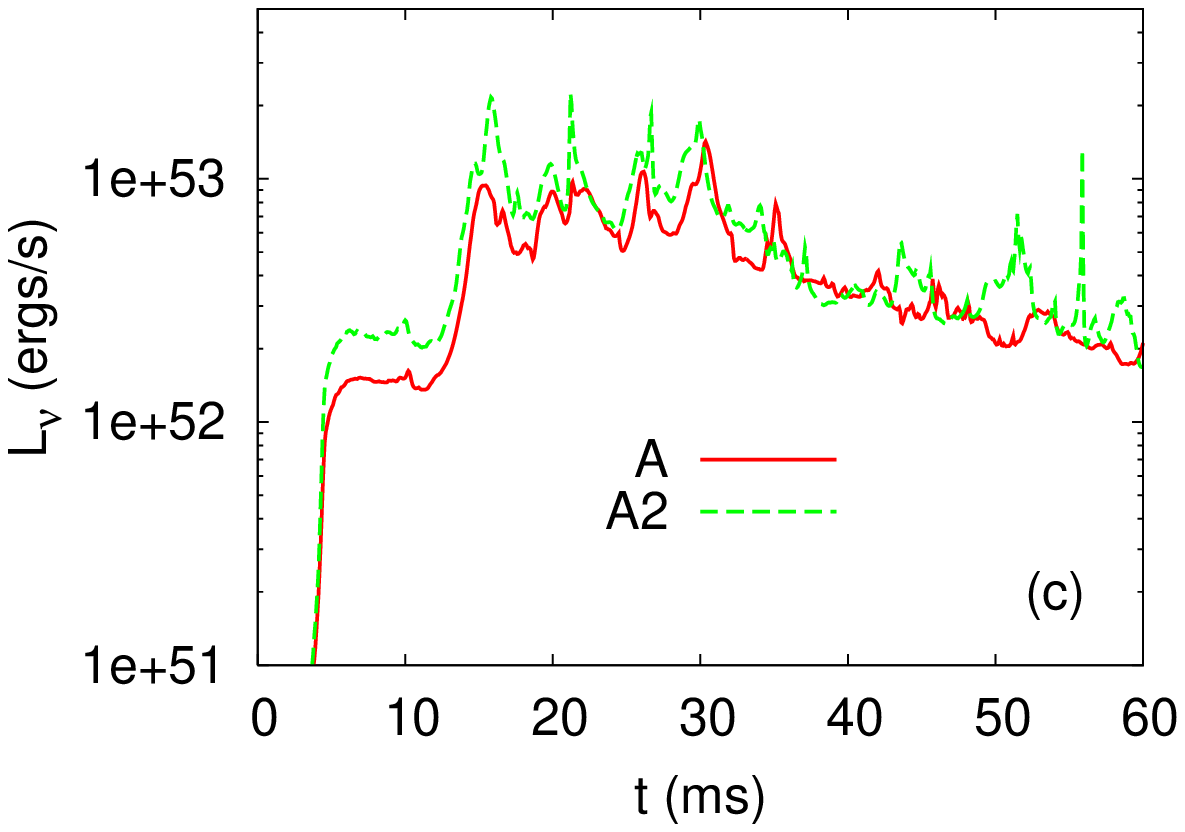}
\end{minipage}
\begin{minipage}[t]{0.47\textwidth}
\includegraphics[width=6.9cm]{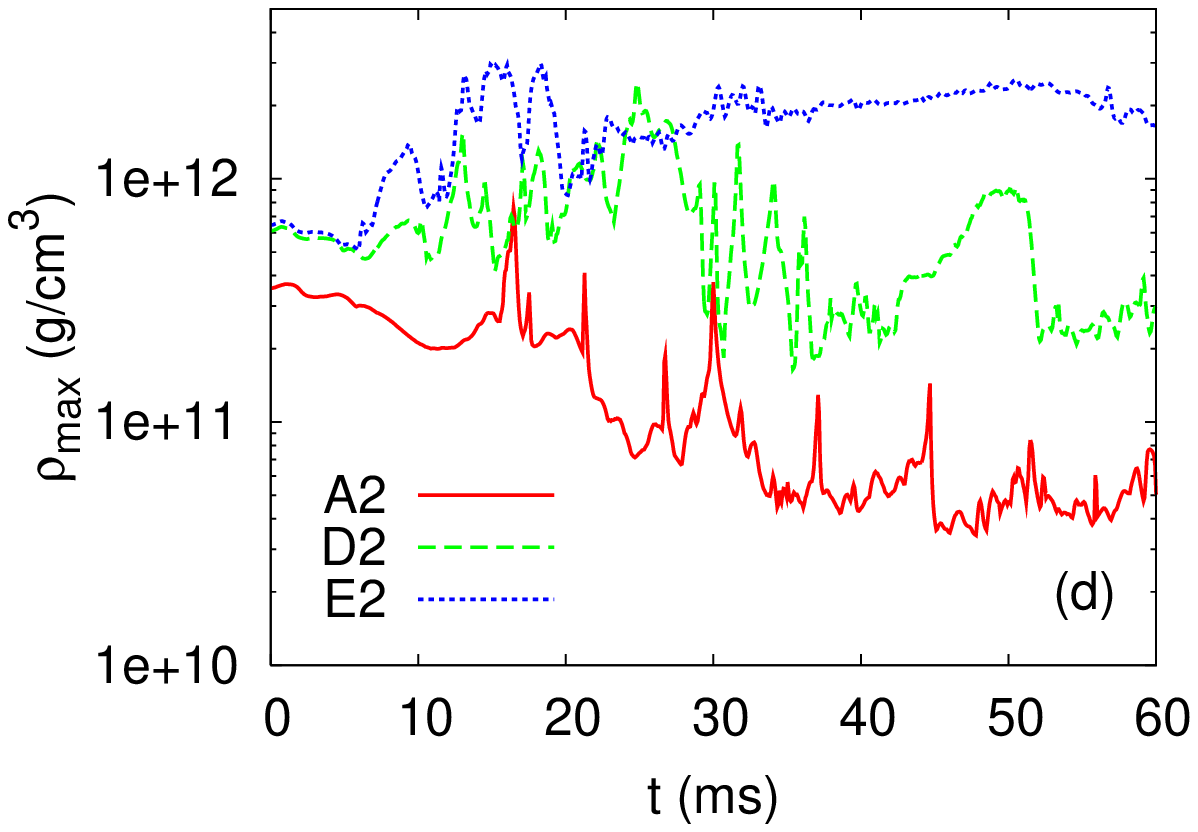}
\end{minipage}\\
\begin{minipage}[t]{0.47\textwidth}
\includegraphics[width=6.9cm]{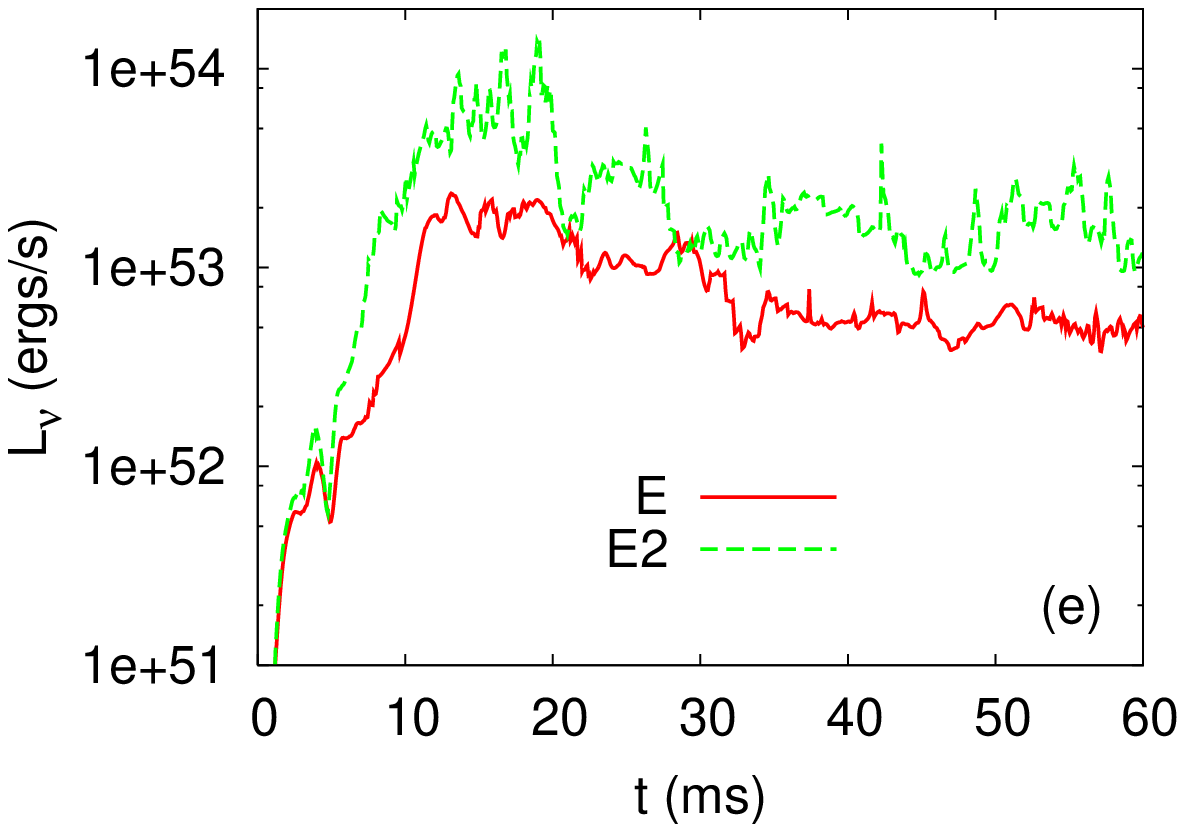}
\end{minipage}
\begin{minipage}[t]{0.47\textwidth}
\includegraphics[width=6.9cm]{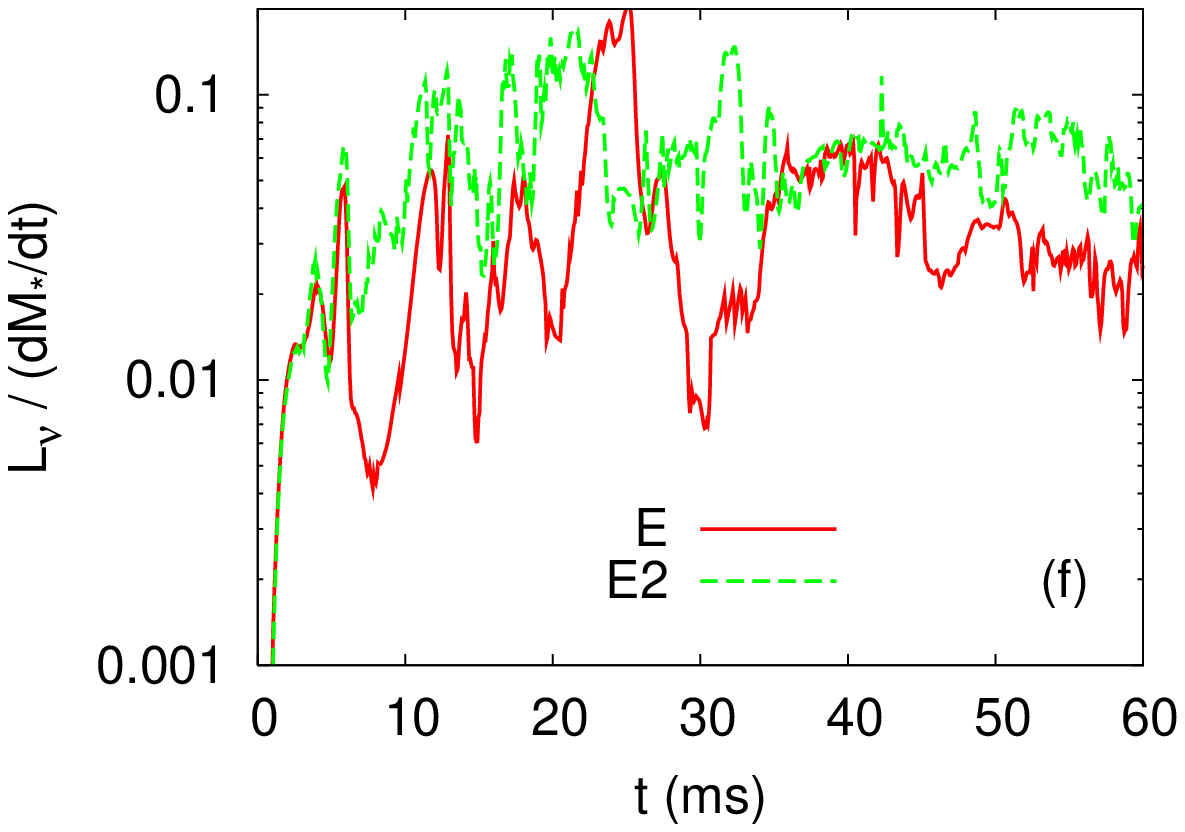}
\end{minipage}
\end{center}
\vspace{0mm}
\caption{Evolution of (a) the neutrino luminosity, $L_{\nu}$, and (b)
the efficiency of the conversion to neutrinos, $L_{\nu}/\dot
M_*$, for models D and D2 with $\Delta/M=0.15$.  (c) The same as (a),
but for models A and A2 with $\Delta/M=0.2$.  (d) Evolution of the
maximum density for models A2, D2, and E2. (e) The same as (a), but for
models E and E2 with $\Delta/M=0.15$.  (f) The same as (b) but for
models E and E2 with $\Delta/M=0.15$.
\label{FIG70}}
\end{figure}

In this subsection, we compare the results for pair models for which
the black hole spin and properties of the torus are identical but the
rule for identifying the optically thick region for neutrinos is
different.  We determine that the region with $\rho_{11} \geq 1$ is
optically thick for models A--E, whereas the region with $\rho_{11}
\geq 3$ is for models A2--E2.

Figure \ref{FIG70}(a) plots the evolution of the neutrino luminosity,
$L_{\nu}$, and the conversion efficiency, $L_{\nu}/\dot M_*$, for
models D and D2. Because the optically thin region is wider, the
neutrino luminosity for model D2 is higher than that for model D, in
particular in the early phase, with $t \alt 30$ ms. The relative
difference is $\sim 30$--40\%. 

The conversion efficiency, $L_{\nu}/\dot M_*$, for model D2 is larger
than that for model D, reflecting the higher luminosity for model D2.
It often reaches 10\% and is $\sim 2.5\% ( \sim \Delta E_{\nu}/\Delta
M_*c^2$) on average.  Nevertheless, it is smaller than the maximum
value, $\sim 11\%$. Recalling that the accretion time scale, $M_*/\dot
M_* \sim 100$ ms, is approximately as long as the neutrino emission
time scale, $E_{\rm int}/L_{\nu}$, this implies that a large fraction
of the neutrinos are still trapped by the matter and advected into the
black hole, thus failing to escape from the torus. Indeed, the maximum
density for model D2 is larger than $3 \times 10^{11}~{\rm g/cm^3}$
for most of the time [see Fig. \ref{FIG70}(d)].

Figure \ref{FIG70}(c) plots the evolution of the neutrino luminosity 
for models A and A2. As in models D and D2, the luminosity
for model A2 is larger than that for model A.  However, the difference
is not as large as that between models D and D2.  The reason for this
smaller difference is that the density of the torus for models A and A2 is
not as large as that for models D and D2 [see Figs. \ref{FIG10}(b) and
\ref{FIG70}(d)]. In particular, for $t \agt 20$ ms, the density is
smaller than $10^{11}~{\rm g/cm^3}$, and hence, no region is optically
thick for models A and A2.  We conclude that for a torus with small
values of $a$ and with mass $\approx 0.25M_{\odot}$, the neutrino
luminosity depends only weakly on the value of $\zeta~(\agt 1/3)$.

The same conclusion is reached for tori of small mass which would have
densities smaller than $10^{11}~{\rm g/cm^3}$. (Compare the maximum
densities for models D, F--J in Table I.) Actually, only a small
fraction of a torus with $M_* \alt 0.1M_{\odot}$ is optically thick
for $a \leq 0.75M$. For $a \sim 0$, the neutrino-trapping effect plays
an important role only for massive tori with $M_* \agt
0.3M_{\odot}$. For large values of $a$, i.e., for $a \agt 0.75M$, 
by contrast, neutrino trapping plays a role even for $M_* \sim
0.1M_{\odot}$.

In Fig. \ref{FIG70}(e), we plot the evolution of the neutrino
luminosity for models E and E2. In contrast to the results for $a \leq
0.75M$, the luminosities for $\zeta=1$ and 1/3 differ
significantly. The reasons are that (i) for $a=0.9M$, the rotation
radii of the torus can be small enough ($\varpi \sim 2.5M$; see Table
II) to constitute a compact torus and that (ii) the accretion time
scale is long enough to halt quick accretion for large-spin black
holes. Consequently, the density can increase to $\agt 10^{12}~{\rm
g/cm^3}$ even with $M \approx 0.25M_{\odot}$ [see
Fig. \ref{FIG10}(b)].  Because the optically thick region is large for
such accretion flow, the decrease of $\zeta$ significantly increases
the size of the optically thin region and the luminosity.

In Fig. \ref{FIG70}(f), the evolution of the conversion efficiency,
$L_{\nu}/\dot M_*$, is plotted for models E and E2. It is seen that
the conversion efficiency for model E2 is about 2.5 times as large as
that for E.  For $a=0.9M$, the hypothetical maximum conversion
efficiency is about 15.5\% (cf. Table II). For model E2, the average
conversion efficiency ($\Delta E_{\nu}/\Delta M_*$) is $\sim 6\%$,
which is thus $\sim 40\%$ of the maximum value.  For $a \sim M$,
approximately $40\%$ of the rest-mass energy may be converted into
thermal energy. \cite{ST} Our results for model E2 suggest that a
conversion efficiency of $\sim 20\%$ may be possible for $a \sim M$. 

We determine the neutrino optical depth in a simple qualitative way,
using the local density. In an actual system, the optical depth
depends on the density distribution, temperature profile, neutrino
energy, geometry of the black hole, and path of neutrinos in the
curved geometry. Although the assumption $\tau_{\nu}=\zeta \rho_{11}$
is not bad qualitatively, the error is not small quantitatively:
Comparing the results for $\zeta=1$ and $1/3$, we infer that the error
in the luminosity is approximately a factor of 2--3 for large values
of the black hole spin. To obtain the neutrino luminosity more
accurately, it is necessary to adopt a more sophisticated method for
determining the optical depth, in particular for large values of
$a$. This is beyond scope of this paper and left for future study.

\section{Summary and discussion}

\subsection{Summary}

We have reported our first numerical results of a GRMHD simulation for
neutrino-cooled accretion tori. We solved the GRMHD equations in the
fixed gravitational field of Kerr black holes of mass $4M_{\odot}$
with a realistic EOS and with neutrino cooling. The simulation was
carried out systematically for a wide range of values of the black
hole spin and the torus mass. Below we summarize the results of the 
numerical simulation.

\begin{itemize}
\item In the presence of both poloidal magnetic fields and
differential rotation, the magnetic field strength is amplified by the
winding of the field lines and by the MRI until the electromagnetic
energy reaches $\sim 10\%$ of the rotational kinetic energy.  Then,
the magnetic stress induces angular momentum transport via magnetic
braking and the MRI, resulting in a quasistationary accretion onto the
black hole. It also drives turbulent motion of matter, which
subsequently generates shocks that convert kinetic energy into thermal
energy. Through these processes, the temperature of the torus
increases typically to $\sim 10^{11}$ K.  The ratio of the
electromagnetic energy to the rotational kinetic energy is maintained
at $\sim 10\%$ in the quasistationary accretion phase. This
electromagnetic energy is comparable to the internal energy in the
quasistationary phase.
\item The maximum density of the torus takes values in a wide range,
between $\sim 10^{10}~{\rm g/cm^3}$ and $\sim 10^{12}~{\rm g/cm^3}$, 
depending on black hole spin $a$ and mass of the torus $M_*$. For larger
values of $a$, the maximum density tends to be higher for a given
value of $M_*$, because (i) the location of the ISCO is closer to the
horizon, leading to a more compact torus, and
(ii) the accretion rate is suppressed by the small ISCO radius,
halting the infall of the matter and resulting in the formation of
a higher-density region near the ISCO. For a larger torus mass, the
maximum density becomes larger.
\item In the case that the density is sufficiently high, some of
the neutrinos are trapped in the accretion flow, which suppresses the neutrino
emission rate. This tends to happen for large values of $a$ and for
large values of the torus mass.
\item Before the torus relaxes to a quasistationary state,
the accretion rate reaches $\sim 10 M_{\odot}$ /s, but after it
relaxes, a typical accretion rate is $\sim M_{\odot}$ /s. The
corresponding accretion time scale is $\sim 50$ ms in the early phase,
but it relaxes to 100--200 ms in the quasistationary state.  This 
accretion rate is in agreement with that in the case that an 
$\alpha$-viscosity in the range $\alpha_v=0.01$--0.1 is 
included.
\item The maximum neutrino luminosity is a few $\times 10^{53}$ ergs/s
for the torus mass $M_* \approx 0.25M_{\odot}$, irrespective of the
value of the black hole spin for $0 \leq a/M \leq 0.9$.
(For $a=0.9M$ and $\zeta=1/3$, the maximum reaches exceptionally
$10^{54}$ ergs/s.) In the
quasistationary phase, it is between $10^{52}$ ergs/s and $10^{53}$
ergs/s, which depends strongly on the black hole spin. The efficiency
of the conversion to neutrinos, $L_{\nu}/\dot M_*$, is between 1 and
10 \%.  This value depends on the black hole spin and the effect of
neutrino trapping. The total emitted neutrino energy is $2 \times
10^{51}$--$2 \times 10^{52}$ ergs for $M_* \approx
0.25M_{\odot}$. This also depends strongly on the black hole spin and
the effect of neutrino trapping.
\item The neutrino luminosity, $L_{\nu}$, and the total emitted
neutrino energy depend strongly on the mass of the torus, $M_*$. The
maximum value of the luminosity and the total emitted energy are 
approximately proportional to $M_*$ for $M_* \alt 0.4M_{\odot}$.
\item Neutrinos are emitted efficiently in the region with $\varpi \alt
10M$ and $|z| \alt 5M$. In particular, the neutrino emissivity is
highest near the inner surface of the accreting torus. 
\item The neutrino luminosity and the conversion efficiency for $a\agt
0.75M$ are larger by a factor of $\agt 2$ than those for $a=0$.
However, moderate values of the spin ($a \alt 0.5M$) do not help to
significantly increase the luminosity for a torus of mass
$M_*=0.1$--0.4$M_{\odot}$.
\item If the accretion flow is optically thin with respect to neutrino
transport, the efficiency of the conversion to neutrinos may be
larger than $\sim 10\%$ for $a \agt 0.9M$. 
\end{itemize}

\subsection{Implications for GRBs}

Because neutrinos are emitted most efficiently near the inner surface
of accreting torus, neutrino-antineutrino pair annihilation is
expected to occur in the vicinity of the rotation axis of black
hole. As a result, a pair plasma of electrons and positrons may be
generated, forming a fireball near the rotation axis. If the energy
density of this fireball is sufficiently high, it can drive
GRBs.\cite{GRB} Setiawan et al. \cite{SRJ} estimated the luminosity of
gamma rays, and found that it could be $E_{\nu\bar\nu}\sim
10^{50}$--$10^{51}$ ergs/s for $L_{\nu} \sim 10^{53}$ ergs/s. Such a
large value would be sufficiently high to generate GRBs if the total
baryon mass in the vicinity of the rotation axis is sufficiently
small. \cite{AJM} We find that the neutrino luminosity is $\sim
10^{53}$ ergs/s for a torus mass of $\agt 0.2M_{\odot}$ and black
hole mass of $4M_{\odot}$, irrespective of the black hole spin. It is
believed that such a system may act as the central engine of GRBs
(specifically, GRBs of short duration). In particular, for large
values of $a$ close to unity, the efficiency of the conversion to
neutrinos could be high, and a strong GRB would then be driven.

In our present treatment, the heating of matter by neutrinos emitted
from the torus and neutrino-antineutrino pair annihilation are not
taken into account. In an actual system, neutrino heating will
help the formation of matter outflow, which will contribute to sweeping
baryons around the rotation axis. 

The neutrino luminosity decreases as a function of the torus mass. For
a torus of mass $\sim 10^{-2}M_{\odot}$, the luminosity is likely to
be at most $\sim 10^{52}$ ergs/s. Because the pair annihilation rate
of a neutrino-antineutrino process is proportional to the square of
the neutrino luminosity, $E_{\nu\bar\nu}$ is likely to be $\alt
10^{49}$ ergs/s, and hence, tori of small mass are unlikely to drive
the observed GRBs. Numerical simulations have shown that the merger of
binary neutron stars of sufficiently large mass can produce a system
consisting of a rotating black hole and a torus of spin $a/M=0.7$--0.8
(e.g., Refs. \citen{STU,OJ}).  However, the torus mass is in general
small, $\ll 0.1M_{\odot}$, unless the mass ratio of the two neutron
stars is sufficiently small, $\alt 0.7$--0.8. Even for a small mass
ratio of $\sim 2/3$, the torus mass is at most $\sim
0.1M_{\odot}$. This indicates that the merger of binary neutron stars
with sufficiently large total mass could result only in weak GRBs in
most cases. If the total mass of binary neutron stars is not large
enough for the direct formation of a black hole, a hypermassive
neutron star is formed.\cite{STU} (See Ref. \citen{BSS} for the
definition of the hypermassive neutron star.) If it is strongly magnetized,
such a hypermassive neutron star will collapse and become a rotating
black hole with a massive torus, due to the transport of angular
momentum induced by magnetic effects. \cite{DLSSS} It is believed that
the resulting system of a black hole and a torus may be the central
engine of GRBs, in contrast to the case that the total mass of the
binary neutron stars is large enough for the direct formation of black
hole. Another possibility as the source of GRBs is the merger of black
holes and neutron stars. If the mass of the black hole is not large
($\alt 4M_{\odot}$), the resulting mass of the torus around the black
hole can be $\agt 0.1M_{\odot}$. \cite{BHNS0,BHNS1,SU06} Therefore, it
is regarded as a strong possibility as the source of GRBs.

A characteristic feature found in the MHD simulation is that the
luminosity varies on a short time scale of a few ms, in contrast to
the results reported in Refs. \citen{SRJ} and \citen{LRP}. This is due
to the fact that shock heating in the magnetized accretion torus is
associated with turbulent matter motion driven by a highly variable
magnetic stress. The light curve of GRBs is often highly
variable.\cite{GRB} If the GRBs are driven by pair annihilation of
neutrino-antineutrino pairs emitted from the magnetized torus
\cite{GRB}, the luminosity curve of neutrinos should also be highly
variable. Thus, the light curve of GRBs may be naturally explained
if the central engine is composed of a rotating black hole and a 
magnetized accretion torus of mass $\agt 0.1M_{\odot}$.

\subsection{Future tasks}

This work is the first step toward obtaining a detailed understanding of 
the evolution of dense, hot magnetized tori surrounding a stellar-mass
black hole. There are several tasks left for the future. One is
to take into account more realistic microphysics, incorporating more
sophisticated EOSs. To this time, two tabulated finite-temperature EOSs
have been published for public use. These incorporate the effects of
heavy nuclei as well as free nucleons and $\alpha$-particles.
\cite{LS,shen} In these EOSs, the pressure, internal energy, and other
various quantities are written as functions of the density, electron
fraction, and temperature, like the EOS used in this paper. Thus, it would 
be straightforward to change the numerical code to use more
realistic tabulated EOSs. Numerical results, such as the rest-mass
accretion rate, temperature, and neutrino luminosity, may depend on the
chosen EOSs. To clarify this dependence, we plan to perform
simulations with such EOSs. 

Another task is to improve the scheme for treating neutrino transfer.
As pointed out in Refs. \citen{GRB1,GRB2,GRB3,SRJ} and \citen{LRP} and
reconfirmed in this paper, the luminosity is significantly affected by
neutrino-trapping effects. In this paper, we determined the neutrino
optical depth by simply using the local density. However, the actual
optical depth depends on the density distribution, temperature
profile, neutrino energy, and geometry of the black hole. As shown in
\S \ref{sec:3.2}, the luminosity depends strongly on the chosen
optical depth for large values of $a$ and $M_*$, because such tori are
composed both of optically thick and thin regions.  To provide a
better estimate of the neutrino luminosity, we have to treat the optical
depth more carefully.

Elucidating the exact neutrino trajectories in the curved spacetime is
also an important problem.  Neutrinos are most efficiently emitted in
the vicinity of the black hole.  This implies that a non-negligible
fraction is swallowed by the black hole due to large curvature
effects. We plan to study this effect in the future.  An associated
problem is to estimate the annihilation rate of neutrino and
antineutrino pairs. As mentioned above, this process generates a pair
plasma for GRBs. Although Setiawan et al. already calculated this rate
in their simulation,\cite{SRJ} they did not take into account the
curvature effect. We consider that estimating the energy generation
rate in curved spacetime is an important problem.

The final problem that warrants further study concerns nonaxisymmetric
effects. The development of the MRI in an axisymmetric system could be
different from that in the nonaxisymmetric, 3D case.\cite{hgb95}
Turbulence tends to persist more in the 3D case due to the lack of
symmetry.  Specifically, according to the axisymmetric anti-dynamo
theorem,\cite{moffatt78} the sustained growth of magnetic field energy
is not possible through axisymmetric turbulence, as demonstrated by
numerical simulations.\cite{hb92} In our present simulation, we found
that the neutrino luminosity decreases with a time scale of $\sim 100$
ms, implying that the efficiency of shock heating in the torus
decreases with time. This may be partly due the anti-dynamo
effect. McKinney and Gammie~\cite{MG} have performed
axisymmetric simulations of magnetized tori accreting onto Kerr black
holes and have found good quantitative agreement with the 3D results
of De Villiers and Hawley~\cite{dvhkh05} for the global quantities
$\dot{E}/\dot{M}_*$ and $\dot{J}/\dot{M}_*$ for $t \leq 2000M$. Thus,
the results presented here likely provide (at least) a good
qualitative picture at least for the short-term evolution. However,
to clarify the longer-term evolution, 3D 
simulations will eventually be necessary.  

\section*{Acknowledgements}

The numerical computations were in part performed on the FACOM VPP5000 at
ADAC at NAOJ, on the NEC SX8 at YITP of Kyoto University, and on the
NEC SX6 at ISAS at JAXA. This work was in part supported by
Monbukagakusho Grant (Nos. 17010519, 17030004, 17540232, and 19540263).

\appendix

\section{Derivation of the equation for $T_{\mu\nu}$}

The total energy-momentum tensor, $T^{\rm T}_{\mu\nu}$, is defined 
by the sum of energy momenta of all the matter fields, and 
obeys the conservation equation 
\beqn
\nabla^{\mu} T^{\rm T}_{\mu\nu}=0. 
\eeqn
In the context of the present paper, $T^{\rm T}_{\mu\nu}$ is
composed of the energy momenta of baryons, electrons, positrons,
radiation, and neutrinos. Then, we split the total energy-momentum
tensor as
\beqn
T^{\rm T}_{\mu\nu}=T_{\mu\nu} + T^{\rm N}_{\mu\nu}, 
\eeqn
where $T_{\mu\nu}$ is written by Eq. (\ref{eq2.9}) which
is composed of baryons, electrons, positrons, radiation,
and thermal neutrinos. In this paper, the energy-momentum
tensor of thermal neutrinos are simply written by 
\beqn
\rho \varepsilon_{\nu} u_{\mu} u_{\nu} + P_{\nu} g_{\mu\nu}. 
\eeqn
On the other hand, $T^{\rm N}_{\mu\nu}$ is the contribution from 
free-streaming neutrinos for which the evolution equation is
assumed to be
\beqn
\nabla_{\mu} T^{{\rm N}\mu}_{~~~\nu}=Q_{\nu}. \label{eqaaa}
\eeqn
Then, the equation for $T_{\mu\nu}$ is
\beqn
\nabla_{\mu} T^{\mu}_{~\nu}=-Q_{\nu}. 
\eeqn

In the present simulation, in which the MHD equations are solve in the
fixed background of a black hole, we do not have to determine $T^{\rm
N}_{\mu\nu}$, because such a term never appears in the basic
equations, and thus, we do not solve Eq. (\ref{eqaaa}) in this
paper. In fully general relativistic simulation in which the Einstein
equation is solved, however, it is necessary to solve this equation,
because it appears as the source terms in the equations of the geometric
variables.

\section{Derivation of equation for electron capture rate}

In this appendix, we derive the equations for the
electron capture rate (\ref{ecap1}) and the associated energy
emissivity (\ref{ecap2}). 

\def\bm#1{{\hbox{\boldmath $#1$}}}

The electron capture rate by nuclei, $\lambda_{e^{-}}$,
is in general given by [e.g., Ref. \citen{LMP00}]
\beq
\lambda_{e^{-}} = \frac{\ln 2}{K}
\sum_{i} \frac{(2J_{i}+1)e^{-E_{i}/(kT)}}
              {G(Z,A,T)}
\sum_{j} M_{ij}f_{ij},
\eeq
where the sums over $i$ and $j$ run over states in the parent and
daughter nuclei, respectively. The constant $K\,(=6146\pm 6 [s])$
is defined as
\beq
K= \frac{2\pi^{3}(\ln 2)\hbar^{7}}
{G_{\rm F}V_{\rm ud}^{2}g_{\rm V}^{2}m_{e}^{5}c^{4}},
\eeq
where $G_{\rm F}$ is the Fermi coupling constant, $V_{\rm ud}$ is the
up-down element in the Cabibbo-Kobayashi-Masukawa quark-mixing matrix, 
and $g_{\rm V}=1$ is the weak vector coupling constant. Further,
the quantity 
$G(Z,A,T) = \sum_{k}\exp(-E_{i}/kT)$ is the partition function of the
parent nucleus, 
$M_{ij}$ is the reduced transition probability, and 
$f_{ij}$ is the phase space integral, given by
\beq
f_{ij} = \left(\frac{kT}{m_{e}c^{2}}\right)^{5}
\int_{\eta_{l}}^{\infty}
x^{2}(x+\zeta_{ij})^{2}
F(Z,x)S_{e^{-}}(x)[1-S_{\nu}(x+\zeta_{ij})]dx. \label{eqb3}
\eeq
Here $x$ is total energy of electrons divided by $kT$,
and $\zeta_{ij}$ is the difference between the 
nuclear mass-energies of the ground states
of the parent and daughter nuclei in units of $kT$: 
\beq
\zeta_{ij}=\frac{1}{kT}(M_{\rm par}c^{2}-M_{\rm dau}c^{2}+E_{i}-E_{j}).
\eeq
In this expression, $M_{\rm par}$ and $M_{\rm dau}$ are the nuclear
masses of the parent and daughter nuclei, respectively, and $E_{i}$
and $E_{j}$ are the excitation energies of the initial and final states.
The lower limit of the integral, $\eta_{l}$, is the capture-threshold
total energy in units of $kT$, which is given by
$\eta_{l}=m_{e}c^{2}/kT$ if $\zeta_{ij} + m_{e}c^{2} > 0$ and
$\eta_{l}=|\zeta_{ij}|$ otherwise.  Also, in (\ref{eqb3}), 
$S_{e}$ and $S_{\nu}$ are the
electron and positron distribution functions, which are assumed to be
described by the Fermi-Dirac distributions with (matter) temperature
$T$ and chemical potential $\eta^{F}=\mu/kT$,
\beq
S_{e^{\pm}}(x) =
\frac{1}{e^{x-\eta_{e^{\pm}}^{F}}+1},~~ S_{\nu}(x) =
\frac{1}{e^{x-\eta_{\nu}^{F}}+1},
\eeq
where $\eta_{e^{\pm}}^{F}$ and $\eta_{\nu}^{F}$ are the electron
(positron) and neutrino chemical potentials in units of $kT$. The
quantity $F(Z, x)$ is the Fermi function, which corrects the phase
space integral for the Coulomb distortion of the electron wave
function near the nuclei.

For actual calculations of the capture rate, we follow
a prescription introduced by Fuller, Fowler, and Newman.\cite{FFN80,FFN85} 
In this prescription, the capture rate is rewritten in terms of the
{\it effective f-t value} $\langle ft \rangle$ as
\beq
\lambda_{e^{-}} = \ln 2 \frac{I_{ij}}{\langle ft \rangle},
\eeq
where we have factored out the contribution of the Fermi function from
the phase space integral [i.e.,
$I_{ij} = f_{ij}/ \langle F(Z, x)\rangle $]
and have put this contribution into the effective f-t value.
The reduced phase space factor is then
\beq
I_{ij} = \left(\frac{kT}{m_{e}c^{2}}\right)^{5}
\int_{\eta_{l}}^{\infty}
x^{2}(x+\zeta_{ij})^{2}
\frac{1}{e^{x-\eta_{e^{-}}^{F}}+1}
\left[1-\frac{1}{e^{x+\zeta_{ij}-\eta_{\nu}^{F}}+1} \right]d x.
\eeq
In terms of the relativistic Fermi integrals $F_{k}(\eta)$, the phase
space factor can be written
\beq
I_{ij} = \left( \frac{kT}{m_{e}c^{2}} \right)^{5}
\frac{1}{1-e^{\eta_{\nu}^{F}-\zeta_{ij}-\eta_{e^{-}}^{F}}} I_{e},
\eeq
where
\beqn
I_{e} &\equiv& \int_{\eta^{L}}^{\infty}
x^{2}(x + \zeta_{ij})^{2}
\left[ \frac{1}{1+e^{x + \eta_{e^{-}}^{F}}}
- \frac{1}{1+e^{x + \zeta_{ij}-\eta_{\nu}^{F}}} \right] dx \nonumber \\
 &=& F_{4}(\eta_{e^{-}}^{F}-\eta^{L})
   - F_{4}(\eta_{\nu}^{F}-\zeta_{ij}-\eta^{L}) \nonumber \\
&&~~~~~~~~~~~~~~~~~~~~~~~~~~~~~~+ (2\zeta_{ij}+4\eta^{L})
\left[ F_{3}(\eta_{e^{-}}^{F}-\eta^{L})
-F_{3}(\eta_{\nu}^{F}-\zeta_{ij}-\eta^{L}) \right] \nonumber \\
&& + [6(\eta^{L})^{2}+6\eta^{L}\zeta_{ij}+\zeta_{ij}^{2}]
\left[ F_{2}(\eta_{e^{-}}^{F}-\eta^{L})
-F_{2}(\eta_{\nu}^{F}-\zeta_{ij}-\eta^{L}) \right] \nonumber \\
&& + [4(\eta^{L})^{3}+6(\eta^{L})^{2}\zeta_{ij}
+2\eta^{L}(\zeta_{ij}^{2})^{2}]
\left[ F_{1}(\eta_{e^{-}}^{F}-\eta^{L})
-F_{1}(\eta_{\nu}^{F}-\zeta_{ij}-\eta^{L}) \right] \nonumber \\
&& + [(\eta^{L})^{4}+2(\eta^{L})^{3}\zeta_{ij}
+(\eta^{L})^{2}(\zeta_{ij}^{2})^{2}]
\left[ F_{0}(\eta_{e^{-}}^{F}-\eta^{L})
-F_{0}(\eta_{\nu}^{F}-\zeta_{ij}-\eta^{L}) \right].
\eeqn

For electron capture by free nucleons $(p+e^{-} \rightarrow n+\nu_e$,
$n+e^{+} \rightarrow p+\bar{\nu}_e)$,
the effective f-t value is $\langle ft \rangle = 10^{3.035}$ s,\cite{FFN85}
and the difference between nuclear mass-energies is
\beq
\zeta_{ij}=\frac{1}{kT}(m_{p}c^{2}-m_{n}c^{2}+E_{p}-E_{n}).
\eeq
In this paper we have assumed that the nuclear mass-energy difference as
well as electron mass is much smaller than the total energy of
relativistic electrons. Accordingly, the phase space factor is
simplified to 
\beq
I_{ij} = \left( \frac{kT}{m_{e}c^{2}} \right)^{5}
\frac{1}{1-e^{\eta_{\nu}^{F}-\zeta_{ij}-\eta_{e^{-}}^{F}}}
\left[
  F_{4}(\eta_{e^{-}}^{F}-\eta^{L}) -
  F_{4}(\eta_{\nu}^{F}-\zeta_{ij}-\eta^{L})
\right].
\eeq
In the region where neutrinos freely stream out, the phase
space factor is further simplified to 
\beq
I_{ij} = \left( \frac{kT}{m_{e}c^{2}} \right)^{5}
  F_{4}(\eta_{e^{-}}^{F}).
\eeq
The electron capture rate by free protons is given by
\beq
\lambda_{e^{-}} = K_c \left( \frac{kT}{m_{e}c^{2}} \right)^{5}
  F_{4}(\eta_{e^{-}}^{F}). \label{apeq2}
\eeq
Similarly, the positron capture rate is 
\beq
\lambda_{e^{+}} = K_c \left( \frac{kT}{m_{e}c^{2}} \right)^{5}
  F_{4}(\eta_{e^{+}}^{F}). \label{apeq3}
\eeq

The energy emission rates of neutrinos in units of $m_{e}c^{2}$ associated
with the electron capture by free nucleons are given by
\beq
\pi_{e^{\pm}} = \ln 2
\frac{J_{ij}}{\langle ft \rangle},
\eeq
where
\beq
J_{ij} = \left(\frac{kT}{m_{e}c^{2}}\right)^{6}
\int_{\eta_{l}}^{\infty}
x^{2}(x+\zeta_{ij})^{3}
\frac{1}{e^{x-\eta_{e^{\pm}}^{F}}+1}
\left[1-\frac{1}{e^{x+\zeta_{ij}-\eta_{\nu}^{F}}+1} \right]dx.
\eeq
Similar calculations to derive equations (\ref{apeq2}) and
(\ref{apeq3}) give
\beqn
\pi_{e^{\pm }} &=& K_c \left( \frac{kT}{m_{e}c^{2}} \right)^{6}
  F_{5}(\eta_{e^{\pm}}^{F}), 
\eeqn

To avoid the direct numerical integration of the relativistic Fermi
integrals $F_{j}(\eta)$, we adopt approximate expressions for them
following Refs. \citen{BvR78} and \citen{FFN85}.
For $j=0$, the relativistic Fermi integrals can be integrated exactly, 
and we obtain $F_{0}(\eta)= \ln(1+e^{\eta}) = \eta + \ln(1+e^{-\eta})$,
which immediately gives
\beq
F_{0}(\eta) - F_{0}(-\eta) = \eta .
\eeq
Then, using the well-known recursion relations
\beqn
F_{j}(\eta) &=& F_{j}(0) + j\int_{0}^{\eta}F_{j-1}(x)dx, \nonumber \\
F_{-j}(\eta) &=& F_{j}(0) - j\int_{0}^{\eta}F_{j-1}(-x)dx,
\eeqn
and noting that $F_{j}(0) = (j!)(1-2^{-j})\zeta(j+1)$, where $\zeta(x)$
is the zeta function, we find
\beqn
F_{4}(\eta) - F_{4}(-\eta) &=&
\frac{7\pi^{4}}{15}\eta + \frac{2\pi^{2}}{3}\eta^{3} +
\frac{\eta^{5}}{5},
\nonumber \\
F_{5}(\eta) + F_{5}(-\eta) &=&
\frac{31\pi^{6}}{126} + \frac{7\pi^{2}}{6}\eta^{2} +
\frac{5\pi^{2}}{6}\eta^{4} + \frac{\eta^{6}}{6}. \label{apeq1}
\eeqn
The standard expansion of the relativistic Fermi integrals
for $\eta\le 0$ is given by (e.g., Ref. \citen{CG68})
\beq
F_{j}(\eta) = (j!)e^{\eta}
\sum_{l=0}^{\infty}\frac{(-1)^{l}e^{l\eta}}{(l+1)^{j+1}}.
\eeq
We adopt the first three terms for our approximation of the
relativistic Fermi integrals for $\eta \le 0$:
\beqn
F_{4}(\eta) &\approx& 24\left[
e^{\eta}-\frac{e^{2\eta}}{2^{5}}+\frac{e^{3\eta}}{3^{5}}
\right], \\
F_{5}(\eta) &\approx& 120\left[
e^{\eta}-\frac{e^{2\eta}}{2^{6}}+\frac{e^{3\eta}}{3^{6}}
\right].
\eeqn
Our approximation of the relativistic Fermi integrals for $\eta > 0$
is obtained as follows by using relations (\ref{apeq1}) and imposing
continuity conditions on the values and their first derivatives:
\beqn
F_{4}(\eta) &\approx&
45.59\eta + \frac{2\pi^{2}}{3}\eta^{3} +
\frac{\eta^{5}}{5}
 + F_{4}(-\eta)
\nonumber \\
F_{5}(\eta) &\approx&
236.65 + \frac{7\pi^{2}}{6}\eta^{2} +
\frac{5\pi^{2}}{6}\eta^{4} + \frac{\eta^{6}}{6}
 - F_{5}(-\eta).
\eeqn

\end{document}